\documentclass[10pt,a4paper]{article}
\usepackage{amsmath,amssymb,amsfonts,amsthm}
\usepackage{graphicx,xcolor,float,textcomp,mathtools,booktabs,array,tabularx,calc,microtype,placeins,etoolbox,xurl}
\newtheorem{theorem}{Theorem}
\newtheorem{proposition}[theorem]{Proposition}
\newtheorem{lemma}{Lemma}
\newtheorem{corollary}[theorem]{Corollary}

\usepackage[margin=25mm,headheight=12pt,headsep=8mm]{geometry}
\usepackage[numbers,sort&compress]{natbib}
\usepackage[hidelinks]{hyperref}
\hypersetup{pdftitle={Conditional stopping guarantees for a folded parametrization of reduced Stokes vectors},pdfauthor={Hiroyasu Usami, Ayato Tsuboi, Keisuke Hara}}
\begin{document}
\raggedright
\begin{center}
{\LARGE\bfseries Conditional stopping guarantees for a folded parametrization of reduced Stokes vectors\par}
\vspace{1em}
{\large Hiroyasu Usami\textsuperscript{1,*}, Ayato Tsuboi\textsuperscript{2}, and Keisuke Hara\textsuperscript{2}\par}
\vspace{0.6em}
{\small\textsuperscript{1}Department of Computer Science, College of Engineering, Chubu University, Kasugai, Aichi 487-8501, Japan\par}
{\small\textsuperscript{2}Department of Computer Science, Graduate School of Engineering, Chubu University, Kasugai, Aichi 487-8501, Japan\par}
\vspace{0.4em}
{\small\textsuperscript{*}Corresponding author: \href{mailto:usami@fsc.chubu.ac.jp}{usami@fsc.chubu.ac.jp}\par}
\end{center}
\begin{abstract}
Stopping decisions made in parametrization coordinates need not certify physically accurate estimates. We establish conditional stopping guarantees for a folded parametrization of reduced Stokes vectors, including partial linear polarization. For strongly convex physical objectives on the bounded-intensity inner sheet, normalized first derivatives and one radial lower-curvature test give an explicit objective-gap bound with sharp tolerance orders $\varepsilon^{4/3}+\nu^2$. The guarantee requires neither prior localization near an optimum nor a positive-semidefinite head Hessian. Its information basis is an exact recovery identity: one signed radial curvature restores the physical-gradient component lost at the fully linearly polarized fold, without stationarity or knowledge of the physical Hessian. For a nominal quadratic model, an outward-rounded implementation certifies a gap of at most $10^{-6}$ for 8,861 of 9,284 stored endpoints; it rejects all 295 historical false successes and 128 accurate endpoints. The full 128-fractional-bit pass takes 7.32 seconds and includes 52 scalar curvature evaluations. Controlled fits and released observations distinguish stopping certification from local fold diagnosis and subsequent coordinate repair. Twenty shared-network fits contain no saved near-fold negative-curvature candidates, and direct physical solvers are cheaper on the tested quadratics.
\end{abstract}

\section{Introduction}
\label{sec:intro}
An optimizer can report convergence while a physically constrained polarization estimate remains inaccurate. We ask when derivatives in a specified smooth output parametrization provide a sufficient certificate for the physical objective gap. Our main result is a conditional stopping guarantee, supported by explicit physical-gradient recovery identities and numerical certification for a nominal quadratic model.

We study the reduced Stokes vector $s=(I,Q,U)$, where $I$ is total intensity and $Q,U$ describe linear polarization. The circular-polarization component $V$ is outside the model and is not estimated. Feasible reduced states satisfy $I\geq\sqrt{Q^2+U^2}$, so the model includes partial polarization. For $I>0$, the degree of linear polarization is $d=\sqrt{Q^2+U^2}/I\in[0,1]$. The rational Stokes parametrization studied here folds at the fully linearly polarized boundary $d=1$; it also represents partially polarized states in the interior.

Physicality-constrained Stokes estimation enforces an admissible polarization state \cite{hu,spc,psara}. Polarization also supplies cues for learned shape and inverse rendering \cite{ba,pandora}, with acquisition noise requiring separate evaluation \cite{polarns}. Smooth output maps offer one way to enforce constraints in differentiable models. Optimization layers \cite{optnet,cvxlayers} and constrained networks \cite{hardnet} make the differential information carried by such maps a relevant design question.

Theorem~\ref{thm:uniformstopping} gives a uniform stopping bound for strongly convex physical objectives. On the inner sheet with bounded intensity, normalized first derivatives and a lower test on one radial curvature bound the gap by an explicit expression of order $\varepsilon^{4/3}+\nu^2$. The result holds throughout this domain, including points far from the optimum. Sharp examples, derivative-error budgets, and counterexamples specify its accuracy and scope. It builds on quantitative second-order transfer for squared-variable formulations \cite{dingwright} and the degenerate behavior described by known Kurdyka--\L{}ojasiewicz (KL) exponents for Hadamard models \cite{ouyang}. The Stokes specialization supplies explicit constants and uses a single curvature direction. Figure~\ref{fig:stopping_certificate} connects this guarantee to the conservatism and coverage of the numerical stopping bounds.

The recovery identities provide the information underlying this stopping analysis. At the fold, first derivatives leave one physical-gradient component undetermined. One signed radial second derivative recovers that component for any $C^2$ objective, without stationarity or knowledge of the physical Hessian (Proposition~\ref{prop:recovery}). Figure~\ref{fig:principle} exhibits objectives with the same value, first head derivatives, and physical Hessian but different physical gradients. Away from the fold, complete first derivatives identify the gradient, but their radial inverse amplifies errors. Propositions~\ref{prop:mixed} and \ref{prop:localcurvature} correct second-order observations using, respectively, one physical-curvature scalar or one Hessian action. Their error formulas distinguish radial and tangential input gains. Corollary~\ref{cor:gap} converts recovered gradients into a quadratic gap bound, with an exact tightness condition and the slack required at interior outputs.

The rational ball map and its first derivative are established \cite[Sec.~V-B, (78)--(79)]{gcopter}. Smooth-lift theory describes how critical-point properties transfer through a parametrization \cite{lifts}; Hadamard parametrizations provide related stationary-point and strict-saddle correspondences \cite{hadamard}. Within this framework, our inverse-information question concerns which signed derivative values determine the physical gradient, their sensitivity, and the stopping implications of finite tolerances. Supplementary Table~\ref{SI-tab:priorart} relates the statements.

The numerical evaluation follows this information-to-certificate structure. Frozen endpoints and outward interval arithmetic assess the stopping bounds. Nonquadratic checks and a common-graph automatic differentiation (AD) experiment test recovery and acquisition cost. Controlled two-frame fits, Fresnel-reflection sequences and released observations then compare local fold diagnosis and coordinate repair with cold starts, physical stopping bounds and direct cone solutions. The shared-network experiment supplies a fixed check of near-fold curvature at stored checkpoints.

\section{Results}
\label{sec:results}

\subsection{Conditional stopping and numerical certification}
\label{sec:certificate_results}
Theorem~\ref{thm:uniformstopping} bounds the gap $\Delta(s)$ to the best physically feasible objective value using normalized first derivatives and one radial lower-curvature test:
\[
 \Delta(s)\leq\frac{52(\overline I L)^{2/3}}{\mu}\varepsilon^{4/3}
             +\frac{3}{\mu}\nu^2.
\]
The physical objective is $C^2$ and $\mu$-strongly convex. The evaluated coordinates satisfy $0<I\leq\overline I$ and $0\leq r\leq1$; $\varepsilon\geq0$ bounds the normalized first derivatives. For $r\geq1/2$, the physical radial curvature is at most $L\geq\mu$, and $\nu\geq0$ bounds $\max\{0,-c_r/I\}$. The small-tolerance condition $\varepsilon\leq\overline I L/8$ applies throughout the domain. No radial-curvature observation is required at $r<1/2$. The coefficient 52 is conservative; the two tolerance orders are sharp. This is a sufficient endpoint test, without a prescribed neighborhood of an optimum. Methods gives the exact conditions, error-budget inflation and proof.

\begin{figure*}[!t]
\centering\includegraphics[width=\textwidth]{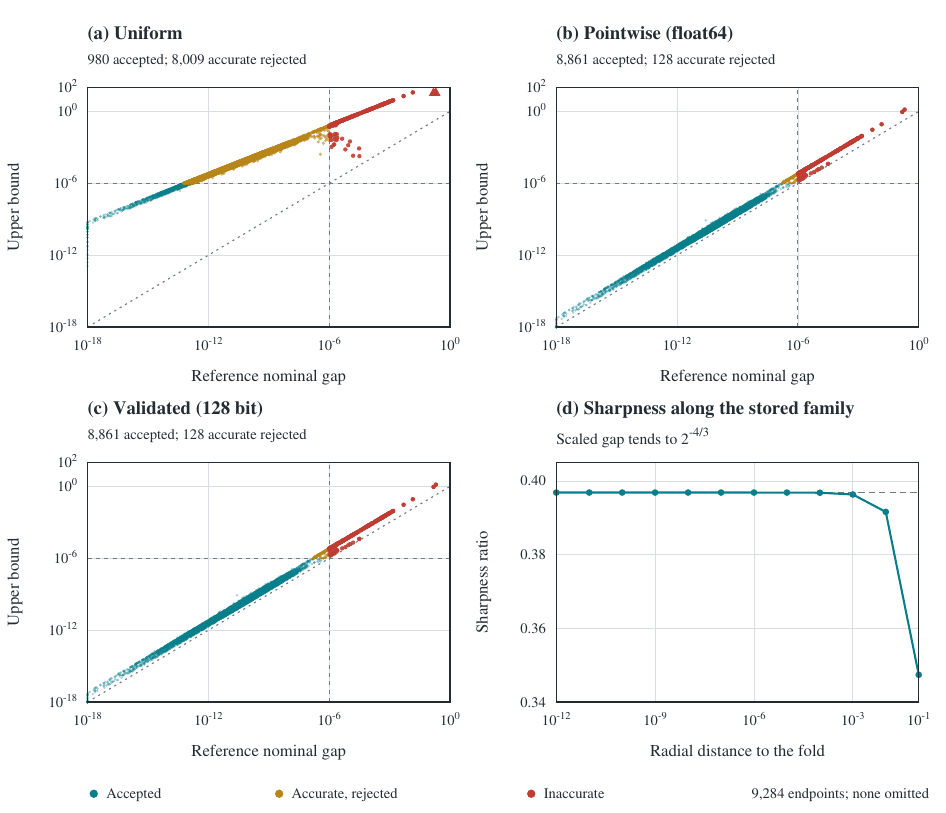}
\caption{Conditional stopping bounds on the same 9\,284 frozen warm-start endpoints. (a) Uniform bound; (b) float64 pointwise bound; (c) outward pointwise bound at 128 fractional bits. The horizontal and vertical dashed lines mark $10^{-6}$; the dotted diagonal is equality. The reference nominal gap is the midpoint of an independently computed 256-bit enclosure. Colors separate accepted states, accurate states rejected by the bound, and inaccurate states; no endpoint is omitted. For display, 35 reference gaps and 21 bounds in each of (b,c) below $10^{-18}$ are placed at that floor. Four unavailable uniform bounds are marked by triangles at the top edge. (d) The ratio $\Delta/\varepsilon^{4/3}$ for the 12 stored sharpness examples approaches $2^{-4/3}$ as the radial offset $1-r$ vanishes; this is a separate deterministic family. All three bounds reject every one of the 295 inaccurate endpoints.}
\label{fig:stopping_certificate}
\end{figure*}

Figure~\ref{fig:stopping_certificate} compares this uniform bound with the tighter pointwise expression and its outward interval evaluation on the same 9\,284 stored warm-start endpoints. The uniform bound accepts 980 endpoints at gap threshold $10^{-6}$; the pointwise and 128-fractional-bit interval bounds accept 8\,861. The interval test rejects all 295 historical false successes and also 128 accurate endpoints. Outer-sheet endpoints are represented by paired inner coordinates with the same physical output before derivatives are evaluated. This derivative-based certificate is distinct from the direct physical-gradient bound used in the repair studies.

The nominal objective is the ideal four-analyzer least-squares loss, $\tfrac12\|As-y\|_2^2$, evaluated per observation. Data retain their specified units: stored normalized DeepSfP values, PolarNS means divided by 4\,096, and burst PNG values divided by 65\,535, without per-pixel renormalization. Consequently, $10^{-6}$ is an absolute optimization-gap target for this scaled model, not a universal imaging-accuracy threshold or a bound on error to noise-free Stokes truth.

At 128 fractional bits, the complete outward pass takes 7.32 s, including 52 scalar curvature evaluations. Decisions agree at 64--256 bits. At 32 bits, three additional accurate endpoints are inconclusive. Supplementary Sections S20--S21 provide the reference calculations, inclusion arithmetic, stress checks and replay instructions.

\subsection{Recovering information lost at full linear polarization}
\begin{figure*}[!t]
\centering\includegraphics[width=\textwidth]{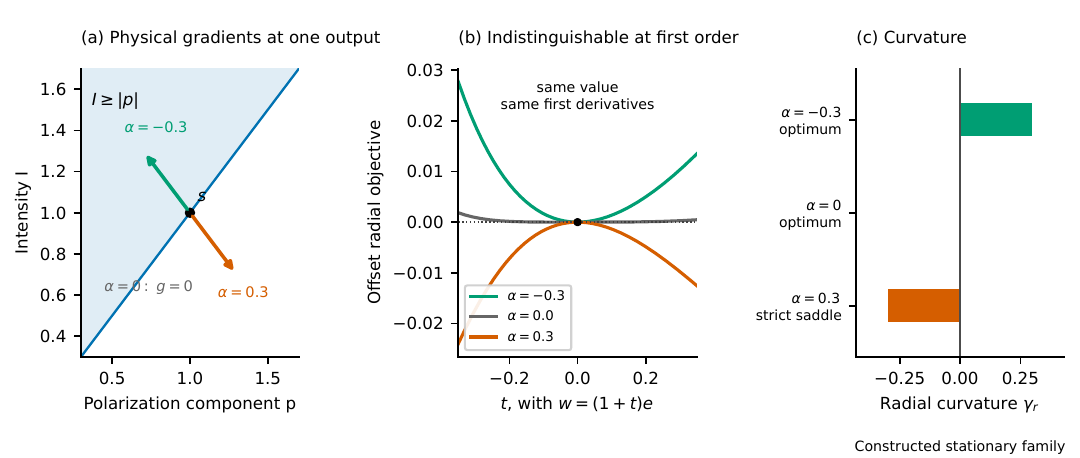}
\caption{Identical first-order information, different physical gradients. Fix $I=1$, $e=(1,0)$, $s=(1,e)$, the ideal four-analyzer metric $W=\operatorname{diag}(1,1/2,1/2)$, and $z_\alpha=s-\alpha W^{-1}n$, $n=(-1,e)$ for $\alpha=-0.3,0,0.3$. The objectives are $f_\alpha(u)=\tfrac12\|u-z_\alpha\|_W^2+c_\alpha$, with $c_\alpha$ chosen so that $f_\alpha(s)=0$. (a) The physical gradients $g_\alpha=\alpha n$ differ at the same output. (b) The objectives have identical value, zero first head derivatives, and the same physical Hessian. Their radial slices separate at second order. (c) In this stationary family, $\gamma_r=c_r=-I\alpha$ distinguishes two constrained optima from a strict saddle. The general gradient-recovery identity also holds without stationarity.}
\label{fig:principle}
\end{figure*}

The rational Stokes map reaches the boundary of the physical polarization cone at a finite raw radius. At that fold, different physical gradients can produce the same first head derivatives. One signed radial second derivative recovers the missing component for any twice continuously differentiable objective, without stationarity or knowledge of its physical Hessian (Proposition~\ref{prop:recovery}). Figure~\ref{fig:principle} exhibits the ambiguity and its resolution. The observation sets and required physical-curvature inputs are kept distinct in Table~\ref{tab:contract}; the exact coordinates, assumptions, and proofs are given in Methods.

\begin{table*}[!t]
\caption{Observations and supplied curvature. Recovery uses any $C^2$ objective with known geometry, $I>0$, $r>0$. The gap row additionally assumes a fixed calibrated quadratic with $W\succ0$.}
\label{tab:contract}
\centering\small\renewcommand{\arraystretch}{1.13}
\begin{tabularx}{\textwidth}{@{}p{.18\textwidth}p{.22\textwidth}p{.24\textwidth}X@{}}\toprule
Route & Head observations & Physical-curvature input & Result and domain\\\midrule
Fold scalar & $b,j,c_r$ & None & Full gradient, \eqref{eq:scalarrecovery}; $r=1$\\
First-order inverse & $b_r,j_r$ & None & Full gradient, \eqref{eq:firstinverse}; $r\ne1$\\
Mixed scalar & $b_r,P_\perp j_r,c_r$ & Scalar $h_{rr}=e^{\mathsf T}H_{pp}e$ & \eqref{eq:mixedrecovery}; $r\ne\sqrt3$; curvature cancels at $r=1$\\
Vector HVP & $b_r,v_r$ & Action $h_p=H_{pp}e$ & \eqref{eq:correctedrecovery}; $r\ne\sqrt3$; curvature cancels at $r=1$\\
Quadratic gap & Recovered gradient & Fixed metric $W$ & Corollary~\ref{cor:gap}; supporting normal and interior slack\\\bottomrule
\end{tabularx}
\end{table*}

Table~\ref{tab:contract} specifies chosen local observation models: the question is what the listed head derivatives establish about physical optimality. For a calibrated quadratic, the polarization Hessian block $H_{pp}=W_{pp}$ is fixed by the instrument and weighting, independently of the observations, so it supplies the stated curvature inputs without supplying the gradient. At a stationary fold, nonnegative radial curvature characterizes the constrained optimum; negative curvature identifies a strict saddle. The experiments expose the full objective and observations and include direct physical gradients as controls. Information sufficiency and computational cost are evaluated separately.

\subsection{Recovery sensitivity near the fold}
\begin{figure*}[!t]
\centering\includegraphics[width=.94\textwidth]{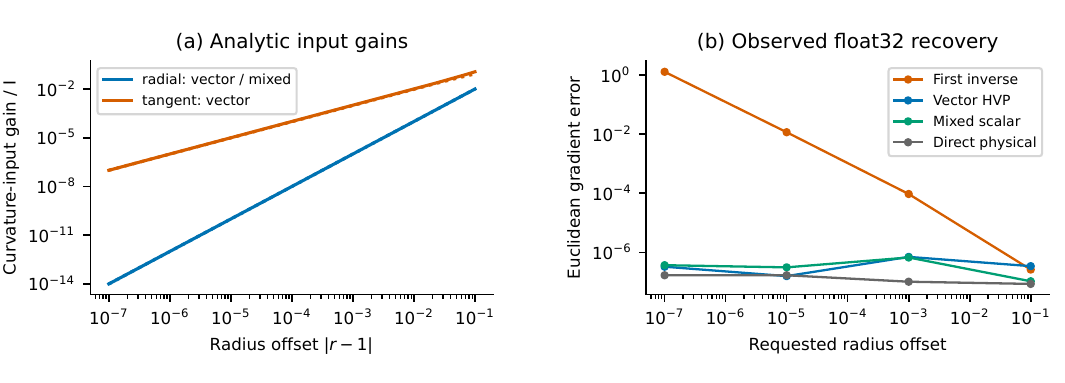}
\caption{Theory and observed finite-precision error. (a) Vector-HVP curvature-input gains, normalized by intensity, on the inner and outer sheets. Mixed recovery uses only the radial coefficient. (b) Observed float32 gradient-recovery errors from the common-graph AD experiment at requested $I=1.2$, maximized over the two sheets at each requested radius offset. References are 80-digit physical gradients at the represented Stokes states. These are measured errors, not deterministic derivative-error envelopes. Prescribed-envelope experiments and the support-slack counterexample are specified in Supplementary Sections S4 and S8--S9.}
\label{fig:sensitivity}
\end{figure*}

Complete first derivatives determine the physical gradient away from the fold, although radial inversion becomes sensitive near it. The mixed-scalar and vector-Hessian routes use different physical-curvature inputs and have different error gains (Propositions~\ref{prop:mixed} and \ref{prop:localcurvature}). The measured finite-precision errors in Figure~\ref{fig:sensitivity} are distinguished from the deterministic input-error bounds.

\subsection{Local diagnostics, controls and subsequent repair}
\label{sec:numerics}
Supplementary Sections S8--S13 document scalar, vector, and mixed recovery checks, controlled input errors, and 274 quadratic primal-dual records. They include three nonquadratic $C^2$ objectives and tests of the local-curvature correction. The main experiments evaluate finite-precision AD and shared-network fitting.

\subsubsection{Acquisition cost and finite-precision recovery}
We also evaluate all three routes and direct physical differentiation on the same eager PyTorch graph \cite{pytorch}. The objective is the quartic in Supplementary Section S9. The test uses one CPU thread on an Apple M4 Max, PyTorch 2.14.1, float32/float64, intensities $0.02,1.2$, and nine radii from $0.9$ to $1.1$, including offsets down to $10^{-7}$. Eighty-digit physical gradients at the represented Stokes states provide the error reference.

Supplementary Table~\ref{SI-tab:adcost} includes the cost of acquiring physical curvature in the corrected routes; Error32 and Error64 denote float32 and float64 gradient errors. This implementation obtains scalar curvatures by contracting vector HVPs. The largest float32 first-inverse error is $1.27$ at requested $I=1.2$, $r=0.9999999$ (represented radius $0.9999998808$), while the maximum vector and mixed errors are $7.07\times10^{-7}$ and $6.75\times10^{-7}$. These errors quantify how accurately the physical gradient is recovered from rounded derivatives. Direct differentiation has the smallest time and maximum-error entries in the table. The mixed route reduces the required curvature input to one scalar; these measurements distinguish that reduction from AD acquisition cost. Supplementary Section S14 gives timings on both sheets and at the fold, plus the warmup and batch protocol.

\subsubsection{Shared-network fitting}
Twenty fixed shared-MLP fits contain no near-fold negative-curvature candidates at the saved checkpoints. The full architecture, training settings, saved-checkpoint diagnostics and constructed positive control are in Supplementary Sections S15 and S23 (Fig.~\ref{SI-fig:training}).

\subsubsection{Same-solver two-frame continuation}
\label{sec:warmexperiment}
We use 60 stored conditions: five noise seeds, three polarization angles $0,0.7,1.3$, two analyzer-noise standard deviations $0.001,0.01$, and next-frame DoLP $0.8,0.99$. The previous true DoLP is one and both true intensities are one. The same seed's noise is reused across conditions. These are dependent controlled conditions, not 60 independent scenes. Each solver estimates the previous frame from $(a,w)=(0,0.2,-0.1)$ with $I=\exp(a)$, then retains its raw coordinates for the next fit.

Before execution we fixed three angle changes $0,0.05,0.2$ radians, three solvers (BFGS, L-BFGS-B, and an exact-Hessian trust-region method), and gradient tolerances $10^{-4},10^{-6},10^{-8}$. This gives 1,620 continuations using the stored noise. A separate PyTorch LBFGS control uses genuine float32 and float64 tensors at tolerance $10^{-6}$, giving 360 continuations. A false success is solver acceptance with cone objective gap exceeding $10^{-6}$. PyTorch has no success flag, so its acceptance here means a finite native gradient infinity norm at most $10^{-6}$. Supplementary Section S17 specifies solver options and keeps nonacceptance separate from objective error.

We compare one unconditional radial restart with the same restart gated by $|r-1|\leq10^{-3}$ and $c_r/I^2<-10^{-7}$. The restart selects the lowest-loss point among offsets $\pm0.01$ and the current point, then reruns the same solver. A direct physical-gradient evaluation of Corollary~\ref{cor:gap}, including support slack, supplies a separate sufficient stopping test with bound threshold $10^{-6}$. A direct cone solution provides the reference objective.

At tolerance $10^{-6}$, BFGS reproduces all 12 preliminary false-success condition IDs and the separate failure. L-BFGS-B and float64 PyTorch LBFGS also produce false successes (Supplementary Table~\ref{SI-tab:warm}). The trust-region arm has none. Changing angle by $0.05$ removes the SciPy false successes at this tolerance, while float64 PyTorch LBFGS retains one; changing by $0.2$ gives none in these arms. The float32 previous-frame fits have 22 nonacceptances among the 60 row references, retained in the comparison.

At zero angle change and tolerance $10^{-6}$, triggering detects all 28 SciPy false successes and no accurate accepted fit. BFGS uses 280 additional function/gradient pairs after triggering, against 479 after unconditional restarting; L-BFGS-B uses 320 against 496. Applying the radius test before curvature would require 17 and 21 full Hessian queries, respectively, plus endpoint diagnostic evaluations. For coverage analysis, the saved run evaluates curvature at every endpoint; Supplementary Section S17 separates that logging work from the required gated queries. Unconditional restarting also repairs the nonfold BFGS failure, which the curvature gate misses.

\begin{figure*}[!t]
\centering\includegraphics[width=\textwidth]{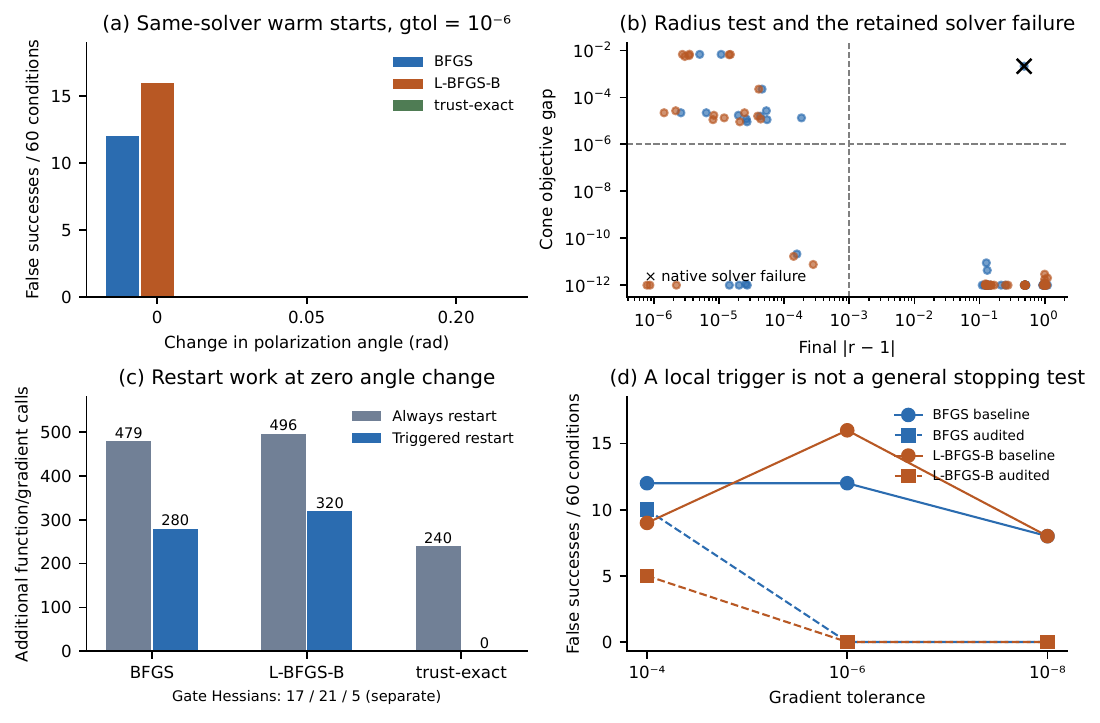}
\caption{Controlled two-frame stopping audit. The same 60 stored conditions are reused throughout. (a) False successes with tolerance $10^{-6}$. (b) Zero-angle-change endpoints for BFGS and L-BFGS-B; dashed lines mark the radius and gap thresholds. The cross is the nonfold BFGS failure. Displayed gaps below $10^{-12}$ and radius distances below $10^{-10}$ are clipped. (c) Additional restart function/gradient calls, including three trial pairs per restart. The separate Hessian counts are queries required after a radius guard; coverage logging evaluates every endpoint. (d) Zero-angle-change false-success counts across the fixed tolerances. Nonacceptances and nonfinite endpoints are retained in the records and are not plotted as false successes.}
\label{fig:warmaudit}
\end{figure*}

At tolerance $10^{-4}$, the 540 SciPy continuations contain 68 false successes; triggering detects 14 and leaves 58 after restart, whereas unconditional restarting leaves 32. At zero angle change and tolerance $10^{-8}$, BFGS and L-BFGS-B each have eight false successes, reduced to zero by triggered restarting (Fig.~\ref{fig:warmaudit}(d)). Separately, BFGS has seven native failures, including six nonfinite terminal losses from exponential-intensity overflow; triggering retains all seven. Native failures and false successes are disjoint categories. Direct physical bounds falsely accept no inaccurate endpoint among the 1,620 SciPy continuations, and reject no endpoint whose finite gap is at most $10^{-6}$.

\subsubsection{A noisy physical-reflection sequence}
\label{sec:fresnel}
We generate a rocking air--dielectric interface from the Fresnel coefficients, with refractive index $1.5$ and incidence angle $\arctan(1.5)+18^{\circ}\cos(2\pi t/40)$ for $t=0,\ldots,40$. Unpolarized incident light gives normalized specular intensity one at Brewster's angle. Two polarization-orientation trajectories, three constant unpolarized background levels, two photon scales, and two seeds give 24 sequences and 984 frames per arm. Four ideal analyzer observations have independent Poisson shot noise and Gaussian read noise of standard deviation two electrons. We fit these observations with unweighted least squares.

Every solver/arm starts from $(0,0.2,-0.1)$ and inherits its own preceding raw endpoint. We use the existing $10^{-6}$ gradient tolerance, gap threshold, and local curvature gate, with at most one restart per frame. BFGS and L-BFGS-B report nine and seven inaccurate successes. Both curvature-triggered and physically audited branches have zero such endpoints, with seven interventions per solver; their corrections propagate through subsequent frames. On the uncorrected branches, the curvature gate detects all 16 false successes and flags no accurate accepted endpoint. The direct physical-gradient bound also rejects all 16. Both solvers have zero native failures in this grid. The simulation exhibits the fold mechanism under a specified physical noise model; direct physical auditing detects the same inaccurate endpoints. Supplementary Section S18 gives the complete protocol and alternative-map control.

\begin{figure*}[!t]
\centering\includegraphics[width=.96\textwidth]{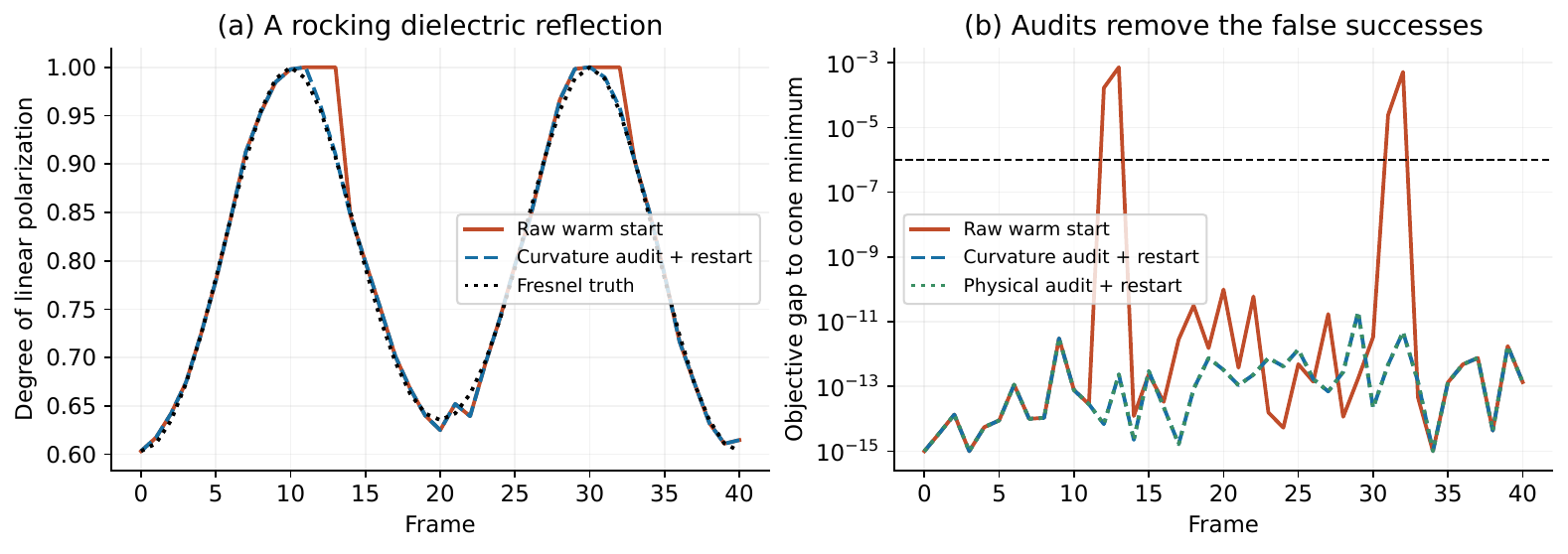}
\caption{Fresnel-reflection continuation with shot and read noise. The displayed BFGS sequence is selected by the largest baseline objective gap in the fixed complete grid, rather than as a typical sequence. (a) Polarization estimate and physical truth. (b) Gap to the direct cone optimum; the dashed line is $10^{-6}$. Both audited branches overlap to plotting accuracy. Gaps below $10^{-15}$ are displayed at that floor. The complete grid has 984 frames per arm, with nine BFGS and seven L-BFGS-B false successes, versus zero in either audited branch.}
\label{fig:fresnel}
\end{figure*}

\subsubsection{Released polarization observations}
\label{sec:real}
We freeze small subsets of DeepSfP \cite{ba} and PolarNS/PolarBurstSR \cite{polarns} before optimization. DeepSfP contributes three indoor test objects and their four-angle observations. PolarNS contributes the lexicographically first object and scene mean/variance records, averaged over 10\,000 and 1\,017 captures. Those means are not treated as individual noisy frames. The first three released PolarBurstSR test crops contribute all 14 raw frames and four nominal color planes, unpacked with the authors' sensor layout.

For each static image, we select 128 seeded uniform eligible pixels and 64 pixels from an observed-high-DoLP stratum. For each burst color plane, a fixed $4\times4$ sensor grid is supplemented by eight high-DoLP positions selected using only the first frame. Finite, nonsaturated observations above a fixed relative intensity floor are eligible. The selections are reported separately in Supplementary Section S18; overlapping inputs and within-image/frame dependencies preclude a prevalence interpretation. Static samples follow a prescribed spatial scan. Burst samples follow fixed sensor coordinates without motion compensation, and are not asserted to follow one material point.

The reference is the cone optimum of the nominal four-analyzer quadratic. Both solvers use $10^{-6}$ tolerance. Each arm carries its own state; a cold-start control resets intensity from the current observation and uses fixed polarization raw coordinates. The fixed protocol yields 576 DeepSfP, 384 PolarNS-mean, and 3\,682 burst entries per solver/arm, totaling 37\,136 fits across two solvers and four arms. The 9\,284 uncorrected warm-start fits form one of these four arms and supply the historical endpoint audit.

\begin{figure*}[!t]
\centering\includegraphics[width=.96\textwidth]{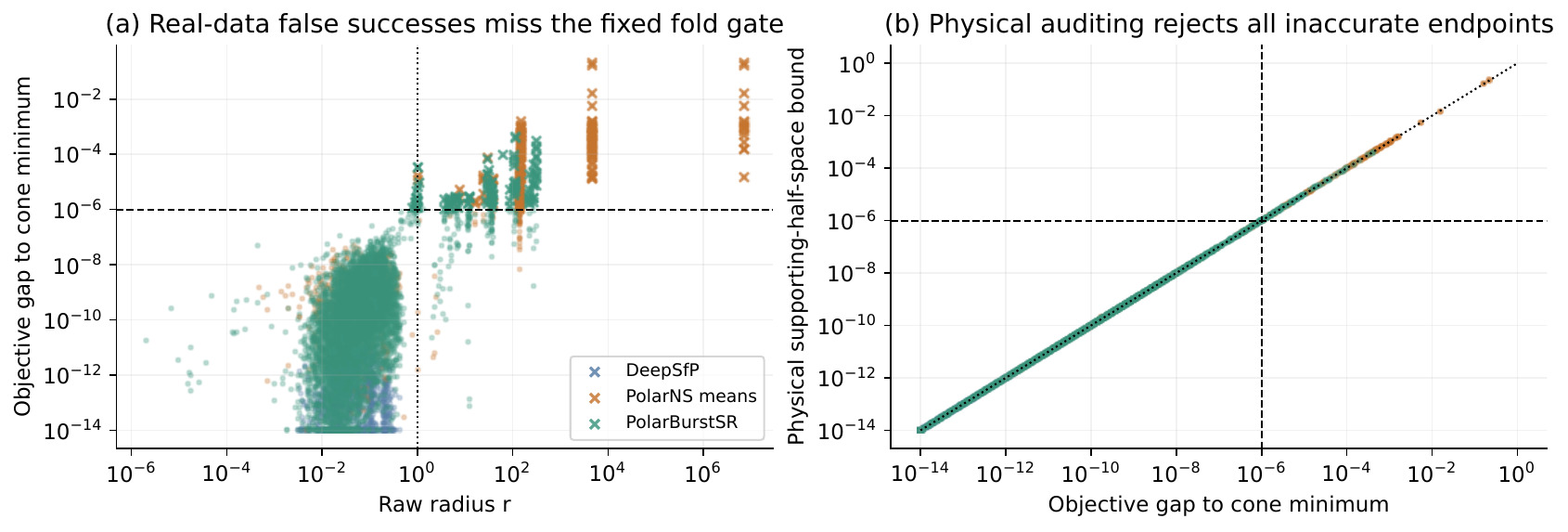}
\caption{All 9\,284 uncorrected real-data warm-start fits across two solvers. The subsets include uniform and enriched selections with repeated observations across distinct histories. (a) Objective gaps against raw radius. Crosses mark native false successes; the dotted line marks $r=1$ and the dashed line marks gap $10^{-6}$. None passes the fixed local fold gate. (b) Physical supporting-half-space bound against the reference gap; the dotted line is equality and dashed lines mark $10^{-6}$. The bound rejects all inaccurate endpoints. Values below $10^{-14}$ are displayed at that floor. The comparisons use float64 arithmetic, without outward-rounded certification.}
\label{fig:realregimes}
\end{figure*}

Supplementary Table~\ref{SI-tab:realdata} shows inaccurate native successes in the PolarNS and burst subsets, while DeepSfP has none. None of the 295 uncorrected warm-start false successes satisfies the predeclared local fold gate, so curvature-triggered restarting leaves those counts unchanged. The direct physical-gradient bound rejects every inaccurate accepted endpoint. For BFGS and L-BFGS-B, respectively, one physically audited radial restart leaves 68 and 64 native false successes in the PolarNS means and 16 and 99 in the bursts; their final bounds still reject them. The audit thus detects errors that a single radial restart leaves unresolved. On these observations, the physical stopping bound has broader coverage than the local fold gate. Figure~\ref{fig:realregimes} places them in the radius and gap-bound regimes.

An independent physical-coordinate SLSQP check of 29 selected reference minima has maximum loss discrepancy $1.31\times10^{-18}$. Independent float64 PyTorch AD audits all 309 cold/warm false successes and 32 control endpoints; maximum head-gradient, Hessian, and physical-gradient discrepancies are $7.77\times10^{-18}$, $8.89\times10^{-16}$, and $1.39\times10^{-17}$. No solver warning or native failure is removed.

\subsubsection{Coordinate repair with physical stopping controls}
\label{sec:chartrepair}
The real-data failures motivate a follow-up on the same observations. We develop the repair rule after inspecting the previous audit, then freeze it before evaluation. The inputs, ordering, and thresholds are retained. Every arm carries its own preceding endpoint. Repeating the raw warm and cold branches reproduces all 18\,568 stored real-data endpoints exactly in raw coordinates.

For $r>1$, replacing $w$ by $C(w)=w/\|w\|^2$ preserves $F(w)$ and moves to the inner sheet. At fixed intensity and data, $\|\nabla_w\widetilde f(C(w))\|_2=r^2\|\nabla_w\widetilde f(w)\|_2$; the Jacobian condition number is unchanged. This relation follows from the paired preimages in Supplementary Section S6. A fresh solver invocation resets quasi-Newton memory. The fold and the dependence on intensity remain.

We compare applying this switch before every warm-start solve with a physically audited repair. On a bound rejection, the latter uses the existing radial trial rule if the local negative-curvature gate holds; otherwise it switches an outer endpoint and restarts once. Inner nonfold failures remain rejected. A control restarts once from the cold initialization instead. A guarded variant follows the coordinate repair with standard projected-gradient steps in physical coordinates when the bound still rejects. This fallback uses step one, at most 64 updates, and the same $10^{-6}$ bound threshold. No reference optimum initializes any iterative arm. A standalone physical projected-gradient solver and the direct cone solution are additional controls. Supplementary Section S19 specifies the algorithm and accounting.

\begin{table*}[!t]
\caption{Two evaluations of repair on the same released observations. Entries are inaccurate outputs for BFGS/L-BFGS-B at nominal gap threshold $10^{-6}$. Fixed-endpoint repair starts from each historical endpoint; intervening in a continuation changes subsequent warm starts. Supplementary Table~\ref{SI-tab:chartrepairdetail} retains all strategies, Fresnel results and native-call accounting.}
\label{tab:chartrepair}
\centering\small\renewcommand{\arraystretch}{1.12}
\begin{tabularx}{\textwidth}{@{}Xrr@{}}\toprule
Evaluation & Fixed historical endpoints & Evolving continuation histories\\\midrule
Uncorrected warm start & 127/168 & 127/168\\
Audited coordinate repair & 7/39 & 7/22\\
Coordinate repair with physical fallback & 0/0 & 0/0\\
Audited cold restart & 0/5 & 0/2\\\bottomrule
\end{tabularx}
\end{table*}

Table~\ref{tab:chartrepair} distinguishes the fixed-endpoint and evolving-history evaluations; Supplementary Table~\ref{SI-tab:chartrepairdetail} gives the full strategy comparison. Switching before each solve reduces released-data inaccurate endpoints from 295 to 25, but increases the Fresnel count from 16 to 22. The audited coordinate arm leaves 29 inaccurate released-data endpoints, all rejected. Adding physical fallback leaves none: BFGS and L-BFGS-B require respectively two and 15 fallback calls, with three and 22 projected updates in total. There are 4\,663 and 4\,704 physical-gradient queries in the BFGS and L-BFGS-B guarded branches, respectively, including stopping audits. No real-data repair invokes the local curvature gate. In the Fresnel grid, each guarded solver uses seven full head-Hessian queries and 21 trial FG evaluations; neither requires physical fallback.

The paired-endpoint audit applies repair to the same 9\,284 historical warm endpoints, rather than changed continuation histories. Coordinate repair leaves 7/39 of the original 127/168 inaccurate endpoints; physical fallback removes those remaining 46. Audited cold restarting leaves 0/5. Thus the coordinate-only step corrects 249 of the 295 original failures, while full correction depends on physical fallback. The fixed-endpoint evaluation leaves 46 residuals; the evolving-history evaluation leaves 29 (Table~\ref{tab:chartrepair}). Independent PyTorch checks cover all 295 historical false successes and 32 controls; the largest output-preservation and gradient-transformation discrepancies are $1.74\times10^{-18}$ and $7.78\times10^{-16}$.

The physical controls are substantially cheaper in this implementation. Across 4\,642 released entries, standalone projected gradient uses 9\,575 updates and 14\,217 physical-gradient queries, with no inaccurate or rejected outputs. Its observed time is 0.203 s, versus 4.64 and 1.35 s for the guarded BFGS and L-BFGS-B head solvers, respectively, and 0.0219 s for the direct cone solution. The timings describe one CPU pass in the stated implementation. All released-data repairs in this follow-up use first derivatives.

\section{Discussion}
The stopping theorem supplies a domain-wide sufficient test for strongly convex physical objectives on the bounded-intensity inner sheet. For the stored endpoints, replacing the uniform formula with the pointwise expression substantially increases acceptance, while outward evaluation at 64--256 fractional bits preserves the pointwise decisions. This comparison identifies the conservatism of the uniform constants separately from the effect of validated arithmetic.

The recovery routes differ in their information requirements and sensitivity. At the fold, one signed radial curvature resolves the one-dimensional physical-gradient ambiguity. Away from it, complete first derivatives suffice; the restricted mixed and vector routes additionally use different physical-curvature inputs. The mixed route has a quadratic radial curvature-input gain, while the vector route also has a first-order tangential gain. Both retain derivative-error coefficients proportional to $1/I$. These error formulas hold at the evaluated geometry; perturbations of the coordinates require their own error analysis.

The real-data endpoints explain why avoiding the fold alone cannot ensure reliable first-order stopping. The inverse in \eqref{eq:firstinverse} has gain $1/(I\min\{|a_r|,t_r\})$, which grows near the fold, at low intensity and as $r^2/(2I)$ on the large-radius outer sheet. The released-data audit contains radii above $10^2$, while the DeepSfP subset remains well inside the fold. The local gate detects controlled fold failures but misses all 295 inaccurate warm-start acceptances in the released observations. On these observations, the direct physical-gap audit rejects the inaccurate endpoints across these regimes.

An endpoint audit therefore begins with the geometry $r,I,I'$ and the available derivative-error budgets. Table~\ref{tab:contract} selects the recovery route. Corollary~\ref{cor:gap} provides the calibrated quadratic test, including support slack; Theorem~\ref{thm:uniformstopping} applies under its strong-convexity and chart conditions. A valid bound below the requested gap supports stopping. When a physical gradient is available, the corresponding bound can be evaluated directly. Table~\ref{tab:chartrepair} illustrates subsequent repair decisions on rejected endpoints.

The numerical certificates concern the specified nominal objective. Extending them to uncertain instrument calibration or noise-free Stokes error requires an additional model of those uncertainties. The 20 shared-network fits contain no saved near-fold negative-curvature candidates. Translating the head-coordinate conditions into guarantees on shared network parameters also requires a separate error analysis: a small upstream parameter gradient alone does not establish the head-derivative conditions. These extensions define the next questions beyond the present endpoint certificate.

\section{Methods}
\label{sec:methods}

\subsection{Stokes geometry and coordinates}
\label{sec:model}
Let $s=(I,p)^{\mathsf T}$ be the reduced Stokes vector, where $I$ is total intensity and $p=(Q,U)^{\mathsf T}\in\mathbb R^2$ contains the linear-polarization components, in one radiometric unit and a fixed polarization basis. The admissible cone for this reduced model is
\begin{equation}
 \mathcal C=\{(I,Q,U):I\geq\sqrt{Q^2+U^2}\}.
 \label{eq:cone}
\end{equation}
For $I>0$, $q=p/I$ has degree of linear polarization (DoLP) $d=\|q\|_2$. The circular-polarization component $V$ is not estimated in this reduced model. The constraint allows $0\leq d\leq1$ and therefore includes partial polarization; $d=1$ is the fully linearly polarized boundary. For an objective $f$, write $g=\nabla_s f=(g_I,g_p)$ and $H_{pp}=\nabla^2_{pp}f\in\mathbb R^{2\times2}$; $W_{pp}$ is the polarization block of metric $W$. For $M\succ0$, $\|x\|_M=(x^{\mathsf T}Mx)^{1/2}$ and $M_{00}$ is its intensity--intensity entry. Bare vector norms are Euclidean; matrix $2$-norms are spectral.

We keep three coordinate spaces distinct:
\begin{equation}
 \theta\longmapsto(a,w)\xmapsto{\ T_F\ }s=(I,Q,U).
 \label{eq:spaces}
\end{equation}
Here $\theta$ denotes network parameters, $a\in\mathbb R$ and $w\in\mathbb R^2$ are independent head (raw) coordinates, and $s$ is the physical Stokes vector. For $r=\|w\|_2>0$, let $e=w/r$ and $P_\perp=\operatorname{Id}_2-ee^{\mathsf T}$, where $\operatorname{Id}_k$ is the $k\times k$ identity. Radial directions are parallel to $e$, tangential directions orthogonal to it. Our derivatives and stationarity tests are in the head space. Define
\begin{equation}
 F(w)=\frac{2w}{1+\|w\|^2},\qquad
 T_F(a,w)=I(a)(1,F(w)).
 \label{eq:foldmap}
\end{equation}
The intensity coordinate is smooth and bijective onto $(0,\infty)$, with $I'(a)=dI/da\ne0$. We use $I(a)=I_0\log(1+e^a)$ with intensity scale $I_0>0$, unless an experiment states another regular coordinate. The regions $r<1$ and $r>1$ are the inner and outer sheets; they meet at the fold $r=1$, where $s=I(1,e)$ is fully linearly polarized. The folded map reaches full linear polarization at finite coordinates. The monotone map $G(w)=w/\sqrt{1+\|w\|^2}$ reaches only the disk interior, whereas disk projection reaches the boundary nonsmoothly. For a radial $C^2$ map admissible on a two-sided neighborhood of a finite boundary radius, Supplementary Section S6 proves a condition-number lower rate proportional to $(1-d)^{-1/2}$; the rational fold attains this rate. Section~\ref{sec:robust} treats recovery at finite radius offsets.

\subsection{What head derivatives identify at the fold}
\label{sec:stationarity}
The observation model in Table~\ref{tab:contract} first leaves one physical component unspecified, then restores it with a radial second derivative.

\subsubsection{A derivative-observation rank condition}
The underlying mechanism is the second-order chain rule, consistent with the differential framework of smooth-lift theory \cite{lifts}. The following lemma states the corresponding observation criterion.
\begin{lemma}[Kernel-contracted observations]
\label{lem:observation}
Let $\mathcal U\subset\mathbb R^{n_x}$ be open, $T:\mathcal U\to\mathbb R^m$ and $f$ be $C^2$, and $h=f\circ T$. At $x\in\mathcal U$, choose $N$ pairs $k_i\in\ker DT(x)$, $u_i\in\mathbb R^{n_x}$, $i=1,\ldots,N$. Put $q_i=D^2T(x)[k_i,u_i]$. Then, for $g=\nabla f(T(x))$,
\begin{equation}
 \nabla h=DT^{\mathsf T}g,\qquad
 D^2h[k_i,u_i]=q_i^{\mathsf T}g.
 \label{eq:observation}
\end{equation}
The gradient is uniquely identifiable from precisely these observations, for arbitrary $C^2$ objectives, if and only if
\begin{equation}
 \operatorname{rank}\mathcal A_x=m,\qquad
 \mathcal A_x=\begin{pmatrix}DT(x)^{\mathsf T}\\q_1^{\mathsf T}\\\vdots\\q_N^{\mathsf T}\end{pmatrix}.
 \label{eq:observationrank}
\end{equation}
\end{lemma}
\begin{proof}
The chain rule adds $D^2f[DTk_i,DTu_i]$ to the second expression; it vanishes because $DTk_i=0$. Full column rank gives uniqueness. If $d_0\ne0$ lies in $\ker\mathcal A_x$, replacing $f(u)$ by $f(u)+d_0^{\mathsf T}(u-T(x))$ changes its gradient but preserves its value and all the stated observations, even with its physical Hessian fixed.
\end{proof}
At the Stokes fold the first-order rank is two; one scalar radial observation raises it to three.

\subsubsection{First-order ambiguity and second-order recovery}
At the fold, radial contraction of the composite Hessian removes the unknown physical-Hessian contribution. At a fixed fold output $s$, let $n=(-1,e)$ and consider
\begin{equation}
 f_\tau(u)=f(u)+\tau n^{\mathsf T}(u-s),\qquad \tau\in\mathbb R.
 \label{eq:ambiguity}
\end{equation}
All these objectives have the same value and first head derivatives at the fold, but their physical gradients differ by $\tau n$. Their physical Hessians also agree, so the ambiguity persists among fixed-metric quadratics. Since the full-head differential has rank two, this is its entire gradient ambiguity. Thus single-point first-order information cannot determine the physical gradient for this class. A radial second derivative resolves precisely this missing component.

\begin{proposition}[Physical-gradient recovery at the fold]
\label{prop:recovery}
Let $f$ be $C^2$ on an open neighborhood of the physical point $s$, including the cone boundary, and let $\widetilde f=f\circ T_F$. At $w=e$, $\|e\|=1$, write $g=(g_I,g_p)=\nabla_s f$, and define
\begin{equation}
 b=\frac{\partial_a\widetilde f}{I'(a)},\qquad
 v=\nabla^2_{ww}\widetilde f\,e.
 \label{eq:hvpinputs}
\end{equation}
With $j=\nabla_w\widetilde f$ and $c_r=e^{\mathsf T}v$, first derivatives and one scalar radial curvature recover the missing component. Here $b,j,v,c_r$ are, respectively, the normalized intensity derivative, head-polarization gradient, radial HVP, and radial second derivative:
\begin{subequations}\label{eq:recoveryroutes}
\begin{equation}
 g_p=\frac{j-c_re}{I},\qquad g_I=b+\frac{c_r}{I}.
 \label{eq:scalarrecovery}
\end{equation}
A vector HVP is an alternative that gives the polarization gradient directly:
\begin{equation}
 g_p=-\frac{v}{I},\qquad g_I=b+\frac{e^{\mathsf T}v}{I}.
 \label{eq:gradientrecovery}
\end{equation}
\end{subequations}
The identities hold at stationary and nonstationary folds.
\end{proposition}
\begin{proof}
At the fold, $DF(e)e=0$ and $D(DF)(e)[e]=-\operatorname{Id}$. The $ww$ Hessian of the composite has a term $I^2 DF^{\mathsf T}(\nabla^2_{pp}f)DF$ and a term from second derivatives of the head. Multiplying by $e$ kills the first term and gives $v=-Ig_p$ from the second. The intensity derivative gives $b=g_I+e^{\mathsf T}g_p$, proving \eqref{eq:gradientrecovery}. Also $j=I(\operatorname{Id}-ee^{\mathsf T})g_p$ and $c_r=-Ie^{\mathsf T}g_p$, giving \eqref{eq:scalarrecovery}.
\end{proof}
The scalar $c_r$ is a directional second derivative at any fold, including nonstationary points. Its interpretation as a Hessian eigenvalue will require stationarity. First derivatives already reveal the tangential physical gradient; second derivatives supply its missing radial component.

\subsection{Recovery at finite radius offsets}
\label{sec:robust}
\subsubsection{First-order inversion at $r\ne1$}
For any $C^2$ physical objective, let $w=re$, $r>0$, $r\ne1$, and $\phi=2r/(1+r^2)$, the output DoLP. The polarization differential has radial and tangential eigenvalues $a_r,t_r$:
\begin{equation}
 \begin{aligned}
 J&=DF(re)=a_ree^{\mathsf T}+t_rP_\perp,\\
 a_r&=\frac{2(1-r^2)}{(1+r^2)^2},\qquad t_r=\frac2{1+r^2}.
 \end{aligned}
 \label{eq:nearjacobian}
\end{equation}
The subscript $r$ indicates evaluation at $w=re$. Writing $j_r=\nabla_w\widetilde f$ and $b_r=\partial_a\widetilde f/I'$, first derivatives alone recover the physical gradient at $s_r=I(1,\phi e)$:
\begin{equation}
 g_p=\frac{ee^{\mathsf T}j_r}{Ia_r}+\frac{P_\perp j_r}{It_r},\quad
 g_I=b_r-\phi e^{\mathsf T}g_p.
 \label{eq:firstinverse}
\end{equation}
Since $a_r=-(r-1)+O((r-1)^2)$, the gradient is identifiable away from the fold, while radial derivative error is amplified by $1/(I|a_r|)$. Supplementary Section S8 gives the induced-norm error bound.

\subsubsection{Mixed recovery from a physical-curvature scalar}
Tangential physical-gradient information remains in the first derivatives near the fold. Combining it with the scalar radial curvature reduces the supplied physical curvature to $h_{rr}=e^{\mathsf T}H_{pp}e$.
\begin{proposition}[Mixed scalar recovery]
\label{prop:mixed}
Let $f$ be $C^2$, $I>0$, $r>0$, and $r\ne\sqrt3$. From $b_r$, $P_\perp j_r$, $c_r=e^{\mathsf T}\nabla^2_{ww}\widetilde f\,e$, and supplied $h_{rr}$, the physical gradient is
\begin{equation}
 \begin{aligned}
 m_r&=\frac{c_r/I-Ia_r^2h_{rr}}{\kappa_r},\\
 g_p&=\frac{P_\perp j_r}{It_r}+e m_r,\qquad
 g_I=b_r-\phi m_r,
 \end{aligned}
 \label{eq:mixedrecovery}
\end{equation}
where $\kappa_r=4r(r^2-3)/(1+r^2)^3$. At $r=1$, this reduces to \eqref{eq:scalarrecovery}: the composite radial second derivative $c_r$ remains an input, but no supplied physical-Hessian scalar $h_{rr}$ is required.
\end{proposition}
\begin{proof}
The first derivative gives $P_\perp j_r=It_rP_\perp g_p$. Radial contraction of the composite Hessian gives $c_r=I^2a_r^2h_{rr}+I\kappa_r e^{\mathsf T}g_p$, so $m_r=e^{\mathsf T}g_p$ in \eqref{eq:mixedrecovery}. At the fold, $a_r=0$, $t_r=1$, and $\kappa_r=-1$.

For $r\ne1$, the triple $(b_r,P_\perp j_r,c_r)$ alone leaves an ambiguity when physical curvature is unknown. Change $g_p$ by $\tau e$, $g_I$ by $-\phi\tau$, and $H_{pp}$ by $-\kappa_r\tau ee^{\mathsf T}/(Ia_r^2)$. A centered linear-plus-quadratic perturbation of $f$ realizes these changes while preserving its value and the observed triple. Supplying $h_{rr}$ removes this ambiguity. In contrast, the complete $j_r$ already identifies $g$ through \eqref{eq:firstinverse}.
\end{proof}

For fixed geometry, put $u_\phi=(-\phi,e)^{\mathsf T}$, $e_0=(1,0,0)^{\mathsf T}$, and $T_\perp=\binom{0_{1\times2}}{P_\perp}/(It_r)$. Errors in the mixed inputs propagate exactly as
\begin{equation}
 \begin{aligned}
 \delta g={}&e_0\delta b+T_\perp\delta j\\
 &+u_\phi\left(\frac{\delta c_r}{I\kappa_r}
                  +\beta_{\parallel}\delta h_{rr}\right),
 \end{aligned}
 \label{eq:mixederror}
\end{equation}
where $\beta_{\parallel}=-Ia_r^2/\kappa_r$. Each $\delta$ is supplied minus exact input. With $|\delta b|\leq\varepsilon_b$, $\|\delta j\|_2\leq\varepsilon_j$, $|\delta c_r|\leq\varepsilon_c$, $|\delta h_{rr}|\leq\varepsilon_h$, and any fixed $M\succ0$, the triangle inequality gives
\begin{equation}
 \begin{aligned}
 \|\delta g\|_M\leq{}&\sqrt{M_{00}}\varepsilon_b
   +\|M^{1/2}T_\perp\|_2\varepsilon_j\\
 &+\|u_\phi\|_M\left(\frac{\varepsilon_c}{I|\kappa_r|}
              +|\beta_{\parallel}|\varepsilon_h\right).
 \end{aligned}
 \label{eq:mixedenvelope}
\end{equation}
Only the radial physical-curvature scalar enters. Near $r=1$, its gain is $O((r-1)^2)$, while the $j_r$ and $c_r$ error coefficients remain of order $1/I$. The geometry $r,e,I,I'$ is obtained from the evaluated head coordinates and held fixed in this input-error calculation.

\subsubsection{Vector-HVP recovery from a curvature action}
For a general $C^2$ objective $f$, set $H_{pp}=\nabla^2_{pp}f(s_r)$ and $h_p=H_{pp}e$ at the same physical point $s_r=I(1,\phi e)$. The correction uses this single Hessian action. With $J=DF(re)$ and $v_r=\nabla^2_{ww}\widetilde f(a,re)e$, the radial HVP satisfies
\begin{equation}
 v_r=I^2a_rJ h_p+IKg_p,\qquad
 b_r=g_I+\phi e^{\mathsf T}g_p.
 \label{eq:nearhvp}
\end{equation}
Here the map-curvature matrix is
\begin{equation}
 \begin{aligned}
 K&=D(DF)(re)[e]=\kappa_r ee^{\mathsf T}+\ell_rP_\perp,\\
 \kappa_r&=\frac{4r(r^2-3)}{(1+r^2)^3},\qquad
 \ell_r=\frac{-4r}{(1+r^2)^2}.
 \end{aligned}
 \label{eq:Keigenvalues}
\end{equation}
\begin{proposition}[Local-curvature recovery]
\label{prop:localcurvature}
Let $f$ be $C^2$, $I>0$, $r>0$, and $r\ne\sqrt3$. Given $h_p=H_{pp}e$, the complete physical gradient is
\begin{equation}
 \begin{aligned}
 g_p&=K^{-1}\left(\frac{v_r}{I}-Ia_rJ h_p\right),\\
 g_I&=b_r-\phi e^{\mathsf T}g_p.
 \end{aligned}
 \label{eq:correctedrecovery}
\end{equation}
The Hessian may be indefinite and state dependent.
\end{proposition}
\begin{proof}
The composite Hessian is $I^2J^{\mathsf T}H_{pp}J$ plus map curvature. Since $J^{\mathsf T}=J$ and $Je=a_re$, radial contraction gives \eqref{eq:nearhvp}. The eigenvalues in \eqref{eq:Keigenvalues} are nonzero under the stated conditions, so subtraction and inversion give \eqref{eq:correctedrecovery}.
\end{proof}
Formula~\eqref{eq:correctedrecovery} holds at every radius in Proposition~\ref{prop:localcurvature}. At $r=1$, $a_r=0$ and $K=-\operatorname{Id}$: physical curvature cancels and \eqref{eq:gradientrecovery} is recovered. Both eigenvalues tend to $-1$, giving a bounded inverse nearby. At $r=\sqrt3$, $K$ is singular while $J$ has eigenvalues $-1/4$ and $1/2$, so first-order inversion remains regular. The inverse $K^{-1}$ also degenerates as $r\downarrow0$ or $r\to\infty$.

\subsubsection{Vector-HVP derivative and curvature-input errors}
Choose any fixed reporting metric $M\succ0$, independently of $H_{pp}$, and write $\|x\|_M^2=x^{\mathsf T}Mx$. Define $e_0=(1,0,0)^{\mathsf T}$, $k_b=\sqrt{M_{00}}$, and
\begin{equation}
 \begin{aligned}
 A_\phi&=\begin{pmatrix}-\phi e^{\mathsf T}\\\operatorname{Id}_2\end{pmatrix},\quad
 R_r=A_\phi K^{-1}/I,\\
 E_r&=-Ia_r A_\phi K^{-1}J.
 \end{aligned}
 \label{eq:correctedoperator}
\end{equation}
With exact $r,e,I,I'$ and errors $\delta b,\delta v,\delta h_p$, the recovery error is exactly
\begin{equation}
 \delta g=e_0\delta b+R_r\delta v+E_r\delta h_p.
 \label{eq:curvatureerror}
\end{equation}
The triangle inequality gives a conservative norm envelope. Supplied bounds $|\delta b|\leq\varepsilon_b$, $\|\delta v\|_2\leq\varepsilon_v$, $\|\delta h_p\|_2\leq\varepsilon_h$ imply
\begin{equation}
 \begin{aligned}
 \|\delta g\|_M\leq\eta_r={}&k_b\varepsilon_b
 +\|M^{1/2}R_r\|_2\varepsilon_v\\
 &+\|M^{1/2}E_r\|_2\varepsilon_h.
 \end{aligned}
 \label{eq:correctederror}
\end{equation}
If $\widehat H_{pp}=H_{pp}+\Delta H_{pp}$ with $\|\Delta H_{pp}\|_2\leq\varepsilon_H$, then $\delta h_p=\Delta H_{pp}e$ and one may set $\varepsilon_h=\varepsilon_H$. In particular, with only curvature input perturbed,
\begin{equation}
 \delta g_p=-Ia_rK^{-1}J\Delta H_{pp}e.
 \label{eq:curvaturebias}
\end{equation}
This perturbation varies the curvature input at a fixed physical state. Its two directional gains have different orders.

\begin{corollary}[Vector-HVP curvature-input sensitivity]
\label{cor:anisotropy}
Under Proposition~\ref{prop:localcurvature}, with $r,e,I,I',b_r,v_r$ exact and only $h_p$ perturbed, the polarization-gradient error is
\begin{equation}
 \begin{aligned}
 \delta g_p&=\beta_{\parallel}ee^{\mathsf T}\delta h_p
             +\beta_{\perp}P_\perp\delta h_p,\\
 \beta_{\parallel}&=\frac{I(1-r^2)^2}{r(3-r^2)(1+r^2)},\\
 \beta_{\perp}&=\frac{I(1-r^2)}{r(1+r^2)}.
 \end{aligned}
 \label{eq:anisotropiccurvature}
\end{equation}
For fixed $I>0$ and $\delta=r-1$, these are $\beta_{\parallel}=I\delta^2+O(\delta^3)$ and $\beta_{\perp}=-I\delta+O(\delta^2)$.
\end{corollary}
\begin{proof}
The matrices $J$ and $K$ have the common projectors $ee^{\mathsf T}$ and $P_\perp$ because the output map is radial; this does not assume an isotropic physical Hessian. Thus $-Ia_rK^{-1}J$ has eigenvalues $-Ia_r^2/\kappa_r$ and $-Ia_rt_r/\ell_r$. Substitution gives \eqref{eq:anisotropiccurvature}; expansion at $r=1$ gives the stated orders. Also $\delta g_I=-\phi e^{\mathsf T}\delta g_p$, so tangential curvature-input error leaves the intensity gradient unchanged.
\end{proof}
Figure~\ref{fig:sensitivity}(a) shows these vector-HVP gains. Mixed recovery shares the radial coefficient and uses no tangential physical-curvature input. The first-order tangential sensitivity is therefore a property of the vector-HVP route.

Reusing the fold formula omits the $I^2a_rJh_p$ term in \eqref{eq:nearhvp}. Supplementary Section S4 separates this recovery bias from the supporting-plane error in an interior gap bound. Figure~\ref{fig:sensitivity} places the analytic input gains beside measured AD recovery errors.

\subsection{A quadratic gap-bound corollary}
\label{sec:gap}
Let $B\in\mathbb R^{m\times3}$ be a full-column-rank calibrated linear instrument, $y\in\mathbb R^m$ its observations, and $\Sigma\in\mathbb R^{m\times m}$ a fixed SPD observation covariance. Its data term can be written
\begin{equation}
 \begin{aligned}
 L(s)&=\tfrac12\|Bs-y\|_{\Sigma^{-1}}^2
       =\tfrac12\|s-z\|_W^2+c,\\
 W&=B^{\mathsf T}\Sigma^{-1}B,\qquad
 z=W^{-1}B^{\mathsf T}\Sigma^{-1}y,
 \end{aligned}
 \label{eq:calibrated}
\end{equation}
where $c$ is the additive least-squares residual independent of $s$. The generalized least-squares estimate $z$ can be unphysical. Define the constrained objective gap $\Delta_{\mathcal C}(s)=L(s)-\min_{u\in\mathcal C}L(u)$.

Set $M=W^{-1}$, $n=(-1,e)$, and $D_n=n^{\mathsf T}Mn$. For any mapped output $s_r=I(1,\phi e)$, the cone lies in the supporting half-space $H_e=\{u:n^{\mathsf T}u\leq0\}$ and
\begin{equation}
 n^{\mathsf T}(u-s_r)\leq\sigma_r,\quad
 \sigma_r=I(1-\phi)=\frac{I(r-1)^2}{1+r^2}.
 \label{eq:supportslack}
\end{equation}
\begin{corollary}[Quadratic gap bound from recovered gradients]
\label{cor:gap}
If $\|\widehat g-g\|_M\leq\eta$, then for any $\lambda\geq0$,
\begin{equation}
 \Delta_{\mathcal C}(s_r)\leq
 \tfrac12\bigl(\|\widehat g+\lambda n\|_M+\eta\bigr)^2
 +\lambda\sigma_r.
 \label{eq:nearcertificate}
\end{equation}
For an exact gradient, the minimizing multiplier is
\begin{equation}
 \lambda_r=\max\{0,-(n^{\mathsf T}Mg+\sigma_r)/D_n\}.
 \label{eq:nearmultiplier}
\end{equation}
\end{corollary}
\begin{proof}
For $u\in\mathcal C$ and $d=u-s_r$, complete the quadratic:
\begin{equation*}
 L(s_r)-L(u)\leq\tfrac12\|g+\lambda n\|_M^2+\lambda\sigma_r.
\end{equation*}
The right-hand side is independent of $u$, so taking the supremum over $u\in\mathcal C$ bounds $\Delta_{\mathcal C}(s_r)$. The triangle inequality gives \eqref{eq:nearcertificate}; minimizing the exact-gradient quadratic in $\lambda$ gives \eqref{eq:nearmultiplier}.
\end{proof}
At a fold $\sigma_r=0$. With $\widehat\lambda=\max\{0,-n^{\mathsf T}Mg/D_n\}$, define
\begin{equation}
 \mathcal B(s)=\tfrac12\|g+\widehat\lambda n\|_M^2.
 \label{eq:residualcertificate}
\end{equation}
This is exactly $L(s)-\min_{H_e}L$, the gap of the supporting-half-space relaxation. Since $\mathcal C\subset H_e$, this relaxation gap is an upper bound on $\Delta_{\mathcal C}(s)$. Its minimizer is $h_s=z-\widehat\lambda Mn$, so $\Delta_{\mathcal C}(s)=\mathcal B(s)$ if and only if $h_s\in\mathcal C$; strict convexity gives both directions. Supplementary Section S10 displays the two objective gaps. Alternatively, $g,W$ determine $z=s-W^{-1}g$, after which the original cone quadratic gives the exact gap. The half-space expression is the closed-form relaxation obtained by standard quadratic duality \cite{cvxbook}.

At an exact stationary fold, $g=\alpha n$ and the raw radial curvature in $w=(1+t)e$ is $\gamma_r=c_r=-I\alpha$. Nonnegative $\gamma_r$ gives the constrained optimum; negative $\gamma_r$ gives a strict saddle and gap at most $D_n\gamma_r^2/(2I^2)$. Supplementary Section S10 proves the classification and equality condition. A regular radial rescaling multiplies $\gamma_r$ by the squared coordinate Jacobian at stationarity. Its sign is invariant; the displayed coefficient uses $w=(1+t)e$.

Support slack remains necessary even with exact recovery; Supplementary Section S4 gives the positive-gap counterexample and Section S10 the relaxation geometry.

\subsection{Sequential initialization and boundary-restricted flow}
\label{sec:sequential}
A previous constrained estimate can be fully linearly polarized. Every finite raw preimage of that nonzero boundary state has $r=1$. Exact optimization followed by a raw-coordinate lift can therefore put mass on the fold, even when the observations have a density. This is a conditional statement about a boundary estimate; it does not assert that a finite solver returns the exact fold. At a fold, the first-order obstruction is already described by smooth-lift theory \cite[Thm.~2.4]{lifts}. The following specialization explains what ordinary continuous flow does after such an initialization.

\begin{proposition}[Flow restricted to a Stokes fold]
\label{prop:restrictedflow}
Let $L(s)=\tfrac12\|s-z\|_W^2+c$, $W=\operatorname{diag}(1,1/2,1/2)$, and $z_I>\rho=\|z_p\|>0$. Use any $C^2$ bijective intensity coordinate onto $(0,\infty)$ with $I'\ne0$. For raw Euclidean gradient flow of $L\circ T_F$, the set $r=1$ is invariant. If $r(0)=1$ and $e(0)\ne-z_p/\rho$, then
\begin{equation}
 e(t)\longrightarrow z_p/\rho,\qquad
 I(t)\longrightarrow I_b=\frac{2z_I+\rho}{3}.
 \label{eq:foldflowlimit}
\end{equation}
The limiting state has gap $(z_I-\rho)^2/6$ and radial head curvature $-I_b(z_I-\rho)/3<0$.
\end{proposition}
\begin{proof}
Writing $a_r=2(1-r^2)/(1+r^2)^2$, the chain rule gives
\begin{equation}
 \frac{d r^2}{dt}=-2Ia_r w^{\mathsf T}g_p.
 \label{eq:radiusflow}
\end{equation}
Thus the smooth vector field is tangent to $r=1$, and uniqueness gives invariance. On that set, put $u=e^{\mathsf T}z_p/\rho$. Then
\begin{equation}
 \begin{aligned}
 \dot I&=-(I')^2(3I/2-z_I-\rho u/2),\\
 \dot u&=I\rho(1-u^2)/2.
 \end{aligned}
 \label{eq:restrictedflow}
\end{equation}
The intensity remains between the positive bounds $\min\{I(0),(2z_I-\rho)/3\}$ and $\max\{I(0),(2z_I+\rho)/3\}$. Hence $u\to1$ unless $u(0)=-1$. Scalar comparison in the first equation gives \eqref{eq:foldflowlimit}. Substitution into the quadratic objective and radial curvature gives the last assertion. Supplementary Section S16 details global existence and the antipodal exception.
\end{proof}

This is a property of continuous raw Euclidean flow. A plain gradient-descent step from the fold instead satisfies
\begin{equation}
 \|w-\eta\nabla_w(L\circ T_F)\|^2
 =1+\eta^2 I^2\|P_\perp g_p\|^2.
 \label{eq:discretefoldstep}
\end{equation}
It can leave the fold through a tangential update. Adam and quasi-Newton updates have different coordinate scalings and are not covered by invariance. The absolutely continuous random initialization and small fixed gradient step in \cite{lee} also differ from a boundary-derived initialization and finite solver termination. The proposition supplies a model-specific dynamical explanation; the numerical experiments in Section~\ref{sec:numerics} evaluate discrete solvers directly.

\subsection{Finite-tolerance stopping guarantees}
\label{sec:stoppingguarantees}
The recovery formulas identify the physical gradient from head derivatives. For stopping, we also need a tolerance that controls the objective gap. For strongly convex physical objectives, a lower test on one radial curvature near the fold provides such a bound.

\subsubsection{Relation to approximate-lift results}
Approximate second-order transfer is established for squared-variable formulations \cite[Thm.~3.6 and Cor.~3.7]{dingwright}. Local gap control through the Kurdyka--\L{}ojasiewicz (KL) exponent is also established for Hadamard difference parametrizations \cite[Thm.~3.11 and Rem.~3.12]{ouyang}. In particular, a KL exponent $3/4$ leads to a gradient-to-gap power $4/3$. Supplementary Table~\ref{SI-tab:stoppingprior} places the Stokes result in this framework. It gives explicit constants over the whole inner sheet with bounded intensity, uses one radial lower-curvature test, and accommodates supplied derivative-error budgets. Its domain is specified by evaluated coordinates and model bounds, independently of the distance to an optimum or nondegeneracy at that optimum.

\subsubsection{One radial test and a uniform gap bound}
Write $\Delta(s)=f(s)-\min_{u\in\mathcal C}f(u)$ for the objective gap and $x_+=\max\{x,0\}$ for the positive part. At the evaluated point, retain $b=\partial_a\widetilde f/I'$ and $j=\nabla_w\widetilde f$. For $r>0$, set $c_r=e^{\mathsf T}\nabla^2_{ww}\widetilde f\,e$. The following gradient tolerance is normalized in physical units; Supplementary Section S20 gives the conversion from a native solver tolerance.

\begin{theorem}[Uniform stopping on the inner Stokes sheet]
\label{thm:uniformstopping}
Let $f:\mathbb R^3\to\mathbb R$ be $C^2$ and $\mu$-strongly convex, $\mu>0$, and fix $L\ge\mu$. At $s=T_F(a,w)$, suppose $0<I\le\overline I$, $0\le r\le1$, and
\begin{equation}
 |b|\le\varepsilon,\qquad \|j\|_2/I\le\varepsilon.
 \label{eq:normalizedstop}
\end{equation}
When $r\ge1/2$, additionally suppose
\begin{equation}
 e^{\mathsf T}H_{pp}(s)e\le L,\qquad c_r\ge-I\nu,
 \quad L\ge\mu,\quad \nu\ge0.
 \label{eq:radialstop}
\end{equation}
No curvature observation is required at $r<1/2$. Put $K_0=\overline I L$ and assume $\varepsilon\le K_0/8$. Then
\begin{equation}
 \boxed{\ f(s)-\min_{u\in\mathcal C}f(u)
 \le \frac{52K_0^{2/3}}{\mu}\varepsilon^{4/3}
       +\frac{3}{\mu}\nu^2.\ }
 \label{eq:uniformstopgap}
\end{equation}
The coefficient 52 is conservative; the two tolerance orders are sharp.
\end{theorem}
\begin{proof}
Write $g_p=me+u$, $u\perp e$, and $\delta=1-r$. Then $g_I=b-\phi m$, $|a_rm|\le\varepsilon$, and $\|u\|\le\varepsilon/t_r$. Strong convexity and the supporting inequality give, for every $\lambda\ge0$,
\begin{equation}
 \Delta(s)\le\frac{\|g+\lambda n\|_2^2}{2\mu}
                  +\lambda I(1-\phi),\quad n=(-1,e).
 \label{eq:strongstopdual}
\end{equation}
For any $0<h\le1/2$, $\delta\ge h$ implies $a_r\ge\delta$ and hence $\Delta(s)\le3\varepsilon^2/(\mu h^2)$. If $\delta<h$, then $1\le|\kappa_r|$, $a_r\le2\delta$, and $1-\phi\le\delta^2$. The radial identity from Proposition~\ref{prop:mixed}, $c_r/I=Ia_r^2 e^{\mathsf T}H_{pp}e+\kappa_r m$, yields
\begin{equation}
 m_+\le\nu+4K_0\delta^2.
 \label{eq:positivegradientbound}
\end{equation}
Choose $\lambda=(-m)_+$. For $m<0$, use $|m|(1-\phi)\le\varepsilon\delta$; for $m\ge0$, use \eqref{eq:positivegradientbound}. The two cases give
\begin{equation}
 \Delta(s)\le\frac{5\varepsilon^2+6\nu^2+96K_0^2h^4}{2\mu}
                  +\overline I\varepsilon h.
 \label{eq:bandstopgap}
\end{equation}
Set $h=(\varepsilon/K_0)^{1/3}$ and use $L\ge\mu$ to obtain \eqref{eq:uniformstopgap}. The zero-tolerance case follows separately or by taking a limit. At $r=0$, $DF=2\operatorname{Id}_2$ gives the first-order bound directly. Supplementary Section S20 gives all inequalities and a tighter pointwise certificate.
\end{proof}

\subsubsection{Input errors and a target stopping rule}
Suppose supplied $\widehat b,\widehat j,\widehat c_r$ have valid error budgets $E_b,E_j,E_c$:
\begin{equation}
 |\widehat b-b|\le E_b,\quad\|\widehat j-j\|\le E_j,
 \quad|\widehat c_r-c_r|\le E_c.
 \label{eq:stopinputerrors}
\end{equation}
With exact evaluated geometry, sufficient tolerance values are
\begin{equation}
 \begin{aligned}
 \varepsilon&=\max\{ |\widehat b|+E_b,
                         (\|\widehat j\|+E_j)/I\},\\
 \nu&=\max\{0,(-\widehat c_r+E_c)/I\}.
 \end{aligned}
 \label{eq:stopinflation}
\end{equation}
The triangle inequality gives this inflation. Here $E_b$ bounds the error in the normalized quantity $b$, not in the raw derivative $\partial_a\widetilde f$. With exact $I'$, a raw-derivative error bound $E_a$ gives $E_b=E_a/|I'|$. Applying the inflation requires valid budgets and justified bounds $\mu,L,\overline I$ for the objective being optimized. Numerical certification would additionally require validated rounding-error bounds; the measured float32/float64 discrepancies alone are insufficient. Equation~\eqref{eq:stopinputerrors} assumes exact evaluated geometry and a fixed instrument model.

For a desired objective gap $\tau>0$, one explicit sufficient rule is
\begin{equation}
 \begin{aligned}
 \varepsilon&\le\min\!\left\{K_0/8,
            \left(\frac{\mu\tau}{104K_0^{2/3}}\right)^{3/4}\right\},\\
 \nu&\le\sqrt{\mu\tau/6}.
 \end{aligned}
 \label{eq:targetstop}
\end{equation}
Theorem~\ref{thm:uniformstopping} then gives $\Delta(s)\le\tau$. A raw gradient tolerance requires conversion through $I$ and $I'$; without positive lower bounds on $I$ and $|I'|$ it cannot uniformly imply \eqref{eq:normalizedstop}. The rule certifies the gap at any point satisfying these conditions. The number of iterations needed to reach them depends on the optimizer.

\subsubsection{Sharp orders and necessary restrictions}
The $4/3$ order cannot be replaced uniformly by a larger power. For the fixed objective $f(s)=\tfrac12\|s-(1,e)\|^2$, take $r\uparrow1$ and $I=(1+\phi)/(1+\phi^2)$. Then $b=0$, $c_r\ge0$, and, with $\delta=1-r$,
\begin{equation}
 \varepsilon\sim\delta^3/4,\qquad
 \Delta(s)\sim\delta^4/16,\qquad
 \Delta/\varepsilon^{4/3}\to2^{-4/3}.
 \label{eq:stopsharpfirst}
\end{equation}
For the curvature order, the fold family in Fig.~\ref{fig:principle}, with Euclidean metric and $g=\alpha(-1,e)$, $\alpha>0$, has $\varepsilon=0$, $\nu=\alpha$ and exact gap $\alpha^2$. Thus a larger power than $\nu^2$ also fails. Both orders are therefore sharp within the stated objective class.

Curvature cannot be omitted at the fold: the same family has zero first derivatives and positive gap. The inner-sheet restriction is also necessary for this uniform implication. On an unbounded outer sheet, the quadratic centered at $(1,1/2,0)$ admits states with both normalized first derivatives and negative radial curvature tending to zero while the gap tends to $1/8$. With exponential intensity, raw first- and second-order tests also tend to zero as $I\downarrow0$ for the quadratic centered at $(1,0,0)$, while its gap tends to $1/2$. Section S20 supplies the constructions. Moving inward preserves the physical output but requires recomputing derivative tests there.

\subsubsection{Numerical verification}
A fixed check set contains 576 quadratic and 24 strongly convex quartic objectives with three supplied derivative-error budgets, giving 1\,800 checks. Of these, 1\,431 satisfy the uniform theorem's small-tolerance hypothesis $\varepsilon\le K_0/8$. The largest computed gap above a bound is $1.24\times10^{-14}$; the smallest signed cone slack $I-\|(Q,U)\|_2$ reported for the independent SLSQP references is $-9.64\times10^{-14}$. All 600 reference solves return success. Independent PyTorch differentiation gives maximum $b$, $j$, and $c_r$ discrepancies $1.78\times10^{-15}$, $5.33\times10^{-15}$, and $1.43\times10^{-14}$. Ninety-digit decimal examples check the sharp-order limits and the counterexamples.

On the existing 9\,284 real-data warm endpoints, we switch outer coordinates inward and recompute derivatives, without running another solver. The pointwise certificate in Section S20 accepts 8\,861 endpoints at nominal gap $10^{-6}$, rejects all 295 historical false successes, and additionally rejects 128 accurate endpoints. The coarse uniform bound alone accepts 980. This retrospective calculation checks the formula's coverage using zero supplied error budgets in float64 and a full analytic head Hessian at every endpoint. Its cost includes these Hessians. Direct physical auditing and the direct solvers in Section~\ref{sec:chartrepair} provide the practical cost comparison.

Section S21 adds outward evaluation of the nominal quadratic bound, including exponential intensity, geometry, and derivative arithmetic. At 128 fractional bits, it certifies gap at most $10^{-6}$ for the same 8\,861 endpoints and rejects all 295 inaccurate endpoints; 128 accurate endpoints remain rejected. It uses 52 scalar curvature evaluations and forms no full head Hessian. The complete pass takes 7.32 s. Decisions agree at 64--256 bits; 32 bits leave three additional accurate endpoints inconclusive. The accompanying standard-library implementation returns an integer-enclosed bound or an explicit inconclusive result.

\section*{Data availability}
The original source images are available from the DeepSfP and PolarNS/PolarBurstSR providers \cite{ba,polarns}. The selected numerical observations, endpoint records and reference outputs used in this study are maintained in the private reproducibility repository at \url{https://github.com/usamihiroyasu/stokes-gradient-recovery}. Access for peer review can be arranged by the corresponding author. The released-observation audit identifies archive members, byte ranges, checksums and selected coordinates. Full source images and model weights are not redistributed. Source attribution and the scope of third-party rights are documented in the repository.

\section*{Code availability}
The same private repository contains the numerical code, interval arithmetic, figure generators, matching TeX sources, fixed protocols, software versions and file hashes. Its README links each result to its input and output files. The nominal interval certificate can be replayed using Python's standard library with \texttt{python3 theory/validated\_stopping/replay.py --output NEW\_DIRECTORY}; the output directory must be new or empty. The reproduction guide lists the separate environments and commands for the other studies. The repository is not publicly accessible; reviewer access is arranged through the corresponding author.

\section*{Author contributions}
H.U. conceived the study, developed the theoretical analysis, and wrote and revised the manuscript. A.T. and K.H. performed the experiments and developed the experimental code.

\section*{Funding}
No funding was received for this research.

\section*{Competing interests}
H.U. is a member of the Editorial Board of Scientific Reports. The authors declare no other competing interests.

\FloatBarrier
\setlength{\bibsep}{0.5em}

\clearpage
\setcounter{section}{0}
\setcounter{equation}{0}
\setcounter{figure}{0}
\setcounter{table}{0}
\setcounter{theorem}{0}
\setcounter{lemma}{0}
\renewcommand{\thesection}{S\arabic{section}}
\renewcommand{\theequation}{S\arabic{equation}}
\renewcommand{\thefigure}{S\arabic{figure}}
\renewcommand{\thetable}{S\arabic{table}}
\renewcommand{\thetheorem}{S\arabic{theorem}}
\renewcommand{\thelemma}{S\arabic{lemma}}
\def\theHsection{supp.\arabic{section}}
\def\theHsubsection{supp.\arabic{section}.\arabic{subsection}}
\def\theHequation{supp.\arabic{equation}}
\def\theHfigure{supp.\arabic{figure}}
\def\theHtable{supp.\arabic{table}}
\def\theHtheorem{supp.\arabic{theorem}}
\def\theHproposition{supp.\arabic{theorem}}
\def\theHcorollary{supp.\arabic{theorem}}
\def\theHlemma{supp.\arabic{lemma}}
\renewcommand{\refname}{Supplementary References}

\raggedright
\begin{center}
{\Large\bfseries Supplementary Information}\par\medskip
{\large Conditional stopping guarantees for a folded parametrization of reduced Stokes vectors}\par\medskip
Hiroyasu Usami, Ayato Tsuboi and Keisuke Hara
\end{center}
\medskip
\noindent Core recovery details and validation are in Sections S4--S5, S8--S9, and S11--S15. Sequential initialization, restricted flow, and the fixed two-frame verification are in Sections S16--S17. Auxiliary stationary geometry, projection, renderer construction, and coordinate dynamics are in Sections S1--S3, S6--S7, and S10. We retain the main paper's conventions: $p=(Q,U)$, $g_p=\nabla_p f$, $H_{pp}=\nabla^2_{pp}f$, $r=\|w\|_2$, $e=w/r$ for $r>0$, and $P_\perp=\operatorname{Id}_2-ee^{\mathsf T}$. For $M\succ0$, $\|x\|_M^2=x^{\mathsf T}Mx$ and $\|A\|_{2\to M}=\|M^{1/2}A\|_2$. The disk-projection map is denoted $P(w)$; $P_\perp$ is the tangential matrix.

\section{Complete ideal-metric critical-point analysis}
\label{SI-sec:ideal}
We use the linear-Stokes cone $\mathcal C=\{(I,Q,U):I\geq\sqrt{Q^2+U^2}\}$ and $T_H(a,w)=I(a)(1,H(w))$, where $I$ is a smooth bijective coordinate onto $(0,\infty)$ with nonzero derivative. The maps are
\begin{equation}
 \begin{aligned}
 F(w)&=2w/(1+r^2),\quad G(w)=w/\sqrt{1+r^2},\\
 P(w)&=w/\max\{1,r\},\qquad r=\|w\|.
 \end{aligned}
\end{equation}
For ideal equal-gain analyzers at $0,45,90,135$ degrees, observations $y\in\mathbb R^4$ have mean $As$, where
\begin{equation}
 A=\frac12\begin{pmatrix}1&1&0\\1&0&1\\1&-1&0\\1&0&-1\end{pmatrix},
 \quad W=A^\mathsf TA=\operatorname{diag}(1,\tfrac12,\tfrac12).
 \label{SI-eq:analyzer}
\end{equation}
The intensity measurements and Stokes components share a unit. Equal independent analyzer variances give this metric up to a common positive scale. The factor $1/2$ is the ideal linear-analyzer transmission; it yields the two polarization weights $1/2$ in $W$. Our loss is
\begin{equation}
 \ell(s;s^*)=\tfrac12\|s-s^*\|_W^2.
 \label{SI-eq:loss}
\end{equation}
The metric is fixed throughout the analysis; a state-dependent covariance defines a different objective.

\subsection{Classification with trainable intensity}
Let $s^*=I^*(1,de)$ with $I^*>0$, $0<d<1$, $\|e\|=1$, and retain $W$ in \eqref{SI-eq:analyzer}. A strict saddle here means a stationary point with at least one negative Hessian eigenvalue. It need not have a nonsingular Hessian.

\begin{theorem}[Complete quadratic classification]
\label{SI-thm:classification}
For $\ell(T_F(a,w);s^*)$, the finite stationary points are the two exact preimages of $s^*$ and two fold states
\begin{equation}
 w_\sigma=\sigma e,\qquad I_\sigma=I^*\frac{2+\sigma d}{3},
 \quad\sigma\in\{-1,1\}.
 \label{SI-eq:criticalstates}
\end{equation}
Both fold states are strict saddles. In local coordinates $(I,\hbox{radial }w,\hbox{tangential }w)$ their Hessian is diagonal, with entries
\begin{equation}
 \left(\frac32,
 -\frac12 I_\sigma(I_\sigma-\sigma I^*d),
 \frac12 I_\sigma\sigma I^*d\right).
 \label{SI-eq:criticalhessian}
\end{equation}
At the aligned fold ($\sigma=1$),
\begin{equation}
 \ell_b=\frac{(I^*)^2(1-d)^2}{6},\qquad
 \partial_{rr}^2\ell_b=-\frac{(I^*)^2(2+d)(1-d)}9.
 \label{SI-eq:boundaryrisk}
\end{equation}
The physical output is not a constrained stationary point.
\end{theorem}
\begin{proof}
Away from $r=1$, $DF$ is invertible and the regular intensity coordinate makes $T_F$ locally invertible. A stationary composite loss must then have $s=s^*$. At $r=1$, the tangential condition forces $w=\sigma e$. The intensity condition is
\begin{equation*}
 (I-I^*)+\tfrac12(I-\sigma I^*d)=0,
\end{equation*}
which gives \eqref{SI-eq:criticalstates}. Differentiating in the stated local coordinates, using the radial second derivative $\phi''(1)=-1$, gives \eqref{SI-eq:criticalhessian}; mixed entries vanish. The radial entry is negative for both signs. A regular change from $I$ to $a$ preserves Hessian inertia at stationarity. Substitution gives \eqref{SI-eq:boundaryrisk}.

At the aligned state the physical gradient is
\begin{equation*}
 \nabla_s\ell=\frac{I^*(1-d)}3(-1,e).
\end{equation*}
Its negative points into the cone and strictly reduces the loss. Yet the head Jacobian kills this normal vector, giving zero raw gradient. Hence the output fails the constrained first-order condition.
\end{proof}
In particular, holding $I=I^*$ at the fold would miss the joint stationary point: its intensity gradient is generally nonzero. Accounting for intensity changes both the location and the classification of the critical states. These saddles are consistent with local-minimum preservation of the folded ball map \cite{supp-gcopter}; a non-optimal critical point need not be a local minimum.

\begin{figure*}[!t]
\centering\includegraphics[width=\textwidth]{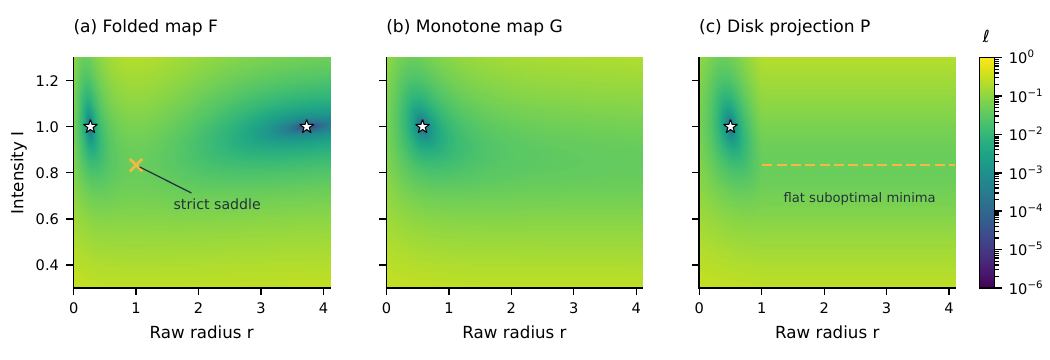}
\caption{Aligned $(I,r)$ slices for $I^*=1$, $d=0.5$. Stars mark exact solutions. The folded cross is a strict saddle at $I=5/6$, $r=1$. The exterior disk-projection family has flat positive loss. Only the target-aligned direction is shown; the theorem also classifies anti-aligned states. A shared logarithmic scale clips values below $10^{-6}$.}
\label{SI-fig:landscapes}
\end{figure*}

\begin{proposition}[Monotone and projection controls]
\label{SI-prop:controls}
For the same interior target, $\ell\circ T_G$ has exactly one finite stationary point, which attains $s^*$. For $T_P$, the unique interior minimum attains $s^*$. Every exterior state $w=\sigma r e$, $r>1$, at $I=I_\sigma$ is stationary. The aligned family consists of suboptimal flat local minima; the anti-aligned family consists of saddles.
\end{proposition}
\begin{proof}
$G$ and the regular intensity coordinate form a diffeomorphism onto the positive-intensity cone interior. An interior stationary point of the strictly convex quadratic is therefore its exact target. For exterior $P$, the radial derivative is zero and the remaining stationarity equations are those above. The Hessian entries are $3/2$, $0$, and $I_\sigma\sigma I^*d/(2r^2)$. For the aligned family, the loss is locally minimized in intensity and direction and is constant in raw radius. At $r=1$, $P$ is nonsmooth and no classical derivative is asserted.
\end{proof}
These examples describe the normalized disk map $P$ inside a trainable head. Figure~\ref{SI-fig:landscapes} compares its raw-coordinate landscape with the folded and monotone maps.

A nonzero cone normal also explains the extension to arbitrary smooth feasible full heads whose open raw neighborhood maps into $\mathcal C$: at a finite fully linearly polarized state $s=(I,Ie)$, the normal $(-1,e)$ is annihilated by the head Jacobian. If all samplewise loss gradients have this form, the chain rule gives zero gradient for any upstream backbone realizing those outputs.

\section{Convex post-hoc projection reference}
We also retain the convex reference. For any $v\in\mathbb R^3$, let $P_W(v)$ be its $W$-projection onto $\mathcal C$. For any $s^*\in\mathcal C$,
\begin{equation}
 \|P_W(v)-s^*\|_W^2\leq\|v-s^*\|_W^2-\|v-P_W(v)\|_W^2.
 \label{SI-eq:projbound}
\end{equation}
This standard projection inequality \cite{supp-prox} applies to arbitrary network outputs as well as least-squares estimates. Separately, when $v$ is the unconstrained least-squares estimate, $P_W(v)$ solves the ideal observation-constrained least-squares problem. For $v=(t,y)$, $\rho=\|y\|_2$, the projection is unchanged if $t\geq\rho$; otherwise, for $\rho>0$,
\begin{equation}
 b=\max\{0,(2t+\rho)/3\},\qquad P_W(v)=(b,by/\rho).
 \label{SI-eq:weightedprojection}
\end{equation}
At $\rho=0$, the answer is $(\max\{t,0\},0,0)$. The inequality measures post-hoc error in $W$.

\section{Saved renderer and constructed stationary field}
\subsection{Reference configuration}
Figure~\ref{SI-fig:renderer} shows the saved $32\times32$ reference, generated with Mitsuba 3.9.1 in its \href{https://mitsuba.readthedocs.io/en/latest/src/rendering/polarized_rendering.html}{documented polarized mode}. The scene is a radius-one GGX rough-conductor sphere, roughness $0.15$, spectral constants $\eta=0.2$ and $k=3$, and a uniform unit environment. A perspective camera at $(0,0,4)$ looks at the origin, with $35$-degree field of view and box pixel filter. Four independent renders, seeds 821--824 and 1,024 samples per pixel each, were averaged.

We use the first RGB output channel of the spectral renderer and the saved Stokes basis. Intensities are relative radiance in arbitrary units. The stored fourth Stokes component is zero. Full-image mean DoLP is $1.311\%$, maximum DoLP $4.744\%$, and intensity ranges from $0.97337$ to $1.21369$. This weak-polarization, finite-sample array is the target used in the stationary-field construction below.

\subsection{A constructed image-based stationary counterexample}
For each of the 1,024 reference pixels, retain $I^*,d,e$ and initialize the folded head at the aligned state in \eqref{SI-eq:criticalstates}: $I_b=I_+=I^*(2+d)/3$, $s_b=I_b(1,e)$. The construction supplies the reference direction to the initialization. At $d=0$, choose any unit direction. Set $I_0=1$ and $a=\log(\exp(I_b)-1)$, so the intensity activation is the same softplus used in the theory.

\begin{figure*}[!t]
\centering\includegraphics[width=\textwidth]{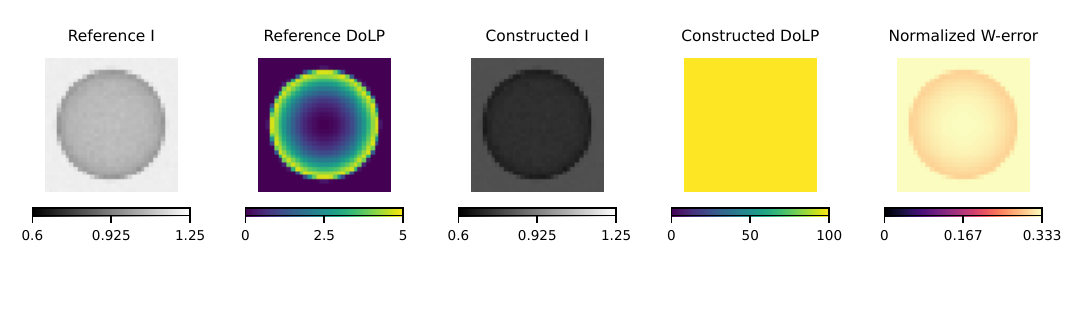}
\caption{A frozen numerical reference and a constructed stationary field on its $32\times32$ grid. The first two panels show the saved first-RGB-channel intensity and DoLP. The next two use the reference directions and the analytically stationary intensity; they are a mathematical counterexample field. Reference and constructed intensity share a scale. Reference DoLP spans 0--5 percent, while the constructed output has DoLP one and uses 0--100 percent. The final panel is normalized weighted error.}
\label{SI-fig:renderer}\label{SI-fig:counterexample}
\end{figure*}

Every initialized output is physically admissible and has DoLP one. Its normalized weighted error is exactly
\begin{equation}
 E_b=\frac{\|s_b-s^*\|_W^2}{(I^*)^2}=\frac{(1-d)^2}{3}.
 \label{SI-eq:imageerror}
\end{equation}
Figure~\ref{SI-fig:counterexample} shows the constructed field. The mean $E_b$ is $0.324735$, with range $0.302460$--$1/3$. Over 1,000 independent-pixel raw-coordinate gradient steps of size $0.1$, the largest raw-gradient component is $2.66\times10^{-16}$ and the largest parameter change is $4.44\times10^{-16}$ in this float64 run. The mean normalized half-loss remains $0.162368$.

The field has zero head gradient and admissible outputs, yet a positive objective gap. Its negative curvature permits escape under perturbations. The archive supplies the reference, initial and final coordinates, and per-step losses.

\section{Finite-radius derivatives and error envelopes}
\label{SI-sec:nearproof}
All quantities in this section are evaluated at known $I>0$, $w=re$, $r>0$, $\|e\|=1$, and $s_r=I(1,\phi e)$ with $\phi=2r/(1+r^2)$. The physical objective is the fixed quadratic $L(s)=\tfrac12\|s-z\|_W^2+c$, $W\succ0$, and $g=\nabla L(s_r)$. The rational head has the exact derivatives
\begin{equation}
 \begin{aligned}
 \phi&=\frac{2r}{1+r^2},& t_r&=\frac{2}{1+r^2},\\
 a_r&=\phi'=\frac{2(1-r^2)}{(1+r^2)^2},&
 \kappa_r&=\phi''(r)=\frac{4r(r^2-3)}{(1+r^2)^3},\\
 \ell_r&=t_r'=-\frac{4r}{(1+r^2)^2},&
 J&=a_ree^{\mathsf T}+t_r P_\perp,\\
 K&=\kappa_ree^{\mathsf T}+\ell_rP_\perp.&&
 \end{aligned}
 \label{SI-eq:suppcoefficients}
\end{equation}
Here $J=DF(re)$ and $K=D(DF)(re)[e]$. In particular, $Je=a_re$, $\|J\|_2=t_r$, and $K=-\operatorname{Id}$ at $r=1$.

\subsection{Recovery bias and its bound}
The composite Hessian splits into physical-Hessian and map-curvature terms. Radial multiplication gives
\begin{equation}
 \begin{aligned}
 v_r&=I^2J^{\mathsf T}W_{pp}Je+IKg_p\\
    &=I^2a_rJW_{pp}e+IKg_p,\\
 b_r&=g_I+\phi e^{\mathsf T}g_p.
 \end{aligned}
 \label{SI-eq:suppnearhvp}
\end{equation}
Set $e_0=(1,0,0)^{\mathsf T}$ and $S_e=(e,-\operatorname{Id}_2)^{\mathsf T}\in\mathbb R^{3\times2}$. Using the exact-fold formula $g^{\mathrm{naive}}=e_0b_r+S_ev_r/I$ at this radius gives the exact bias
\begin{equation}
 \begin{aligned}
 g^{\mathrm{naive}}-g={}&(\phi-1)(e^{\mathsf T}g_p)e_0\\
 &+S_e\{Ia_rJW_{pp}e+(K+\operatorname{Id})g_p\},
 \end{aligned}
 \label{SI-eq:suppbias}
\end{equation}
Apply the $M=W^{-1}$ norm, with $k_b=\|e_0\|_M$, $k_v=\|S_e\|_{2\to M}$, and supplied bound $\|g_p\|_2\leq G_p$. Adding derivative errors $|\delta b|\leq\varepsilon_b$, $\|\delta v\|_2\leq\varepsilon_v$ gives the scalar naive-reuse envelope used in the prescribed-envelope comparison:
\begin{equation}
 \begin{aligned}
 \eta_r={}&k_b\{(1-\phi)G_p+\varepsilon_b\}\\
 &+k_v\{I|a_r|\|J\|_2\|W_{pp}\|_2
       +\|K+\operatorname{Id}\|_2G_p+\varepsilon_v/I\}.
 \end{aligned}
 \label{SI-eq:nearenvelope}
\end{equation}
Its scalar coefficients are given above. The physical model in this retained bound is quadratic.

For a tighter envelope define
\begin{equation}
 \begin{aligned}
 k_r&=Ia_rS_eJW_{pp}e,\\
 A_r&=(\phi-1)e_0e^{\mathsf T}+S_e(K+\operatorname{Id}).
 \end{aligned}
 \label{SI-eq:suppoperator}
\end{equation}
Then
\begin{equation}
 \|g^{\mathrm{naive}}-g\|_M\leq
 \|k_r\|_M+\sqrt{\lambda_{\max}(A_r^{\mathsf T}MA_r)}G_p.
 \label{SI-eq:supptightenvelope}
\end{equation}
This refinement follows directly from \eqref{SI-eq:suppbias}; the prescribed-envelope plots in Section S7 use the scalar envelope.

A known model gives one conservative choice without a fitted derivative error:
\begin{equation}
 G_p=I\|W_{pI}\|_2+I\phi\|W_{pp}\|_2+\|(q_W)_p\|_2,
 \qquad q_W=Wz.
 \label{SI-eq:suppG}
\end{equation}
This choice uses the exposed model quantity $q_W$. Otherwise, the application must supply a bound $G_p$.

Writing $\delta=r-1$ gives $1-\phi=\delta^2/(1+r^2)$, $a_r=-\delta+O(\delta^2)$, $1+\kappa_r=3\delta+O(\delta^2)$, and $1+\ell_r=\delta+O(\delta^2)$. The physical-Hessian bias is generally first order. For a general $C^2$ objective, replace $W_{pp}$ by its local physical Hessian block; a bound on this block is additionally needed. Radius alone does not bound the recovery error.

\subsection{Why support slack cannot be omitted}
Set $n=(-1,e)$ and $\sigma_r=I(1-\phi)\geq0$. Since $n^{\mathsf T}u\leq0$ for every $u\in\mathcal C$ and $n^{\mathsf T}s_r=-\sigma_r$, completion of the quadratic gives, for every $\lambda\geq0$,
\begin{equation}
 \begin{aligned}
 L(s_r)-L(u)&\leq-(g+\lambda n)^{\mathsf T}(u-s_r)\\
 &\quad-\tfrac12\|u-s_r\|_W^2+\lambda\sigma_r\\
 &\leq\tfrac12\|g+\lambda n\|_M^2+\lambda\sigma_r.
 \end{aligned}
 \label{SI-eq:suppslackproof}
\end{equation}
Gradient-error inflation gives the main paper's near-fold bound. Its support plane is $n^{\mathsf T}u\leq0$, not the plane through the interior $s_r$.

For the counterexample take $e=(1,0)$, $r=4/5$, $I=1$, $W=\operatorname{diag}(1,1/2,1/2)$. Then $\phi=40/41$ and
\begin{equation}
 z=(4/5,282/205,0),\quad
 P_{\mathcal C}^W(z)=(122/123,122/123,0).
 \label{SI-eq:suppcounterexample}
\end{equation}
The exact gradient at $s_r$ is $-n/5$, so its distance to the negative normal ray is zero. The rational coordinates above are chosen to give this gradient while keeping $s_r$ interior; direct substitution gives the positive gap $241/50430$. The corrected half-space minimizer is feasible and attains the bound. The executable independent check retains the analytic expressions, sequential least-squares programming (SLSQP) comparison, and floating-point tolerances.

\section{Numerical audit definitions and access model}
\label{SI-sec:audit}
For a candidate cone quadratic-program (QP) solution $x$, write $h(x)=x_I-\|x_p\|_2$ and $g_x=W(x-z)$. In the 84 stationary and 108 nonstationary retained checks the solver starts at $(1.5,0,0)$, uses analytic first derivatives, SLSQP objective tolerance $10^{-13}$, and at most 1,000 iterations. On the non-apex boundary, define $n_x=(-1,x_p/\|x_p\|_2)$ and the multiplier and Karush--Kuhn--Tucker (KKT) residuals by
\begin{equation}
 \begin{aligned}
 \lambda_x&=\max\{0,-g_x^{\mathsf T}n_x/\|n_x\|_2^2\},\\
 R_{\mathrm{stat}}&=\|g_x+\lambda_xn_x\|_\infty,\\
 R_{\mathrm{comp}}&=|\lambda_xh(x)|,\qquad
 R_{\mathrm{feas}}=\max\{0,-h(x)\}.
 \end{aligned}
 \label{SI-eq:suppkkt}
\end{equation}
The retained implementation uses $10^{-7}$ as its numerical boundary-classification tolerance: it classifies $h(x)\leq10^{-7}$ as boundary and otherwise sets the multiplier to zero. The smallest solution norms in the 84 and 108 retained cases are 1.452 and 0.4142, respectively. All 84/108-case solutions are away from the cone apex; the sensitivity implementation checks the apex separately. At the apex the full normal cone must be used: optimality is $g_0\in\mathcal C$, rather than a single fixed normal. Euclidean projection onto $\mathcal C$ gives a cone-distance residual there. The single-normal residual above applies away from the apex.

The constructed derivative-only wrapper exposes composite value, head first derivatives, normalized intensity derivative, radial HVP, and the fixed metric, while its verification harness retains a physical-gradient oracle. The harness compares recovered and direct gradients; the restricted interface is an implementation of a stipulated information model. Sensitivity perturbations use deterministic envelopes fixed before their errors are evaluated. Finite-difference HVP checks use central differences of the analytic raw gradient; their observed discrepancies test the implementation but are not universal error envelopes.

For the 82-case sensitivity block, the SLSQP starting point is $(1.5,0.1,-0.1)$ and the iteration limit is 2,000; the objective tolerance remains $10^{-13}$. All returned solutions are non-apex. The three separate apex checks are exact self-duality tests, not additional SLSQP solves. The radius block uses 15 log-spaced distances on each sheet. The intensity block spans $10^{-3}$ to $10^1$; the metric block spans $\kappa_2(W)=1$ to $10^4$ with a fixed orthogonal eigenbasis. Derivative envelopes are log-spaced from $10^{-8}$ to $10^{-1}$.

\section{Auxiliary geometry and direct-coordinate dynamics}
\label{SI-sec:conditioning}
\begin{figure*}[!t]
\centering\includegraphics[width=.96\textwidth]{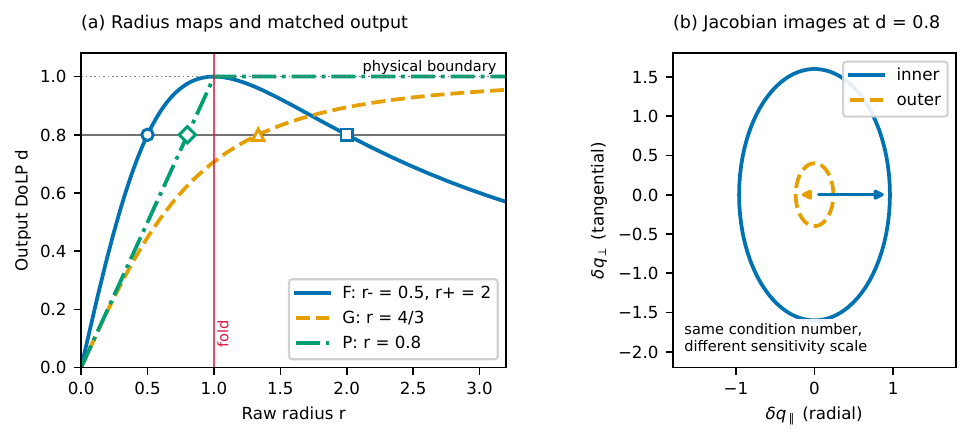}
\caption{Matched outputs and different raw-coordinate geometry. (a) Radial output degree for the folded map $F$, monotone map $G$, and disk projection $P$. At $d=0.8$, the raw radii are $0.5$ and $2$ for $F$, $4/3$ for $G$, and $0.8$ for $P$, matching the descent diagnostic's initialization. (b) The differential images of a unit raw perturbation ball at the two $F$ preimages are concentric ellipses in physical polarization-increment coordinates. Their aspect ratios agree, but the outer-sheet singular values are one quarter of the inner-sheet values. Arrows show the opposite signs of the radial eigenvalues. These are local Jacobian images, rather than finite perturbation regions.}
\label{SI-fig:geometry}
\end{figure*}

For raw polarization coordinates $w\in\mathbb R^2$, $r=\|w\|_2$, consider
\begin{equation}
 \begin{aligned}
 F(w)&=\frac{2w}{1+r^2},\qquad G(w)=\frac{w}{\sqrt{1+r^2}},\\
 P(w)&=\frac{w}{\max\{1,r\}}.
 \end{aligned}
 \label{SI-eq:maps}
\end{equation}
$F$ reaches the closed disk and folds at $r=1$. $G$ is a smooth bijection onto the open disk. $P$ is Euclidean disk projection and is nonsmooth at $r=1$. Figure~\ref{SI-fig:geometry} distinguishes a finite fold from asymptotic boundary approach and shows how duplicated physical outputs can have different local sensitivity.

\subsection{Exact comparison at matched polarization}
For a radial map, use an orthonormal raw basis consisting of $w/r$ and its perpendicular. Write $\kappa_2(A)=\sigma_{\max}(A)/\sigma_{\min}(A)$ for the spectral condition number. Direct differentiation gives Table~\ref{SI-tab:spectra}.
\begin{table*}[!t]
\caption{Exact raw-coordinate Jacobian eigenvalues and image. Here $r>0$; the origin follows by continuity. The condition numbers below concern the normalized-polarization block at fixed intensity.}
\label{SI-tab:spectra}
\centering\small\renewcommand{\arraystretch}{1.15}
\begin{tabular}{@{}llll@{}}\toprule
Map & Image & Radial eigenvalue & Tangential eigenvalue\\\midrule
$F$ & Closed disk & $2(1-r^2)/(1+r^2)^2$ & $2/(1+r^2)$\\
$G$ & Open disk & $(1+r^2)^{-3/2}$ & $(1+r^2)^{-1/2}$\\
$P$, $r<1$ & Disk interior & $1$ & $1$\\
$P$, $r>1$ & Boundary circle & $0$ & $1/r$\\\bottomrule
\end{tabular}
\end{table*}

\begin{proposition}[Matched-DoLP spectra]
\label{SI-prop:spectra}
For $0<d<1$, let $s_d=\sqrt{1-d^2}$. A fixed output $de$ has two $F$ preimages,
\begin{equation}
 r_-=\frac{d}{1+s_d},\qquad r_+=\frac{1+s_d}{d}=r_-^{-1}.
 \label{SI-eq:inverse}
\end{equation}
Their signed radial/tangential eigenpairs are
\begin{equation}
 \begin{aligned}
 F_- &: \bigl(s_d(1+s_d),\,1+s_d\bigr),\\
 F_+ &: \bigl(-s_d(1-s_d),\,1-s_d\bigr),\\
 G &: (s_d^3,s_d).
 \end{aligned}
 \label{SI-eq:matchedspectra}
\end{equation}
Consequently,
\begin{equation}
 \kappa_2(DF)=\frac1{\sqrt{1-d^2}},\qquad
 \kappa_2(DG)=\frac1{1-d^2}.
 \label{SI-eq:conditionnumbers}
\end{equation}
\end{proposition}
\begin{proof}
Solve $d=2r/(1+r^2)$ for $r_\pm$, and $d=r/\sqrt{1+r^2}$ for $G$. Substitution into Table~\ref{SI-tab:spectra} gives \eqref{SI-eq:matchedspectra}. Taking singular-value ratios proves \eqref{SI-eq:conditionnumbers}.
\end{proof}
Thus the folded map attains the nondegenerate boundary exponent in Theorem~\ref{SI-thm:rate}. Removing its finite fold with $G$ avoids finite boundary critical points but worsens this raw-coordinate condition number near $d=1$.

The two $F$ sheets have equal condition numbers, yet their singular-value scale ratio is $(1-s_d)/(1+s_d)$. For $q=H(w)$ and $J=DH(w)$, direct raw-coordinate Euclidean gradient flow gives $\dot q=-JJ^\mathsf T\nabla_q\ell$. The radial mobilities, the squared radial Jacobian eigenvalues, are
\begin{equation}
 \begin{aligned}
 m_-&=(1-d^2)(1+s_d)^2,\\
 m_+&=(1-d^2)(1-s_d)^2,\qquad m_G=(1-d^2)^3.
 \end{aligned}
 \label{SI-eq:mobility}
\end{equation}
As $d\downarrow0$, $m_-\to4$ while $m_+\sim d^4/4$. Their ratio is asymptotic to $d^4/16$. Low-polarization outputs can therefore be nearly isotropically conditioned on both sheets while having very different absolute optimization scales. A backbone or optimizer can precondition these coordinates; these formulas are not network condition numbers.

\subsection{A finite-boundary conditioning bound}
At a smooth physical boundary, a finite smooth parametrization loses a normal derivative. For the disk this follows immediately: if $H(w_0)=e$, $\|e\|=1$, then $e^\mathsf TH(w)\leq1$ has an interior maximum at $w_0$, so $DH(w_0)^\mathsf Te=0$. This is a supporting-normal form of the first-order obstruction in \cite{supp-lifts}. We now quantify the conditioning for radial heads.

\begin{theorem}[Finite-boundary rate]
\label{SI-thm:rate}
Let $H(w)=\phi(r)w/r$. Suppose $R>0$ has a two-sided open neighborhood on which $\phi$ is $C^2$, $0\leq\phi\leq1$, $\phi(R)=1$, and $|\phi''|\leq L_\phi$ for some scalar $L_\phi>0$. For $r\to R$ with $d=\phi(r)<1$,
\begin{equation}
 \kappa_2(DH(w))\geq\frac{\phi(r)}{r\sqrt{2L_\phi(1-d)}}.
 \label{SI-eq:lowerbound}
\end{equation}
A singular Jacobian has infinite condition number. If $\phi''(R)=-c<0$, then
\begin{equation}
 \kappa_2(DH(w))\sim\frac{1}{R\sqrt{2c(1-d)}}.
 \label{SI-eq:rateasymptotic}
\end{equation}
\end{theorem}
\begin{proof}
Put $h=1-\phi$. Fermat's rule gives $h'(R)=0$. In a sufficiently small neighborhood, $z=r-h'(r)/L_\phi$ remains in the original neighborhood: $|h'(r)|\leq L_\phi|r-R|$. Taylor's upper bound and $h(z)\geq0$ give
\begin{equation*}
 0\leq h(z)\leq h(r)-|h'(r)|^2/(2L_\phi).
\end{equation*}
This local self-bounding smoothness argument is related to standard nonnegative-loss estimates \cite{supp-selfbound}. Near $R$, the singular values satisfy $|\phi'(r)|\le\phi(r)/r$, since they tend to $0$ and $1/R>0$, respectively. Their ratio gives \eqref{SI-eq:lowerbound} at regular points; singular points have infinite condition number. If $\phi''(R)=-c$, Taylor expansion gives $1-d\sim c(r-R)^2/2$ and $|\phi'(r)|\sim c|r-R|$, proving \eqref{SI-eq:rateasymptotic}.
\end{proof}
The two-sided neighborhood is essential: the map $\phi(r)=r/R$ defined only for $r\leq R$ is a counterexample without it. The exponent also requires bounded second derivative. A $C^1$ local profile $1-c|r-R|^p$ with $1<p<2$ has a smaller exponent. The theorem identifies an attainable rate for this regularity class; the rational map is one example, not a uniquely optimal head.

\subsection{An exact stationary example}
For ideal four-analyzer measurements, $W=\operatorname{diag}(1,1/2,1/2)$. Let $s^*=I^*(1,de)$ with $I^*>0$ and $0<d<1$. The aligned folded stationary output has
\begin{equation}
 \begin{aligned}
 I_b&=I^*(2+d)/3,\qquad s_b=I_b(1,e),\\
 L(s_b)-L(s^*)&=(I^*)^2(1-d)^2/6,\\
 \gamma_r&=-(I^*)^2(2+d)(1-d)/9.
 \end{aligned}
 \label{SI-eq:boundaryexample}
\end{equation}
Its gradient is $I^*(1-d)(-1,e)/3$. The negative gradient points into the cone, yet the head's first derivative annihilates it. Supplementary Section S1 gives the complete ideal-metric critical-point classification, including the anti-aligned saddle and the monotone and disk-projection controls. Convex post-hoc projection remains a separate baseline \cite{supp-prox}; its standard error guarantee is recalled in Supplementary Section S2.

\begin{figure*}[!t]
\centering\includegraphics[width=.96\textwidth]{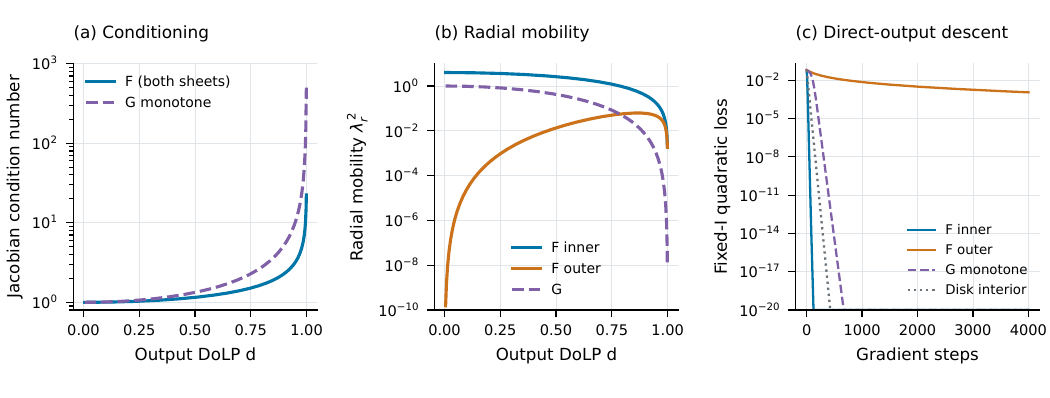}
\caption{Auxiliary conditioning and direct-output dynamics. (a) Exact normalized-polarization Jacobian condition numbers at matched output DoLP. (b) Radial mobilities: the two folded sheets have equal condition numbers and different absolute scales. (c) Fixed-intensity descent to $(I,Q,U)=(1,0.3,0)$, with initial normalized output $(0.8,0)$ in every arm, quadratic polarization weight $1/2$, step $0.1$, and 4,000 gradient updates. Values below $10^{-20}$ are clipped for display. The trajectories use the prescribed raw coordinates and a common step size.}
\label{SI-fig:conditioning}
\end{figure*}

\subsection{Derivative checks and matched initialization}
The numerical study uses NumPy 2.3.5, SciPy 1.18.1, and Matplotlib 3.11.2 in float64; it has no random optimization seed. Central differences with step $10^{-5}$ check 36 Jacobians at matched output degrees $0.05,0.3,0.8,0.99$ and directions $0,0.71,2.1$ radians, plus eight full-head fold Hessians. The maximum Jacobian error is $2.00\times10^{-10}$, singular-value discrepancy $6.66\times10^{-16}$, stationary-gradient magnitude $1.67\times10^{-16}$, and Hessian error $7.42\times10^{-11}$. These checks supplement the proofs.

Figure~\ref{SI-fig:conditioning}(c) uses independent raw polarization coordinates and fixed intensity. The folded inner and outer initial radii are $0.5$ and $2$, the monotone radius is $4/3$, and the disk radius is $0.8$. Their physical initial prediction is identical. Plain gradient descent uses step $0.1$ for 4,000 updates of $\tfrac14\|H(w)-(0.3,0)\|^2$. The outer folded arm ends at loss $1.124\times10^{-3}$; the other three end below $10^{-20}$. The prescribed raw coordinates and their Jacobians determine these untuned trajectories.

\section{Retained sensitivity to intensity and metric conditioning}
\begin{figure*}[!t]
\centering\includegraphics[width=.94\textwidth]{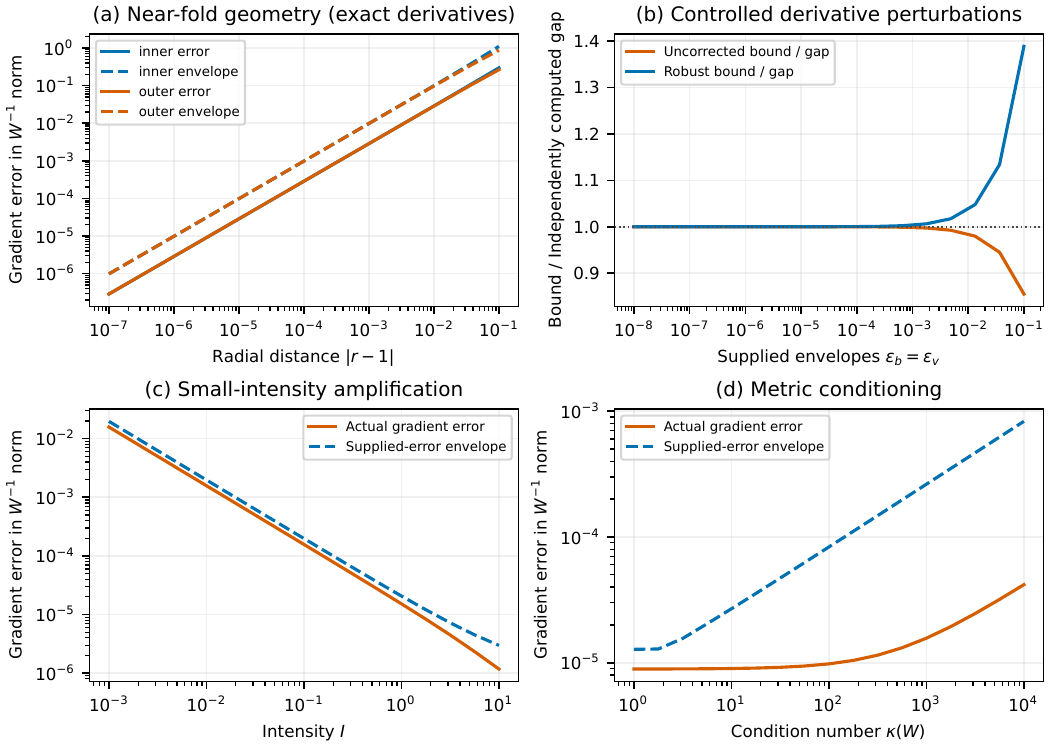}
\caption{Retained controlled sensitivity. The scalar envelopes are supplied before evaluating recovered errors. Panel (b) compares bounds with an independently computed objective-gap estimate. The metric norm changes with $W$ in panel (d).}
\label{SI-fig:retainedsensitivity}
\end{figure*}

The 82-case sensitivity study supplies controlled perturbations and their error envelopes. Figure~\ref{SI-fig:retainedsensitivity} shows its finite-radius error envelope, derivative-error perturbations, low-intensity amplification, and metric-conditioning sweep. The denominator in panel (b) is an independently computed gap estimate. Sections S8--S9 report a separate primal-dual numerical width for that estimate.

\section{First-order inversion and numerical gap brackets}
\subsection{Derivative-information regimes}
For $w=re$, $r>0$, the folded Jacobian has radial eigenvalue $a_r=2(1-r^2)/(1+r^2)^2$ and tangential eigenvalue $t_r=2/(1+r^2)$. It is singular only at $r=1$. Off the fold, $j_r=IJ^{\mathsf T}g_p$ gives the main paper's first-order inverse. If the first-derivative errors are $\delta b,\delta j$, put
\begin{equation}
 A_\phi=\begin{pmatrix}-\phi e^{\mathsf T}\\\operatorname{Id}_2\end{pmatrix},\qquad
 T_r=A_\phi J^{-1}/I.
 \label{SI-eq:inverseerroroperator}
\end{equation}
The exact recovery error and an induced-norm envelope are
\begin{equation}
 \begin{aligned}
 \delta g^{(1)}&=e_0\delta b+T_r\delta j,\\
 \|\delta g^{(1)}\|_M&\leq k_b\varepsilon_b+
 \sqrt{\lambda_{\max}(T_r^{\mathsf T}MT_r)}\,\varepsilon_j.
 \end{aligned}
 \label{SI-eq:inverseenvelope}
\end{equation}
A simpler bound follows from $\|J^{-1}\|_2=1/\min(|a_r|,t_r)$ and $\|A_\phi\|_2=\sqrt{1+\phi^2}$. The prescribed-envelope comparison uses \eqref{SI-eq:inverseenvelope}, the finite-radius naive-reuse envelope, and the model-corrected envelope described in Section S9, on the same prescribed inner radii and derivative perturbations. The two derivative orders have equal numeric envelope magnitudes for illustration; this does not impose equivalent acquisition cost or error statistics.

\subsection{Primal and dual numerical candidates}
For $\mathcal Q(u)=\tfrac12\|u-z\|_W^2$, the Lorentz cone is self-dual. For every $q_{\mathrm d}\in\mathcal C$ and feasible $u$, $q_{\mathrm d}^{\mathsf T}u\geq0$, and therefore
\begin{equation}
 \begin{aligned}
 \mathcal Q(u)&\geq \mathcal Q(u)-q_{\mathrm d}^{\mathsf T}u\\
 &\geq\inf_x\{\mathcal Q(x)-q_{\mathrm d}^{\mathsf T}x\}\\
 &=-q_{\mathrm d}^{\mathsf T}z-\tfrac12q_{\mathrm d}^{\mathsf T}W^{-1}q_{\mathrm d}.
 \end{aligned}
 \label{SI-eq:dualgapderivation}
\end{equation}
This is the standard weak-duality construction \cite{supp-cvxbook}, specialized to the fixed quadratic. Denote its dual value by $\mathcal D(q_{\mathrm d})$. For a feasible primal point $\bar u$, the objective gap lies between $\mathcal Q(s)-\mathcal Q(\bar u)$ and $\mathcal Q(s)-\mathcal D(q_{\mathrm d})$; the interval width is the primal-dual objective gap.

For each saved SLSQP candidate $\widehat u$, the audit forms $\bar u=\Pi_{\mathcal C}(\widehat u)$ using Euclidean Lorentz-cone projection,
\begin{equation}
 \Pi_{\mathcal C}(t,p)=
 \begin{cases}
 (t,p),&\|p\|\leq t,\\
 0,&\|p\|\leq-t,\\
 \tfrac12(t+\|p\|)(1,p/\|p\|),&\text{otherwise}.
 \end{cases}
 \label{SI-eq:euclideancone}
\end{equation}
This is a feasibility repair for the numerical audit, distinct from the physical-metric estimate. It has maximum norm $6.80\times10^{-14}$ across the 274 saved cases. For the dual, the code selects the largest objective among $q_{\mathrm d}=0$, the cone projection of $W(\bar u-z)$, and a cone-feasible supporting-normal ray candidate. For $\bar u_p\ne0$, the latter is $q_{\mathrm d}=\gamma d_{\mathrm{ray}}$, with direction $d_{\mathrm{ray}}=(1,-\bar u_p/\|\bar u_p\|)$ and coefficient
\begin{equation}
 \gamma=\max\{0,-d_{\mathrm{ray}}^{\mathsf T}z/(d_{\mathrm{ray}}^{\mathsf T}W^{-1}d_{\mathrm{ray}})\}.
 \label{SI-eq:dualcandidate}
\end{equation}
When $\bar u_p=0$, the ray candidate is omitted and the other two candidates remain; the apex optimality test uses the full self-dual cone as in Section S5. Cone feasibility of the primal and dual candidates and every recorded width are checked. Small negative slacks caused by floating-point rounding are repaired by raising the intensity coordinate to at least the polarization norm, with the numerical guard recorded in the code. Float64 widths are tested with tolerance $2\times10^{-12}$; all signed values are retained. These float64 results support numerical consistency; outward rounding and interval error bounds would be needed for rigorous computed enclosures.

\section{Purpose-specific numerical checks}
\subsection{Certificate conditions and tightness}
Four synthetic analyzer configurations induce one ideal and three non-diagonal SPD metrics, each normalized to $\operatorname{tr}W=2$. The nonideal configurations vary analyzer angle, gain, efficiency, and observation variance; their exact matrices are retained. At $I=1.2$, three directions $0,0.7,1.9$ radians and seven values $\alpha\in\{-0.6,-0.1,0.01,0.1,0.4,1,2\}$ give 84 fold states by setting $z=s-\alpha W^{-1}n$. This construction tests the certificate's conditions directly. We use $I=\exp(a)$ for this block, a regular positive coordinate covered by Theorem~\ref{SI-thm:certificate}.

An independent Stokes-coordinate SLSQP solve, with analytic first derivatives, cone constraint $I-\sqrt{Q^2+U^2}\geq0$, tolerance $10^{-13}$, and limit 1,000 iterations, computes an approximation to each constrained minimum. All 84 solves succeed. Maximum cone violation is $7.88\times10^{-14}$ and KKT stationarity residual is $2.18\times10^{-7}$. Raw stationarity error is $9.77\times10^{-16}$. Central differences of the raw gradient along the radial direction, step $10^{-4}$, give curvature error $2.13\times10^{-8}$.

\begin{figure*}[!t]
\centering\includegraphics[width=\textwidth]{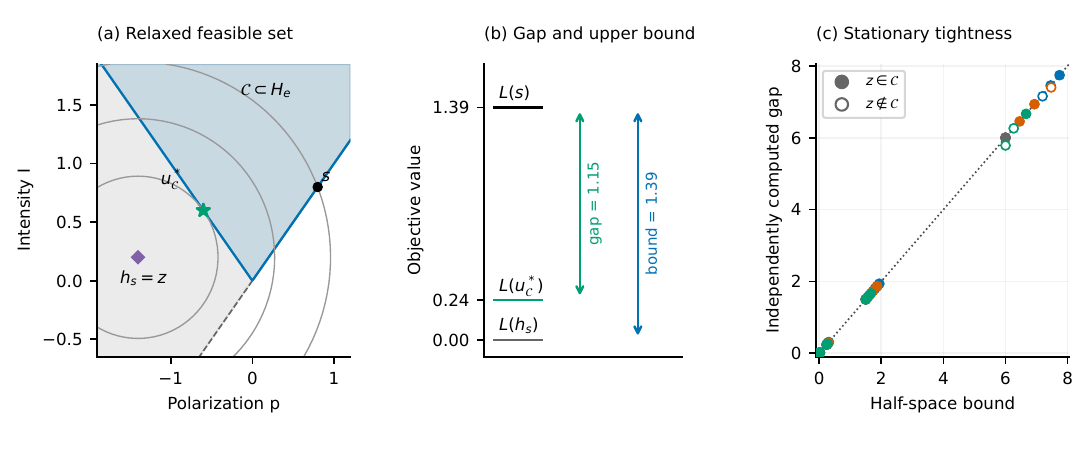}
\caption{From recovered gradients to a quadratic gap bound. (a) In the $(p,I)$ section, $\mathcal C\subset H_e$; relaxing the cone moves the minimum from $u^*_{\mathcal C}=(-0.6,0.6)$ to $h_s=z=(-1.4,0.2)$. The illustrative metric is $\operatorname{diag}(1/2,1)$ and $s=(0.8,0.8)$. (b) Objective heights make the two differences explicit: the true gap is $1.39-0.24=1.15$, while the half-space bound is $1.39-0=1.39$. Equality holds exactly when $h_s$ is cone-feasible. (c) The 60 negative-curvature stationary cases, with independently computed gap estimates. Filled markers have feasible $z$; hollow markers have infeasible $z$. The numerical primal-dual widths are reported separately from the analytic equality statement.}
\label{SI-fig:certificate}
\end{figure*}

Theorem~\ref{SI-thm:certificate} classifies the 24 cases with $\alpha\leq0$ as constrained optima; their computed gaps have magnitude at most $4.73\times10^{-14}$. All 60 positive-$\alpha$ cases have negative radial curvature. Of those, 56 have feasible $z$ and attain the certificate within $4.80\times10^{-14}$; the four infeasible-$z$ cases have a strict upper bound. Figure~\ref{SI-fig:certificate} shows these checks. These finite-case tests supplement the proof.

For the nonstationary extension, keep the same four metrics, three directions, and $I=1.2$, and use nine fixed estimates $z$: three interior, three boundary, and three unphysical vectors. All 108 raw gradients are nonzero. A central difference of $\nabla_w\widetilde L$ along $e$, step $10^{-4}$, measures the vector HVP independently. The maximum error in $v=-Ig_p$ is $1.36\times10^{-8}$, and complete gradient recovery has error $1.13\times10^{-8}$. The independent cone solves all succeed with maximum KKT residual $1.37\times10^{-7}$. Using the finite-difference HVP, the largest numerical excess of the solver-computed gap over the main paper's residual bound is $1.01\times10^{-9}$; using exact derivatives it is $2.22\times10^{-16}$. The main paper's inflated bound applies when a deterministic derivative-error envelope is supplied. These 108 cases test gradient recovery away from raw stationarity, while still exactly on the fold.

\subsection{Controlled numerical sensitivity}
The retained sensitivity record tests the error terms with 82 deterministic cases: 30 radii $r=1\pm\delta$, $\delta\in[10^{-7},10^{-1}]$; one exact-fold baseline; 17 derivative envelopes; 17 intensities; and 17 metric condition numbers. Each case has a fixed quadratic. The supplied perturbations are $\delta b=-0.75\varepsilon_b$ and $\delta v=0.8\varepsilon_ve$, chosen before evaluating the recovery error. The original radius study uses exact derivatives and Section S4's scalar naive-reuse envelope; the other panels use the exact fold. A conservative model-derived $G_p$ is supplied independently of the observed recovery error. The formulas and all metric matrices are retained.

All 82 independent SLSQP solves succeed, with maximum KKT residual $1.07\times10^{-7}$ and cone violation $9.43\times10^{-14}$. Central differences of the raw gradient, step $2\times10^{-5}$, give maximum HVP error $1.14\times10^{-8}$, or $2.14\times10^{-10}$ after division by $1+\|v\|_\infty$. Every recovered-gradient error lies within its supplied envelope; every robust bound exceeds the independently computed gap estimate. The primal-dual audit in Section S8 gives its numerical uncertainty. In all 17 deliberately directed derivative-perturbation cases, the uninflated bound falls below the computed gap estimate. These are directed perturbations of the derivative observations.

Separate checks use the analytic ideal-cone projection in 28 cases, verify the support-slack counterexample, and test the apex condition in three cases. The affine-perturbation family in the main paper's first-order ambiguity equation is checked at $\tau=-1,0,1$; its indistinguishable values and first raw derivatives are resolved by scalar-curvature recovery within $2.23\times10^{-16}$. Section S5 defines the solver residuals and the constructed derivative interface.

\subsection{Retained first-order and naive-reuse comparison}
We reuse the 30 inner/outer radius locations, now prescribing equal numerical envelopes $\varepsilon_b=\varepsilon_j=\varepsilon_v=10^{-6}$ for the two recovery routes. The imposed errors are $\delta b=-0.75\varepsilon_b$, $\delta j=0.8\varepsilon_je$, and $\delta v=0.8\varepsilon_ve$. All 30 errors lie within their analytic envelopes. With exact first derivatives, inverse recovery has maximum weighted error $1.05\times10^{-9}$ at these radii. The retained radius records include both sheets; the main paper now plots observed AD errors separately.

\subsection{Known-metric second-order correction}
The main paper subtracts $I^2a_rJW_{pp}e$ before inverting $K$. The triangle inequality applies to all errors in the prescribed norm balls. We test the same 30 radius points and absolute errors as the first/naive comparison. The fixed model quantities $r,e,I,I',W$ are treated as exact; uncertainty in them is not included in the displayed envelope. Independent 80-digit decimal central differences of the raw polarization gradient, radial step $10^{-18}$, agree with the analytic HVP to $4.75\times10^{-16}$ in Euclidean norm after float64 conversion. The exact-gradient recovery discrepancies and the prescribed-error envelopes are retained in \nolinkurl{corrected_recovery_checks.json}. The correction uses the retained derivative records.

\subsection{Nonquadratic objectives and independent derivative differences}
For the test objectives, let $s=(I,Q,U)$, $\chi=Q+0.4U$, $c_2=(0.3,-0.2,0.5)$, $c_3=(-0.4,0.7,0.2)$, and $\rho=\|s\|_2^2$. The fixed coefficients specify three $C^\infty$ functions with different physical Hessians:
\begin{equation}
 \begin{aligned}
 f_1(s)={}&0.2I^4+0.1\chi^4+0.3IQU+0.05U^3\\
          &+0.4I+0.2Q-0.3U,\\
 f_2(s)={}&\exp(c_2^{\mathsf T}s)+0.03\rho^2,\\
 f_3(s)={}&\log(1+\exp(c_3^{\mathsf T}s))\\
          &+0.07IQU+0.05\rho.
 \end{aligned}
 \label{SI-eq:nonquadraticlosses}
\end{equation}
Their physical gradients are differentiated analytically in the script and independently checked by 80-digit decimal central differences, step $10^{-18}$; the largest discrepancy is $4.00\times10^{-36}$ in infinity norm. The intensities are $0.2,1.2,5$ and directions $0,0.7,1.9$ radians. Decimal direction vectors are normalized at working precision to put each evaluation on the exact fold to that precision.

For each of the 27 cases, a central second difference of the composite objective along $w=(1+t)e$, step $10^{-12}$, estimates $c_r$. Composite first derivatives with that scalar give maximum gradient-recovery discrepancy $1.24\times10^{-23}$ in infinity norm. Separately, a central difference of the composite polarization gradient, radial step $10^{-18}$, estimates the vector HVP; its recovery discrepancy is at most $6.75\times10^{-35}$. The first derivative observation matrix has numerical rank two and augmentation by the radial second-output row gives rank three in all cases. The state-dependent physical Hessians are not supplied to either fold recovery route. These examples test the general recovery identity.

For completeness, with $j=\nabla_w\widetilde f$ and prescribed scalar-curvature error $|\delta c_r|\leq\varepsilon_c$, define
\begin{equation}
 J_0=\begin{pmatrix}0_{1\times2}\\\operatorname{Id}_2\end{pmatrix},\qquad
 q_e=(1,-e)^{\mathsf T}.
 \label{SI-eq:scalaroperators}
\end{equation}
The scalar fold formula has error $e_0\delta b+(J_0\delta j+q_e\delta c_r)/I$ and hence
\begin{equation}
 \begin{aligned}
 \|\widehat g-g\|_M\leq{}&k_b\varepsilon_b
 +\frac{\sqrt{\lambda_{\max}(J_0^{\mathsf T}MJ_0)}}{I}\varepsilon_j\\
 &+\frac{\|q_e\|_M}{I}\varepsilon_c.
 \end{aligned}
 \label{SI-eq:scalarerror}
\end{equation}
Here $I,e,I',M$ are exact and the derivative-error bounds are prescribed.

\section{Retained stationary classification and exact-fold errors}
A strict saddle means a stationary point with at least one negative Hessian eigenvalue; the Hessian may be singular.

\begin{theorem}[Curvature certificate at a fold]
\label{SI-thm:certificate}
For the fixed quadratic $L(s)=\tfrac12\|s-z\|_W^2+c$, $W\succ0$, let $(a,e)$, $\|e\|=1$, be an exact stationary point of $L\circ T_F$, with $I=I(a)>0$. Set $s=T_F(a,e)=I(1,e)$, $n=(-1,e)$, and $D_n=n^{\mathsf T}W^{-1}n$. The raw radial direction is an eigenvector of the composite Hessian. Its signed eigenvalue $\gamma_r$ satisfies
\begin{equation}
 \nabla_s L=\alpha n,\qquad \gamma_r=-I\alpha.
 \label{SI-eq:normalcurvature}
\end{equation}
If $\gamma_r\geq0$, $s$ is the global constrained optimum. If $\gamma_r<0$, the raw point is a strict saddle, and
\begin{equation}
 0<\Delta_{\mathcal C}(s)\leq
 \frac{D_n}{2I^2}\,[\max\{-\gamma_r,0\}]^2.
 \label{SI-eq:curvaturecertificate}
\end{equation}
The same upper bound is zero when $\gamma_r\geq0$. For $\gamma_r<0$, equality holds exactly when $z\in\mathcal C$.
\end{theorem}
\begin{proof}
At the fold, the full-head differential has image spanned by $(1,e)$ and $(0,e_\perp)$; its orthogonal complement is $n$. Raw stationarity therefore gives $\nabla_sL=\alpha n$. Along $w=(1+t)e$, the first output derivative vanishes and the second is $(0,-Ie)$. Hence $\gamma_r=-I\alpha$. Mixed radial derivatives vanish after contraction with $\alpha n$, so this is a Hessian eigenvalue, not only a directional test.

The normal cone of $\mathcal C$ at $s$ is $\{\tau n:\tau\geq0\}$. If $\alpha\leq0$, $-\nabla L$ lies in this normal cone, which is the sufficient optimality condition for the convex problem. If $\alpha>0$, the negative gradient has a strictly inward component and $s$ is not constrained-stationary. Finally, $s-z=\alpha W^{-1}n$ gives
\begin{equation*}
 L(s)-\min_{u\in\mathbb R^3}L(u)=\tfrac12\alpha^2D_n.
\end{equation*}
The constrained minimum is at least the unconstrained minimum, proving the upper bound. Since $W$ is positive definite, the minima coincide exactly when $z\in\mathcal C$.
\end{proof}
For $\varepsilon\geq0$ and exact stationary folds with $\gamma_r\geq-\varepsilon$, the gap is at most $D_n\varepsilon^2/(2I^2)$. Under ideal analyzers $D_n=3$. The general implication from second-order stationarity to constrained stationarity belongs to smooth-lift theory \cite{supp-lifts}. The expression above makes the coefficient and equality condition explicit for calibrated linear-Stokes data.

\subsection{Error-aware certificate at an exact fold}
Put $M=W^{-1}$, $e_0=(1,0,0)^{\mathsf T}$, and define
\begin{equation}
 S_e=\begin{pmatrix}e^{\mathsf T}\\-\operatorname{Id}_2\end{pmatrix},\quad
 k_b=\sqrt{M_{00}},\quad k_v=\sqrt{\lambda_{\max}(S_e^{\mathsf T}MS_e)}.
 \label{SI-eq:errorconstants}
\end{equation}
\begin{proposition}[Derivative-error inflation]
\label{SI-prop:robust}
At the exact fold, let $|\widehat b-b|\leq\varepsilon_b$ and $\|\widehat v-v\|_2\leq\varepsilon_v$, with $I,e,I',W$ exact. Recover $\widehat g$ by the main paper's vector fold formula and compute $\widehat{\mathcal B}$ by the main paper's fold residual expression using $\widehat g$. Then
\begin{equation}
 \|\widehat g-g\|_M\leq\eta,
 \qquad\eta=k_b\varepsilon_b+k_v\varepsilon_v/I,
 \label{SI-eq:gradienterror}
\end{equation}
and
\begin{equation}
 \Delta_{\mathcal C}(s)\leq
 \tfrac12\bigl(\sqrt{2\widehat{\mathcal B}}+\eta\bigr)^2.
 \label{SI-eq:robustcertificate}
\end{equation}
\end{proposition}
\begin{proof}
The recovery error is $e_0(\widehat b-b)+S_e(\widehat v-v)/I$. The induced norms and triangle inequality give \eqref{SI-eq:gradienterror}. Use the nonnegative multiplier minimizing the residual for $\widehat g$ in the quadratic bound for $g$, and apply $\|g+\lambda n\|_M\leq\|\widehat g+\lambda n\|_M+\eta$. This proves \eqref{SI-eq:robustcertificate}.
\end{proof}
The proof uses the standard Lipschitz property of distance to a convex set. The specialized content is the explicit propagation from head derivatives. If only $|\widehat{\partial_a\widetilde L}-\partial_a\widetilde L|\leq\varepsilon_a$ is supplied, then $\varepsilon_b=\varepsilon_a/|I'|$. A measured finite-difference discrepancy is not itself a guaranteed error envelope. The constants expose the $1/I$ and $1/|I'|$ amplification.

\section{Local-curvature input checks}
The nonquadratic objective is $f_1$ in \eqref{SI-eq:nonquadraticlosses}, with $\chi=Q+0.4U$. Its local polarization Hessian is
\begin{equation}
 H_{pp}=\begin{pmatrix}1.2\chi^2&0.48\chi^2+0.3I\\0.48\chi^2+0.3I&0.192\chi^2+0.3U\end{pmatrix}.
 \label{SI-eq:localHessian}
\end{equation}
The reporting metric is $M=\operatorname{Id}_3$, not this Hessian. Intensities and directions are $I=1.2$ and $e=(\cos0.7,\sin0.7)$. The radii are $0.9,0.99,0.9999,1,1.0001,1.01,1.1$. Eighty-digit decimal central differences, step $10^{-18}$, check $H_{pp}$ from the physical gradient and obtain the radial HVP from the raw gradient. Geometry is normalized at working precision. The Hessian check discrepancy is at most $4.00\times10^{-37}$; the composite HVP identity discrepancy after float64 conversion is at most $6.87\times10^{-16}$.

For curvature-input error, use $r=0.9,1,1.1$ and $\Delta H=\varepsilon_H B/\|B\|_2$, where
\begin{equation}
 B=\begin{pmatrix}0.6&0.3\\0.3&-0.4\end{pmatrix},\qquad
 \varepsilon_H\in\{0,10^{-8},10^{-5},10^{-2}\}.
 \label{SI-eq:curvatureperturbation}
\end{equation}
The physical loss, radius, direction, intensity, and reporting norm remain fixed. The curvature-only difference is compared with the analytic linear term. Separately, the combined-error check adds $\delta b=-0.75\times10^{-6}$ and $\delta v=0.8\times10^{-6}e$ to the same curvature perturbation. Each combined bound uses supplied $\varepsilon_b=\varepsilon_v=10^{-6}$ and $\varepsilon_h=\varepsilon_H$; no independence assumption is used. The maximum curvature-identity discrepancy is $4.23\times10^{-16}$, and all combined bounds hold with the recorded $2\times10^{-13}$ float64 guard. These checks evaluate the identities and error coefficients in finite precision. The records are in \nolinkurl{local_curvature_checks.json}; the script is \nolinkurl{local_curvature_checks.py}.

\section{Directional curvature-input checks}
\label{SI-sec:anisotropychecks}
The check uses $I=1.2$, $e=(0.6,0.8)$, $e_\perp=(-0.8,0.6)$, and fixed observations $b=0.31$, $v=(0.7,-0.2)$, $h_p=(0.4,0.6)$. Radii are $r=1$ and $r=1\pm10^{-k}$, $k=1,\ldots,8$. At each radius, perturb the supplied action by $0.1e$ or $0.1e_\perp$, giving 34 directional checks while keeping $b,v$ fixed.

Eighty-digit decimal arithmetic builds $J$ directly from the Cartesian derivatives of $F$ and $K$ from their radial derivative, then inverts the $2\times2$ matrix by its determinant. The before/after recovery difference is compared with the closed coefficients in the directional-sensitivity corollary. The largest component discrepancy is $1.20\times10^{-80}$; both fold changes are exactly zero in this arithmetic. The corresponding full-gradient gains are
\begin{equation}
 \frac{\|\delta g\|_2}{\|\delta h_p\|_2}=
 \begin{cases}
 |\beta_{\parallel}|\sqrt{1+\phi^2},&\delta h_p\parallel e,\\
 |\beta_{\perp}|,&\delta h_p\parallel e_\perp.
 \end{cases}
 \label{SI-eq:fullanisotropicgain}
\end{equation}
The gain discrepancy is below $1.26\times10^{-79}$. Component and gain checks use a $10^{-65}$ decimal tolerance.

A log--log fit of the measured polarization gain over $k=4,\ldots,8$ gives radial exponents $2.00000000109$ on both sheets, and tangential exponents $1.00001367279$ (inner) and $0.99998632787$ (outer). Figure 2(a) in the main paper plots the closed-form gains; these directional records check the same coefficients. The executable check and full decimal coefficient records are in \nolinkurl{anisotropic_curvature_checks.py} and \nolinkurl{anisotropic_curvature_checks.json}.

\section{Mixed scalar observations and input-error checks}
The mixed check uses the three objectives defined in Section S9, intensities $0.2,1.2$, direction $e=(0.6,0.8)$, and radii $0.8,0.99,1,1.01,1.2$, giving 30 recovery cases. The scalar $c_r$ is obtained by an 80-digit decimal central second difference of the raw objective, radial step $10^{-12}$. The physical curvature $h_{rr}$ is obtained by a central difference of the physical gradient along $(0,e)$, step $10^{-18}$, then contracted with $e$. Recovery agrees with the analytic physical gradient within $1.04\times10^{-24}$ in maximum component norm.

For each case, perturb $b$ by $-0.75\varepsilon$, $j$ by $0.8\varepsilon e_\perp$, $c_r$ by $0.6\varepsilon$, and $h_{rr}$ by $-0.5\varepsilon_h$, where $\varepsilon=10^{-6}$ and $\varepsilon_h=10^{-3}$. The 30 before/after changes agree with the exact mixed error identity within $1.01\times10^{-79}$. The Euclidean norm envelope holds with the recorded $10^{-18}$ allowance for the differenced baseline.

The ambiguity check fixes $I=1.2$, $g=(0.4,0.2,-0.3)$, $H_{pp}=\left(\begin{smallmatrix}0.5&0.1\\0.1&0.8\end{smallmatrix}\right)$, radii $0.8,0.99,1.01,1.2$, and $\tau=\pm0.1$. It adds the centered affine/quadratic perturbation in Proposition~2 of the main paper. All eight pairs preserve $b,P_\perp j,c_r$ to $10^{-80}$ while changing $g$; the complete first derivative changes by $Ia_r\tau e$. These pairs test the specified observation restriction, with physical curvature allowed to change. The implementation and inputs are in \nolinkurl{mixed_recovery_checks.py} and its JSON report.

\section{Common-graph AD acquisition and accuracy}
The microbenchmark uses CPU eager PyTorch 2.14.1, Python 3.12.14, NumPy 2.3.5, and one intra/inter-op thread on an Apple M4 Max. Sections S17--S19 use separate continuation and repair environments. All four routes use the same quartic objective and raw head $I=\exp(a)$, $F(w)=2w/(1+\|w\|^2)$. Requested inputs $I_{\mathrm{req}},r_{\mathrm{req}}$ initialize $a=\log I_{\mathrm{req}}$ and $w=r_{\mathrm{req}}(0.6,0.8)$; geometry for recovery is evaluated from the represented coordinates.

The direct route differentiates the objective with respect to the intermediate physical Stokes tensor. The first route acquires the head gradient. The vector and mixed routes acquire head derivatives and the physical Hessian action by reverse-over-reverse AD. The mixed implementation contracts the vector HVPs to form $c_r$ and $h_{rr}$. The physical-curvature query is omitted at the exact fold. Required curvature acquisition, graph construction, tensor allocation, and recovery are included in elapsed time; importing the framework and reference calculations are excluded.

Accuracy uses two precisions, intensities $0.02,1.2$, and nine radii: $1$ and $1\pm10^{-k}$ for $k=1,3,5,7$. There are 144 route evaluations, including four singular first-inverse cases at the fold. The other 140 return finite gradients. The reference is the analytic physical gradient evaluated in 80-digit decimal arithmetic at the represented Stokes state, so state-rounding is accounted for in the reference. The measured errors also include finite-precision geometry arithmetic within the implementation.

Timing uses $I_{\mathrm{req}}=1.2$, radii $0.999,1,1.001$, and both precisions. Each route has five warmup calls and seven batches of 20 fresh-graph calls. Method order rotates between batches. The 22 valid timing groups retain every batch mean, median, and interquartile range. At float64 and $r=0.999$, median [interquartile] microseconds are: direct $62.9$ [$62.5,68.2$], first inverse $114.4$ [$113.4,116.9$], vector HVP $274.5$ [$264.5,279.7$], and mixed $277.6$ [$273.5,284.5$]. These are observations of this acquisition implementation. The saved hardware, inputs, gradients, and batch times are in \nolinkurl{ad_route_benchmark.json}. Running \nolinkurl{ad_route_benchmark.py}\ \nolinkurl{--check-only} verifies the deterministic accuracy records without repeating wall-clock measurements.

\section{Shared-network fitting and diagnostic protocol}
\begin{figure*}[!t]
\centering\includegraphics[width=.94\textwidth]{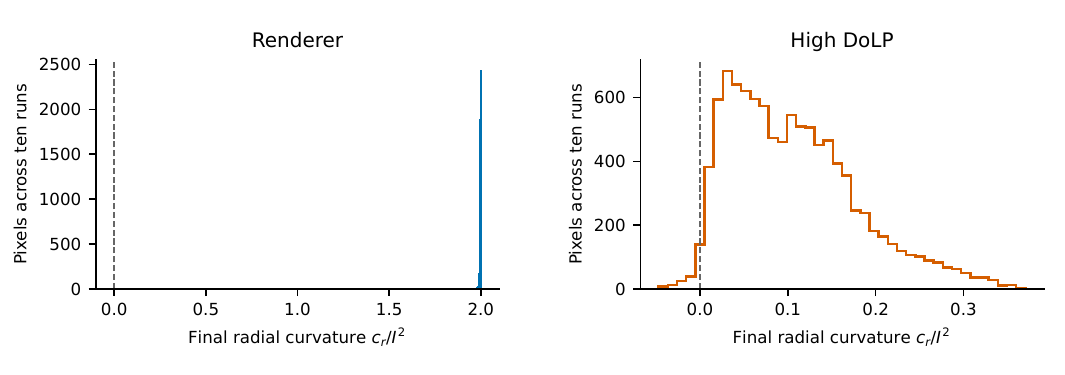}
\caption{Final independent-head radial-curvature histograms across ten fixed runs per field, normalized by squared intensity. The vertical line marks zero. These are directional second derivatives at the actual trained radii, not Hessian eigenvalues of stationary folds or of the network parameter space. The near-fold incidence additionally requires the radius criterion stated in Section S15.}
\label{SI-fig:trainedcurvature}
\end{figure*}

The four analyzer rows are $A_k=\tfrac12(1,\cos2\theta_k,\sin2\theta_k)$ for $\theta_k=0,\pi/4,\pi/2,3\pi/4$, so $A^{\mathsf T}A=\operatorname{diag}(1,1/2,1/2)$. Each input pixel is $y=As^*+\epsilon$, with independent Gaussian components of standard deviation $0.01$. We fit the mean of $\tfrac12\|As-y\|_2^2$ over all 1,024 pixels. This is a single-field fit without a held-out split.

The synthetic field uses grid coordinates $x_g,y_g\in[-1,1]$, $I^*=1+0.2\cos(\pi x_g)\cos(\pi y_g)$, $d=0.90+0.09(x_g+1)/2$, and polarization angle $\theta_{\mathrm{pol}}=\pi(0.35x_g+0.2y_g)$. Its target is $I^*(1,d\cos2\theta_{\mathrm{pol}},d\sin2\theta_{\mathrm{pol}})$. These fixed coefficients define the synthetic intensity, DoLP, and angle fields. The second field is the frozen renderer array in Section S3. The network has input dimension four, two width-32 SiLU hidden layers, and three head coordinates. Intensity is softplus plus $10^{-6}$, a positive intensity floor; the two remaining coordinates use the rational map. The final intensity bias is inverse-softplus of one, and polarization biases are zero. The standard and broad initializations multiply the final polarization weights by $0.2$ and $3$, respectively. Seeds are $0$--$4$; noise uses a separate seed offset of 1,000. Float32 CPU Adam uses learning rate $0.003$, moment parameters $(0.9,0.999)$, denominator offset $10^{-8}$, zero weight decay, one thread, and 1,500 full-batch updates. All 20 runs and their final weights are retained.

Diagnostics use the per-pixel loss before averaging and float64 evaluation at stored float32 head coordinates. The radius and radial curvature are evaluated without replacing a nearby radius by one. For each of the three radius thresholds $10^{-3},10^{-4},10^{-5}$, a candidate additionally requires $c_r/I^2<-10^{-7}$. The radius thresholds define neighborhoods; the curvature threshold excludes values within $10^{-7}$ of zero. We separately record the independent-head stationarity measure $\|\nabla_{a,w}L\|_2/\max\{I^2,10^{-12}\}$ with threshold $10^{-5}$; the denominator floor prevents division by zero. Negative radial curvature at a nearby point alone does not identify a stationary saddle. Checkpoints are updates $0,100,500,1000,1500$; 25-step loss traces and all checkpoint predictions are saved. No radius/negative-curvature candidates occur at these checkpoints. Two high-DoLP pixels meet the $10^{-3}$ radius criterion at update 500, with nonnegative curvature. The closest final radius distances across the ten runs are $0.957049$ and $0.058588$ for the renderer and high-DoLP fields.

For each stored state we evaluate both signed radial offsets $0.001,0.01,0.05,0.1$ at fixed intensity. Their best nonnegative loss decrease is compared with the general supporting-half-space bound including interior slack. The cone minimum uses the analytic ideal-metric solution in Section S2. No stationary-fold specialization is applied to nearby trained points. This action changes independent head outputs with the shared network fixed.

The constructed control uses $z=(1,0.6,0.3)$, its aligned unit direction, and stationary intensity $(2z_I+\|z_p\|)/3$. It gives $\gamma_r=-0.0976866$ and exact quadratic gap and bound $0.0180599$. Its float64 Adam trajectories start either at these represented coordinates or at radial offset $0.01$. Both decrease the objective over 300 updates; the nominally stationary start also escapes through finite-precision perturbations. The control checks the saddle calculation and feasible loss reduction; it is excluded from observed incidence and does not demonstrate an advantage of a deliberate perturbation over Adam.

Records are in \nolinkurl{trained_fold_diagnostic.json} and \nolinkurl{results/trained_fold/}. Each NPZ contains noisy observations, clean target, final and checkpoint outputs, raw coordinates, weights, and loss trace. Separate diagnostic NPZ files contain physical gradients, radial curvatures, stationarity measures, gaps, and bounds. Weight replay reproduces all 20 final predictions exactly. Independent AD checks of 80 head Hessian contractions have maximum discrepancy $1.78\times10^{-15}$. The command \nolinkurl{trained_fold_diagnostic.py}\ \nolinkurl{--check-only} recomputes diagnostics from saved records without retraining.

\section{Boundary-derived initialization and restricted flow}
\label{SI-sec:suppflow}
For a nonzero boundary Stokes state $s=I(1,e)$, any finite preimage under $T_F$ has $F(w)=e$. The scalar relation $2r/(1+r^2)=1$ gives $(r-1)^2=0$, hence $w=e$. If an exact preceding cone estimate has positive probability of lying on this boundary, every measurable choice of its finite raw preimage puts the same probability on $r=1$. This conditional pushforward statement does not assume an absolutely continuous raw initialization or identify the output law of a finite solver.

For Proposition~\ref{SI-prop:dummyflow} below we give a self-contained statement, so this supplement can be read independently.
\begin{proposition}[Restricted ideal-metric flow]
\label{SI-prop:dummyflow}
Let $z_I>\rho=\|z_p\|>0$, $W=\operatorname{diag}(1,1/2,1/2)$, and use any $C^2$ bijective intensity coordinate onto $(0,\infty)$ with $I'\ne0$. Raw Euclidean gradient flow for $\tfrac12\|T_F(a,w)-z\|_W^2$, started with $r=1$, exists for all forward time and stays there. Except for the antipodal direction $e(0)=-z_p/\rho$, it converges in physical coordinates to
\begin{equation}
 (I_b,I_b z_p/\rho),\quad I_b=(2z_I+\rho)/3.
\end{equation}
Its objective gap is $(z_I-\rho)^2/6$ and radial head curvature is $-I_b(z_I-\rho)/3$.
\end{proposition}
\begin{proof}
The polarization Jacobian has radial eigenvalue $a_r=2(1-r^2)/(1+r^2)^2$. Therefore $d r^2/dt=-2Ia_r w^{\mathsf T}g_p$, and uniqueness of the $C^1$ vector field preserves $r=1$. On this set,
\begin{equation}
 \begin{aligned}
 \dot e&=I P_\perp z_p/2,\\
 \dot I&=-(I')^2(3I/2-z_I-e^{\mathsf T}z_p/2).
 \end{aligned}
\end{equation}
Write $I_-=(2z_I-\rho)/3>0$ and $I_+=(2z_I+\rho)/3$. The vector field points upward below $I_-$ and downward above $I_+$. Thus
\begin{equation}
 \begin{aligned}
 0<m&=\min\{I(0),I_-\},\\
 M_0&=\max\{I(0),I_+\}<\infty,\\
 m&\leq I(t)\leq M_0.
 \end{aligned}
\end{equation}
The intensity coordinate $a$ lies in the compact preimage of $[m,M_0]$; $w$ lies on the unit circle. The vector field is smooth on a neighborhood of this compact set, so the solution extends to every forward time. In particular, $|I'|$ is bounded above and away from zero along it.

Set $u=e^{\mathsf T}z_p/\rho$. Then $\dot u=I\rho(1-u^2)/2$. If $-1<u(0)<1$,
\begin{equation}
 \operatorname{artanh}u(t)=\operatorname{artanh}u(0)
        +\frac{\rho}{2}\int_0^t I(\tau)\,d\tau.
\end{equation}
The integral grows at least as $mt$, so $u(t)\to1$. Initial $u=1$ is already aligned. Since $u\to1$, for any $\varepsilon>0$ the intensity derivative eventually points toward the interval $[I_+-\varepsilon,I_++\varepsilon]$ and has magnitude bounded away from zero outside it on the compact intensity range. Scalar comparison yields $I\to I_+$. The case $u(0)=-1$ stays antipodal and instead gives $I\to I_-$. Its gap and radial curvature are $(z_I+\rho)^2/6$ and $-I_-(z_I+\rho)/3$.

Because $z$ is cone-interior, its quadratic loss is the cone minimum. For the aligned limit,
\begin{equation}
 \begin{aligned}
 L(I_b,I_b e)-L(z)
 &=\tfrac12(I_b-z_I)^2+\tfrac14(I_b-\rho)^2\\
 &=\frac{(z_I-\rho)^2}{6}.
 \end{aligned}
\end{equation}
At $r=1$, the radial second derivative is $-I e^{\mathsf T}g_p=-I(I-\rho)/2$, yielding the displayed curvature at $I_b$.
\end{proof}

This result specializes the differential geometry of the rational ball lift \cite{supp-gcopter,supp-lifts}. Theorem 2.4, Lemma 3.24, Proposition 3.26, and Theorem 3.23 of \cite{supp-lifts} describe the first-order obstructions and second-output derivatives. Here the calibrated quadratic bound, including its support slack, supplies the approximate stopping test. Section S20 extends gap control to strongly convex objectives and compares it with quantitative squared-variable transfer.

The random-initialization theorem of \cite{supp-lee} assumes an absolutely continuous initialization, a $C^2$ objective with Lipschitz gradient, and a fixed step smaller than the reciprocal Lipschitz constant. Those hypotheses do not follow from a boundary-derived law, and the theorem concerns asymptotic gradient descent rather than finite BFGS or LBFGS termination. For plain discrete gradient descent, $\nabla_w L=I P_\perp g_p$ at $r=1$ is tangent, so one step gives $r_{\rm new}^2=1+\eta^2 I^2\|P_\perp g_p\|^2$. Coordinatewise or quasi-Newton preconditioning need not preserve that tangency.

\section{Fixed two-frame verification protocol and complete controls}
\label{SI-sec:suppwarm}
The machine-readable protocol, its registration receipt, raw input records, solver outputs, and independent audits are in \nolinkurl{repro/warm_start/}. No optimization condition was added after viewing the grid results. The 60 input rows use seeds $0$--$4$, angles $0,0.7,1.3$, noise SD $0.001,0.01$, and next DoLP $0.8,0.99$. Each seed starts the generator at $10000+\mathrm{seed}$ for each condition, so its two four-dimensional noise vectors are reused. The previous clean state is $(1,\cos2\theta,\sin2\theta)$; the next is $(1,d\cos2\theta,d\sin2\theta)$. For angle changes $\delta=0,0.05,0.2$, only the next clean direction changes to $\theta+\delta$, retaining the stored next-frame noise exactly.

\subsection{Solvers, precision, and termination}
Every previous fit starts at $(0,0.2,-0.1)$ with exponential intensity. BFGS, unbounded L-BFGS-B, and \nolinkurl{trust-exact} use analytic head gradients, maximum 500 iterations, and gradient tolerances $10^{-4},10^{-6},10^{-8}$. L-BFGS-B additionally uses \nolinkurl{ftol=0}, maximum 5,000 function calls and 40 line-search steps. The trust-region method acquires the full analytic $3\times3$ Hessian. It uses its native gradient norm; the quasi-Newton methods use their native stopping rules. Previous fits with identical observations, method, and tolerance are cached, giving 270 distinct previous fits for the 1,620 continuations. Previous failures would be retained; none occurs in these SciPy arms.

The SciPy run uses macOS arm64, Python 3.12.3, NumPy 2.5.4, and SciPy 1.18.1. Initialization of a separate BFGS probe with float32 coordinates returns float64 coordinates and float64 oracle gradients; incoming oracle coordinates remain float32 in that probe. The oracle explicitly evaluates in float64. This mixed computation is not used as evidence of full float32 optimization.

The separate genuine-precision control uses Python 3.11.6, NumPy 2.3.5, and PyTorch 2.9.1, CPU, one thread, with all optimizer parameters, objective values, and gradients in float32 or float64. LBFGS has learning rate 1, history size 20, maximum 500 iterations and 1,000 evaluations, gradient tolerance $10^{-6}$, zero change tolerance, and strong-Wolfe line search. It exposes no native success flag. We therefore report native-gradient acceptance, nonacceptance, and objective gap separately. Its diagnostic controller, radial trial selection, and final reference gap use float64 at the represented native coordinates with the original observations. The 22 float32 previous-fit nonacceptances are row counts representing 11 of 30 distinct previous fits. They remain in the table rather than being excluded. Float64 previous fits all meet the operational acceptance test.

\subsection{Reference, direct stopping, and curvature stopping}
The reference uses the ideal-metric cone optimum: for unconstrained estimate $z$, $\rho=\|z_p\|$, return $z$ when $z_I\geq\rho$, the apex when $2z_I+\rho\leq0$, and otherwise $(I_c,I_c z_p/\rho)$ with $I_c=(2z_I+\rho)/3$. An independent physical-coordinate SLSQP calculation starts at $(1.1,0.1,0.1)$ and uses $I\geq0$, $I^2-\|p\|^2\geq0$, objective tolerance $10^{-13}$, and at most 1,000 iterations. Across all 180 distinct next-frame observation vectors, all solves succeed; maximum objective discrepancy is $5.00\times10^{-14}$ and maximum cone violation is $7.67\times10^{-15}$.

The direct physical stopping test minimizes the supporting-half-space bound over its nonnegative multiplier and accepts a finite bound at most $10^{-6}$. We use the support slack at the actual radius, including interior outputs. Every inaccurate SciPy endpoint is rejected; no finite accurate endpoint is rejected in this grid. This does not make the half-space relaxation exact for arbitrary data. The expression is evaluated in float64; the empirical acceptance counts are not outward-rounded numerical certificates. At the 149 near-fold SciPy endpoints, mixed recovery using the known $h_{rr}=1/2$ agrees with the direct physical gradient within $4.17\times10^{-17}$ componentwise. No physical-gradient availability restriction is imposed.

The local curvature gate uses $|r-1|\leq10^{-3}$ and $c_r/I^2<-10^{-7}$ at the represented endpoint. Both signed radial trials and the current point are evaluated; the lowest loss is restarted once with the same solver and settings. The three trial evaluations obtain both function and gradient in this implementation. For each of the 60 zero-angle-change fits at tolerance $10^{-6}$, the direct and curvature controllers also evaluate the endpoint state. Gated BFGS adds 280 function/gradient pairs; a radius-first implementation requires 17 full Hessians. Unconditional restart adds 479 pairs. The corresponding L-BFGS-B counts are 320 pairs and 21 required Hessians versus 496 pairs. The trust-region arm has no trigger, five required audit Hessians, and 240 unconditional additional pairs plus 60 unconditional solver Hessians. Coverage logging instead computes a Hessian at all 60 endpoints in each of these groups, including points outside the gate. The saved accounting separates actual analysis work, optimizer calls, final checks, and queries required by a deployed gate. These counts are not interchangeable units of work.

At tolerance $10^{-4}$, the gate detects 14 of 68 false successes among the 540 SciPy continuations and also triggers at one accurate accepted endpoint. After restarting, 58 false successes remain, compared with 32 under unconditional restarting. At tolerance $10^{-8}$ and zero angle change, BFGS reports seven native failures, including six nonfinite terminal losses; unconditional restarting leaves six failures and the gate leaves seven. All failures are retained. Nonfinite numbers are serialized as null together with failure flags, warnings, and evaluation counts; a null maximum is never interpreted as zero gap. The file \nolinkurl{audited_summary.json} explicitly separates finite and nonfinite errors, and \nolinkurl{failure_records.json} preserves their conditions and terminal states.

\subsection{Independent reproduction and scope of the evidence}
A new Cartesian Jacobian/Hessian implementation reproduces the original BFGS false-success IDs $1,2,3,21,25,29,40,41,45,47,53,55$ (zero-based), the separate failure at ID 52, and the post-trigger counts zero and one. The largest previous-coordinate discrepancy is $5.62\times10^{-14}$ and the largest objective-gap discrepancy is $4.86\times10^{-12}$. Independent PyTorch AD at all 60 frozen endpoints gives maximum gradient and Hessian discrepancies $9.49\times10^{-18}$ and $4.45\times10^{-16}$. Thus the reference arithmetic, derivative identities, and solver reproduction are checked separately.

These results quantify how physical or recovered derivatives audit a local head stopping decision. The direct cone solution solves this three-variable calibrated quadratic without a raw warm start, while the direct physical bound rejects failures away from the fold that the local curvature gate misses. The 20 MLP fits in Section S15 and their zero-candidate result are unchanged and are not pooled with the two-frame conditions.

\section{Physical-reflection and released-observation pilots}
\label{SI-sec:realprotocol}
\subsection{Fresnel sequence and controls}
The additional fixed physical simulation uses a single air-to-dielectric interface with real refractive index $1.5$. For incidence $\theta$, Snell's law gives $\sin\theta_t=\sin\theta/1.5$. The squared Fresnel amplitude coefficients give reflected fractions $R_s,R_p$. For unpolarized illumination, specular intensity and linear-polarization magnitude are proportional to $(R_s+R_p)/2$ and $(R_s-R_p)/2$. We normalize specular intensity to one at Brewster's angle and add constant unpolarized backgrounds $0,0.02,0.1$. These backgrounds are not geometry-dependent diffuse-shading models.

Incidence follows $\arctan(1.5)+18^{\circ}\cos(2\pi t/40)$ for 41 frames. Polarization orientation is either zero or $0.05\sin(2\pi t/40)$ radians. Photon scales are 1\,000 and 10\,000, and seed indices are zero and one. Each configuration has its own NumPy seed sequence $[20261022,\text{configuration index},\text{seed index}]$. Analyzer electron counts have means $NAs$, independent Poisson fluctuations, and independent Gaussian read noise of standard deviation two electrons; dividing by $N$ gives the fitted observations. No clipping, shared frame noise, or initialization from physical truth is used. Each solver/arm begins at raw coordinates $(0,0.2,-0.1)$ and propagates its own subsequent estimate.

The 24 sequences contain 984 observations per arm. BFGS and L-BFGS-B use the options in Section S17 at tolerance $10^{-6}$. Native false successes number nine and seven, with zero native failures. Curvature-audited and physically audited branches each have zero false successes and seven interventions per solver. A radius-first curvature controller requires 22 initial endpoint Hessians per solver. Optimizer function/gradient pairs, including restart trials, total 11\,394 and 11\,292 in the curvature-audited branches, versus 11\,813 and 11\,575 without intervention. These counts do not equate a Hessian query with a first-derivative query or establish wall-clock superiority. Direct physical auditing requires no Hessian; the script's extra diagnostic logging is separately identified.

The direct-cone truth Stokes RMSE is $0.0251904$, compared with $0.0252375$ for raw BFGS and $0.0251904$ after either audit. Mean error is dominated by measurement noise. False successes occur with background zero or $0.02$, in both orientation modes; none occurs with background $0.1$ in this grid. The displayed sequence is selected by the largest raw BFGS objective gap. It is an illustrative worst case, not a typical case.

A monotone-map BFGS control under the same absolute head-gradient tolerance reports 160 native false successes. The first recorded version shares the raw initial coordinate, hence not the physical initial state. A separately recorded correction uses $w_G=q/\sqrt{1-\|q\|^2}$ at the same initial physical polarization $q=F(0.2,-0.1)$; it also reports 160. Both controls and their protocols are retained. This finite-tolerance comparison involves different map conditioning, with the monotone boundary attainable only at infinite radius; it does not establish a universal ranking. Independent physical-coordinate SLSQP checks of 20 minima have maximum loss discrepancy below $2\times10^{-14}$. PyTorch float64 AD at all 16 raw false successes and 48 initial control endpoints has maximum normalized radial-curvature discrepancy below $7\times10^{-16}$.

\subsection{Released data, selection, and observation units}
The acquisition plan fixes the data before any solver outcomes are inspected. From the DeepSfP archive linked by its data documentation, we select the first three distinct indoor test objects: \nolinkurl{box_f}, \nolinkurl{father_christmas_f}, and \nolinkurl{horse_l}. The archive's test list identifies all three. Each MAT file contains four-angle images, a foreground mask, and reference surface normals; those normals are not Stokes truth and are not used. The selected archive's README and current repository documentation describe different full split totals. We therefore identify the exact archive and selected members, rather than treating the small subset as the complete current benchmark.

PolarNS contributes \nolinkurl{object/241015104249} and \nolinkurl{scene/241021103114}, the lexicographically first records in each category. Their \nolinkurl{img_mean}, \nolinkurl{img_variance}, and \nolinkurl{image_count} arrays describe 10\,000 and 1\,017 captures. We use their means as released observations, without synthesizing new noisy frames. Means are divided by 4\,096, matching the authors' MATLAB convention. The $2\times2$ polarizer layout gives $90^{\circ},45^{\circ}$ on the first row and $135^{\circ},0^{\circ}$ on the second. This spatial sampling is not a calibrated co-location claim.

The PolarBurstSR subset is \nolinkurl{test/0000000} through \nolinkurl{test/0000002}, with \nolinkurl{frame_00.png} through \nolinkurl{frame_13.png}. The authors' real-data configuration points to this released test directory. Each raw image is $192\times192$ uint16; division by 65\,535 and unpacking of each $4\times4$ sensor block follow the provided loader exactly. Four nominal color planes, R, G1, G2, and B, each yield $48\times48$ four-analyzer observations without demosaicing. A track uses one fixed sensor coordinate across the 14 frames. Because no registration is performed, it need not follow one scene point. The released high-resolution references are not used for aligned reconstruction metrics.

For static images, eligible pixels are finite, nonnegative, have every analyzer channel below $0.995$, and have nominal intensity at least $5\%$ of the foreground/sample 99th percentile. DeepSfP additionally requires its foreground mask and retains its stored normalized double units. A seeded uniform selection chooses 128 eligible pixels per image; the seed sequence is $[20261022,\text{sample index}]$. The enriched selection takes 64 evenly spaced ranks from the top $1\%$ by observed DoLP, with a minimum pool of 64. Each selection is sorted in raster order and forms a separate spatial continuation stream. This is an implementation stress test on static observations, not a video trajectory.

For each burst color plane, the same finite, saturation, and relative-intensity conditions are evaluated only on the first frame. A fixed $4\times4$ spatial grid contributes its eligible positions. Eight positions are taken at evenly spaced ranks from the top $1\%$ first-frame observed-DoLP pool, with a minimum pool of eight. All 14 frames at each selected position are retained. After eligibility, there are 167 uniform and 96 enriched tracks, totaling 3\,682 burst entries per solver/arm. Uniform and enriched selections may overlap, and color planes and frames share a capture. They are separate conditional selections, not independent scene samples or estimates of population incidence.

\subsection{Fitting, diagnostics, and complete counts}
The nominal ideal analyzer matrix and $W=\operatorname{diag}(1,1/2,1/2)$ define the same least-squares objective and exact cone reference as Section S17. The units above are fixed per data source; no per-pixel normalization or clipping is introduced. No Poisson-likelihood, real-camera calibration, denoising, super-resolution, or normal-estimation improvement is inferred from objective-gap accuracy.

Each solver uses four arms: cold initialization, uncorrected warm start, curvature-triggered restart, and physical-bound-triggered restart. A stream starts at $[\log\max(\widehat I,10^{-6}),0.2,-0.1]$, with $\widehat I=\sum_k y_k/2$. Cold starts recompute this point for each observation. Other arms propagate their own preceding raw state, including native failures. Solver settings, local radius/curvature thresholds, objective gap $10^{-6}$, and the one-restart rule remain those in Section S17. The protocol is recorded before optimization; no additional conditions are selected after seeing the results.

\begin{table*}[!t]
\caption{Complete released-data selection counts. Each pair is BFGS/L-BFGS-B; $n$ is the number of entries per solver/arm. ``Physical post'' counts inaccurate \emph{native} successes after the physically audited restart branch. Every such final endpoint is rejected by its physical bound; it is not an accepted output of the monitor. All native failure counts are zero.}
\label{SI-tab:realstrata}
\centering\small\renewcommand{\arraystretch}{1.15}
\begin{tabularx}{\textwidth}{@{}Xrrrrr@{}}\toprule
Subset and selection & $n$ & Cold & Warm & Curvature post & Physical post\\\midrule
DeepSfP, uniform & 384 & 0/0 & 0/0 & 0/0 & 0/0\\
DeepSfP, high observed DoLP & 192 & 0/0 & 0/0 & 0/0 & 0/0\\
PolarNS means, uniform & 256 & 0/0 & 43/0 & 43/0 & 0/0\\
PolarNS means, high observed DoLP & 128 & 0/7 & 68/64 & 68/64 & 68/64\\
PolarBurstSR, uniform & 2\,338 & 0/1 & 0/9 & 0/9 & 0/9\\
PolarBurstSR, high observed DoLP & 1\,344 & 0/6 & 16/95 & 16/95 & 16/90\\\bottomrule
\end{tabularx}
\end{table*}

All 37\,136 fits are retained. The fixed curvature controller has no intervention and detects none of the raw warm-start false successes. The physically audited branch intervenes 69/64 times in the PolarNS means and 16/101 times in the bursts. It rejects all inaccurate final endpoints, although a single radial restart does not repair all of them. Cold starts reduce the counts in this subset but retain 14 L-BFGS-B false successes, so changing initialization is not a general physical stopping criterion.

Among raw warm-start false successes, PolarNS radii range from approximately $0.996$ to $6.74\times10^6$, and burst radii from $0.752$ to $314$. No inaccurate endpoint meets the fixed local fold gate. The physical bound rejects all 295 raw warm-start false successes and has zero inaccurate acceptance across the recorded arms. Independent SLSQP checks of 29 reference minima have maximum objective discrepancy $1.31\times10^{-18}$. Independent PyTorch float64 AD checks all 309 cold/warm false successes and 32 initial controls; maximum head-gradient, Hessian, and physical-gradient differences are $7.77\times10^{-18}$, $8.89\times10^{-16}$, and $1.39\times10^{-17}$. No native solver failures, warnings, or nonfinite oracle calls occur in this fixed run.

The curvature controller requires a Hessian only after its radius guard. The scripts also evaluate Hessians for full endpoint coverage and independent final diagnostics; these are verification overhead. Physical-gradient auditing requires zero Hessians. Function/gradient, restart-trial, and diagnostic queries remain separate; optimizer-only timings are not end-to-end runtime comparisons. Exact archive-member names, verified CRC32 and SHA-256 values, byte-range receipts, selected coordinates, input observations, native flags, endpoint states, bounds, query counts, protocols, and figure generators are included in the reproduction records. Raw data are obtained from the original providers; no full multi-gigabyte archive or pretrained model is required for this pilot.

\section{Frozen coordinate-repair follow-up}
\label{SI-sec:schartrepair}
This follow-up was designed after the released-data results in Section S18 were inspected. The code and protocol were hashed before the new algorithm outcomes were evaluated. All 4\,642 released observations, 273 streams, sampling strata, nominal measurement units, and ordering are reused unchanged. All 24 Fresnel sequences and 984 frames are also reused; they are a previously inspected mechanism control, not a held-out test set. No additional seeds, thresholds, data, or parameter grids are introduced. The experiment uses float64 NumPy 2.5.4 and SciPy 1.18.1 on one local CPU process.

\subsection{Sheet symmetry and its scope}
Let $C(w)=w/\|w\|^2$ for $w\ne0$, $r=\|w\|$, $e=w/r$, and $H_e=\operatorname{Id}_2-2ee^{\mathsf T}$. Direct substitution and differentiation give
\begin{equation}
 F(C(w))=F(w),\qquad DC(w)=r^{-2}H_e.
 \label{SI-eq:chartidentity}
\end{equation}
For $h(a,w)=f(T_F(a,w))$, at the same intensity and observations,
\begin{equation}
 \nabla_w h(a,C(w))=r^2H_e\nabla_w h(a,w).
 \label{SI-eq:chartgradient}
\end{equation}
Indeed $h(a,w)=h(a,C(w))$, so the chain rule gives $\nabla_w h(a,w)=r^{-2}H_e\nabla_w h(a,C(w))$, and $H_e^2=\operatorname{Id}_2$. The normalized-output condition number is unchanged, while the polarization-gradient norm increases by $r^2$ when an outer state is moved inward. The intensity derivative is unchanged. These identities follow from the two-sheet map and the matched spectra in Section S6. They improve derivative scale at outer coordinates while retaining the fold and intensity dependence.

\subsection{Fixed algorithm and controls}
All head solvers use gradient tolerance $10^{-6}$ and at most 500 iterations. L-BFGS-B additionally uses zero objective tolerance, at most 5\,000 function evaluations, and 40 line-search steps. Every new solve resets its approximate inverse Hessian and line-search history. At the first released observation of each stream, and at every released cold start, the initialization is $[\log(\max\{S_{0,\mathrm{LS}},10^{-6}\}),0.2,-0.1]$. The corresponding Fresnel initialization is $[0,0.2,-0.1]$. Every continuation arm retains its own preceding result, including a rejected result.

The switch-before-solve arm replaces an outer preceding state by $(a,C(w))$ before the next solve; it otherwise uses native acceptance. The audited coordinate arm first solves from the preceding state and evaluates Corollary~5's physical bound with the exposed physical gradient. On rejection, it uses the following fixed choices. If $|r-1|\le10^{-3}$, one full analytic head Hessian supplies $c_r$. If $c_r/I^2<-10^{-7}$, the candidates are the current point and radial offsets $-0.01,+0.01$; the lowest-loss candidate is moved to the inner sheet if necessary and starts one fresh solve. If the negative-curvature condition does not hold and $r>1$, the endpoint is moved to the inner sheet and starts one fresh solve. Otherwise no head correction is made. A final physical audit retains any rejection. There is at most one head restart per observation. The audited cold control instead uses one fresh cold solve on the same type of rejection.

The guarded coordinate arm applies standard projected gradient only if that coordinate repair remains rejected. For the ideal quadratic, write $g(s)=W(s-z)$ with $W=\operatorname{diag}(1,1/2,1/2)$. Its update is
\begin{equation}
 s^{k+1}=\Pi_{\mathcal C}\bigl(s^k-g(s^k)\bigr).
 \label{SI-eq:physicalpg}
\end{equation}
Here $\Pi_{\mathcal C}(v)=\arg\min_{u\in\mathcal C}\|u-v\|_2^2$ is the Euclidean projection in \eqref{SI-eq:euclideancone}. The $W$-projection $P_W$ in Section S2 instead supplies the direct least-squares reference. For example, $\Pi_{\mathcal C}(0,1,0)=(1/2,1/2,0)$, whereas $P_W(0,1,0)=(1/3,1/3,0)$. The projected-gradient step size is one because $\lambda_{\max}(W)=1$. Constrained optimality gives $s^*=\Pi_{\mathcal C}(s^*-g(s^*))$. Euclidean nonexpansiveness therefore yields
\begin{equation}
 \|s^{k+1}-s^*\|_2\le\|\operatorname{Id}_3-W\|_2\|s^k-s^*\|_2
 =\tfrac12\|s^k-s^*\|_2.
 \label{SI-eq:pgcontraction}
\end{equation}
This standard contraction follows from nonexpansiveness of projection and the fixed quadratic metric. The implementation audits before and after updates, reuses the acquired physical gradient, stops only at bound at most $10^{-6}$, and permits at most 64 updates. An unresolved bound is reported as a rejection, never as a repaired accepted solution. The finite budget alone is not an unconditional accuracy guarantee. In exact arithmetic, Corollary~5 implies gap at most $10^{-6}$ whenever this bound accepts; the recorded float64 calculation is not outward-rounded certification.

A positive-intensity physical correction is lifted by the inner inverse $w=q/(1+\sqrt{1-\|q\|^2})$, $q=p/I$, $a=\log I$. Only the square-root argument is clamped to remove feasibility roundoff. The observed lift discrepancy in the independent audit is at most $1.12\times10^{-16}$. An exact apex has no finite exponential lift: the specified implementation would retain its physical output and cold-start the next head solve. No experimental output requires this case. The standalone projected-gradient control starts from the same first physical state and subsequently carries its own physical result; it uses the same updates, bound, and budget without any head solve. The direct cone reduction is timed separately and supplies the reference gap; its answer is never used to initialize an iterative arm.

\subsection{Full accounting and paired audit}
The follow-up stores 78\,764 sequential output records: 67\,512 head-arm outputs across six strategies and two solvers, and 11\,252 outputs from the two physical controls. These counts exclude extra internal restarts. All native head invocations return success, with no recorded warnings or nonfinite queries. Monitored acceptance and native head status are separate fields; a physical fallback does not inherit a native head success flag.

Table~\ref{SI-tab:chartstrata} retains the released sampling strata. The standalone physical controls and the guarded coordinate arm have no inaccurate outputs in any stratum. In the released grid the guarded arm makes 16/25 outer restarts and two/15 physical fallbacks; its total projected updates are three/22. The audited cold control makes 18/35 cold restarts. The audited coordinate arm makes 18/33 outer restarts and leaves 7/22 inaccurate outputs, all rejected. The switch-before-solve arm performs 26/41 sheet changes and leaves 9/16 inaccurate native successes. No real-data correction requires a Hessian query. In the Fresnel grid each guarded arm makes seven fold restarts, acquires seven full Hessians and 21 trial FG pairs, and uses no projected fallback. Those full-Hessian queries are not reported as cheaper directional-HVP implementations.

\begin{table*}[!t]
\caption{Inaccurate released outputs by original stratum. Pairs are BFGS/L-BFGS-B. ``Switch'' applies before every warm solve; ``chart'' means audited coordinate repair without physical fallback. Guarded coordinate repair, standalone physical projected gradient, and the direct cone control are zero in every row.}
\label{SI-tab:chartstrata}
\centering\small\renewcommand{\arraystretch}{1.12}
\begin{tabularx}{\textwidth}{@{}Xrrrrr@{}}\toprule
Subset and stratum & Entries & Warm & Switch & Audited cold & Audited chart\\\midrule
DeepSfP, uniform & 384 & 0/0 & 0/0 & 0/0 & 0/0\\
DeepSfP, high DoLP & 192 & 0/0 & 0/0 & 0/0 & 0/0\\
PolarNS means, uniform & 256 & 43/0 & 3/0 & 0/0 & 2/0\\
PolarNS means, high DoLP & 128 & 68/64 & 1/2 & 0/0 & 0/9\\
Burst, uniform & 2\,338 & 0/9 & 0/2 & 0/0 & 0/1\\
Burst, high DoLP & 1\,344 & 16/95 & 5/12 & 0/2 & 5/12\\
\bottomrule\end{tabularx}
\end{table*}

The paired audit keeps all 9\,284 historical raw warm endpoints as a common starting population. Its repair selection is the physical bound rejection, not the reference objective gap. All 295 rejections in this population are the previously inaccurate native successes. Audited coordinate repair leaves 7/39, using 124/157 outer restarts and 1\,919/1\,259 additional native FG pairs. Guarding this procedure adds seven/39 physical fallbacks and nine/56 projected updates, leaving zero inaccurate outputs. Audited cold restarting leaves 0/5, using 127/168 restarts and 3\,421/1\,407 FG pairs. These conditional endpoint counts differ from the sequential experiment because an intervention changes later initialization.

Independent PyTorch 2.9.1 float64 differentiation checks all 295 historical false-success endpoints and 32 evenly spaced warm controls. Their maximum output-preservation discrepancy is $1.74\times10^{-18}$ and gradient-transformation discrepancy $7.78\times10^{-16}$. A separate physical-step implementation checks those same 327 states: all reach the bound threshold within nine updates, no objective increase is observed, and the maximum excess over the half-distance contraction is $5.56\times10^{-17}$. The separate implementations check the derivatives and physical updates numerically.

Native FG pairs, physical-gradient queries, projected updates, full Hessians, trial evaluations, coordinate changes, internal solver calls, and observed wall times are retained separately. Post-hoc reference-gap calculations are excluded from algorithm timings and query counts. Timings are one sequential CPU pass with fixed arm order; the protocol does not support statistical timing claims. The results quantify partial coordinate repair and the effectiveness of standard physical controls on these previously inspected quadratics. Evaluation on unseen scenes and aligned imaging-quality metrics would address different questions. Exact frozen-summary replay and source hashes are included in \nolinkurl{application/chart_repair/}.

\section{Proofs and checks for finite-tolerance stopping}
\label{SI-sec:sstopproof}
Write $\Delta(s)=f(s)-\min_{u\in\mathcal C}f(u)$ and $x_+=\max\{x,0\}$.
The guarantee concerns a $C^2$, $\mu$-strongly convex physical objective on $\mathbb R^3$. The evaluated Stokes state has $0<I\le\overline I$ and $0\le r\le1$. The radial physical Hessian is bounded above by $L\ge\mu$ at the evaluated point when $r\ge1/2$. A global Hessian upper bound is sufficient but not necessary for this pointwise assumption. The intensity coordinate may be any regular coordinate specified in the main text. All identities below use its exact evaluated $I,I'$ and exact $r,e$. Uncertain geometry requires an additional error model.

\subsection{Supporting-gap inequality from strong convexity}
Let $s=I(1,\phi e)$, $n=(-1,e)$ and $\sigma=I(1-\phi)$. For every $u\in\mathcal C$, $n^{\mathsf T}(u-s)\le\sigma$. Strong convexity gives, with $d=u-s$,
\begin{equation}
 f(s)-f(u)\le-g^{\mathsf T}d-\frac\mu2\|d\|^2
 \le-(g+\lambda n)^{\mathsf T}d-\frac\mu2\|d\|^2+\lambda\sigma.
 \label{SI-eq:sstrongsupport}
\end{equation}
Taking the supremum over $d\in\mathbb R^3$ yields $\Delta(s)\le\|g+\lambda n\|^2/(2\mu)+\lambda\sigma$ for every $\lambda\ge0$. This is the standard strong-convexity/dual-residual argument. The original quadratic corollary is sharper when the full metric is known; this isotropic inequality extends to the present strongly convex class.

\subsection{A pointwise bound using one-sided radial curvature}
For $r>0$ write $g_p=me+u$, $u\perp e$. The first derivatives satisfy
\begin{equation}
 b=g_I+\phi m,\qquad j/I=a_rm e+t_ru.
 \label{SI-eq:sstopdecomposition}
\end{equation}
Suppose $|b|\le\varepsilon$ and $\|j\|/I\le\varepsilon$. Then $|a_rm|\le\varepsilon$, $\|u\|\le\varepsilon/t_r$. For $r<1$, setting $\lambda=0$ in \eqref{SI-eq:sstrongsupport} gives
\begin{equation}
 \begin{aligned}
 U_1=\frac1{2\mu}\bigl[&(\varepsilon+\phi\varepsilon/a_r)^2\\
              &+(\varepsilon/a_r)^2+(\varepsilon/t_r)^2\bigr],\qquad
 \Delta(s)\le U_1.
 \end{aligned}
 \label{SI-eq:sstopfirst}
\end{equation}
At $r=1$ define $U_1=+\infty$. At the origin $\phi=0$ and $DF=2\operatorname{Id}_2$; the displayed first-order estimate remains a conservative bound with $a_r=t_r=2$ and does not require a radial direction.

For $1/2\le r\le1$ let $k_r=|\kappa_r|$ and assume $c_r\ge-I\nu$. Since $\kappa_r<0$, the chain rule implies
\begin{equation}
 m\le M_r:=\frac{\nu+ILa_r^2}{k_r}.
 \label{SI-eq:sstopmplus}
\end{equation}
Define the continuous slack-to-derivative ratio
\begin{equation}
 D_r=\frac{(1-r)(1+r^2)}{2(1+r)}.
 \label{SI-eq:sstopD}
\end{equation}
For $r<1$ this equals $(1-\phi)/a_r$, while $D_1=0$. If $m\ge0$, choose $\lambda=0$ and bound $m$ by $M_r$. If $m<0$, choose $\lambda=-m$; then $g+\lambda n=(b+(1-\phi)m,u)$ and $\lambda\sigma=-mI(1-\phi)$. Consequently,
\begin{equation}
 \begin{aligned}
 U_+&=\frac{(\varepsilon+\phi M_r)^2+M_r^2+(\varepsilon/t_r)^2}{2\mu},\\
 U_-&=\frac{(\varepsilon+\varepsilon D_r)^2+(\varepsilon/t_r)^2}{2\mu}
             +I\varepsilon D_r,\\
 \Delta(s)&\le\min\{U_1,\max(U_+,U_-)\}.
 \end{aligned}
 \label{SI-eq:sstoplocal}
\end{equation}
At $r=1$ the negative-$m$ residual needs no bound on $|m|$: its slack and radial residual are exactly zero. At $r<1/2$ only $U_1$ is used. These gap bounds use an upper bound on one physical curvature. Complete gradient recovery requires the signed derivative values specified in the main text.

\subsection{Uniform constants and the tolerance exponents}
Put $\delta=1-r$ and $K_0=\overline I L$. The following bounds hold:
\begin{equation}
 \begin{aligned}
 &t_r\ge1,\quad \phi\le1,\quad a_r\ge\delta &&(0\le r\le1),\\
 &a_r\le2\delta,\quad k_r\ge1 &&(1/2\le r\le1),\\
 &1-\phi=\delta^2/(1+r^2),\quad D_r\le\delta.
 \end{aligned}
 \label{SI-eq:sstopgeometrybounds}
\end{equation}
For example, $a_r\ge\delta$ follows from
\begin{equation*}
 2(1+r)-(1+r^2)^2=(1-r)(r^3+r^2+3r+1)\ge0.
\end{equation*}
For $r\ge1/2$, $a_r\le2\delta$ follows from $(1+r^2)^2-(1+r)=r(r^3+2r-1)\ge0$. Also
\begin{equation*}
 k_r-1=\frac{(1-r)(r^5+r^4+4r^3+8r^2+11r-1)}{(1+r^2)^3}\ge0
\end{equation*}
on that interval. The bound on $D_r$ is immediate; in fact $D_r\le\delta/2$ also holds, but is not needed for the stated constant.

Fix any $0<h\le1/2$. If $\delta\ge h$, first-order inversion gives $\|g\|^2\le3\varepsilon^2+3\varepsilon^2/a_r^2\le6\varepsilon^2/\delta^2$, so
\begin{equation}
 \Delta(s)\le\frac{3\varepsilon^2}{\mu h^2}.
 \label{SI-eq:sstopfar}
\end{equation}
If $\delta<h$, then $r>1/2$ and \eqref{SI-eq:sstopmplus} gives $m_+\le\nu+4K_0\delta^2$. For $m\ge0$, the zero-multiplier residual obeys
\begin{equation}
 \|g\|^2\le3\varepsilon^2+3m^2
 \le3\varepsilon^2+6\nu^2+96K_0^2\delta^4.
 \label{SI-eq:sstoppositive}
\end{equation}
For $m<0$ and $r<1$, $|m|(1-\phi)\le\varepsilon D_r\le\varepsilon\delta$. With $\lambda=-m$, this gives $\|g+\lambda n\|^2\le5\varepsilon^2$ and $\lambda\sigma\le\overline I\varepsilon\delta$. At the exact fold these claims hold because the corresponding slack terms vanish. Combining both signs,
\begin{equation}
 \Delta(s)\le\frac{5\varepsilon^2+6\nu^2+96K_0^2h^4}{2\mu}
                   +\overline I\varepsilon h.
 \label{SI-eq:sstopnear}
\end{equation}
For $0<\varepsilon\le K_0/8$, set $h=(\varepsilon/K_0)^{1/3}\le1/2$. Let $X=K_0^{2/3}\varepsilon^{4/3}/\mu$. Then $\varepsilon^2/\mu\le X$ and $\overline I\varepsilon h\le X$, the latter because $\mu\le L$. The far bound is $3X$ and the near bound is at most $(5/2+48+1)X+3\nu^2/\mu$. Thus 52 is a conservative common coefficient. When $\varepsilon=0$, first-order inversion gives $g=0$ away from the fold; at the fold the same multiplier argument yields the remaining $\nu^2$ bound.

\subsection{Noisy observations and native tolerances}
The main-text error inflation follows directly from the supplied absolute/norm budgets. In particular, $c_r\ge\widehat c_r-E_c\ge-I\nu$, including when the supplied curvature is positive. The proof applies to any valid supplied budgets. Establishing rounding-error budgets is a separate numerical task. Model constants and geometry must correspond to the objective actually optimized.

Suppose a native infinity-norm gradient test satisfies $\|\nabla_{a,w}\widetilde f\|_\infty\le t_g$, with $t_g\ge0$, and $I\ge I_{\min}>0$, $|I'|\ge J_{\min}>0$. Then
\begin{equation}
 \begin{aligned}
 \varepsilon_{\mathrm{actual}}&:=\max\{|b|,\|j\|_2/I\}\\
 &\le\varepsilon_{\mathrm{native}}
 :=\max\{t_g/J_{\min},\sqrt2\,t_g/I_{\min}\}.
 \end{aligned}
 \label{SI-eq:sstoprawconversion}
\end{equation}
Thus $\varepsilon=\varepsilon_{\mathrm{native}}$ is a valid tolerance value for the theorem. Similarly, a scalar-curvature test $c_r\ge-t_H$, with $t_H\ge0$, gives
\begin{equation*}
 \nu_{\mathrm{actual}}:=\max\{0,-c_r/I\}
 \le\nu_{\mathrm{native}}:=t_H/I_{\min}.
\end{equation*}
When curvature is required, use $\nu=\nu_{\mathrm{native}}$. The other theorem assumptions, including $\varepsilon_{\mathrm{native}}\le K_0/8$, must also hold. Supplied derivative errors require the error-inflation formulas in the main text. For an exponential intensity coordinate $J_{\min}=I_{\min}$. Native objective-reduction tolerances or successful line-search flags alone do not supply these derivative bounds.

For target gap $\tau$, the two main-text thresholds allocate half of $\tau$ to each term. The example in the code has $\mu=L=1/2$, $\overline I=8$, $\tau=10^{-6}$. Rounding its exact thresholds downward gives the sufficient choices $\varepsilon\le2.886\times10^{-7}$ and $\nu\le2.886\times10^{-4}$ in the stated normalized units. The pointwise bound can accept points rejected by the coarse uniform threshold.

\subsection{Sharp examples and failure of unrestricted transfer}
For the first tolerance take the single objective $f(s)=\tfrac12\|s-(1,e)\|^2$ with fixed unit $e$. Let $r\uparrow1$, $\phi=2r/(1+r^2)$, $I=(1+\phi)/(1+\phi^2)$. These intensities are bounded by two. Define $m=I\phi-1=(\phi-1)/(1+\phi^2)$. The intensity derivative satisfies $b=(I-1)+\phi m=0$. Exact formulas are
\begin{equation}
 \begin{aligned}
 \varepsilon&=a_r|m|
 =\frac{2(1-r)^3(1+r)}{(1+r^2)(r^4+6r^2+1)},\\
 \Delta(s)&=\frac{(1-r)^4}{2(r^4+6r^2+1)}.
 \end{aligned}
 \label{SI-eq:ssharpexact}
\end{equation}
Here $c_r/I=Ia_r^2+\kappa_rm\ge0$ for $0<r<1$: $I a_r^2\ge0$, $m<0$ and $\kappa_r<0$. Hence $\nu=0$, while $\varepsilon\sim(1-r)^3/4$ and $\Delta\sim(1-r)^4/16$. Any putative uniform upper bound $C\varepsilon^p$ with $p>4/3$ and fixed $C$ fails. The exact intensity adjustment is essential; keeping $I=1$ would leave a first intensity derivative of order $(1-r)^2$ and would not establish this claim.

For the curvature tolerance let $s=(1,e)$, $n=(-1,e)$, $0<\alpha<1$, and $f_\alpha(u)=\tfrac12\|u-z_\alpha\|^2$ with $z_\alpha=s-\alpha n$. The center $z_\alpha=(1+\alpha,(1-\alpha)e)$ is in $\mathcal C$. At the fold, $g=\alpha n$, $b=0$, $j=0$, $c_r=-\alpha$, and $\Delta=\alpha^2$. Thus the $\nu^2$ order cannot be improved. This is the Euclidean-metric version of the fold ambiguity family used earlier. It also proves that first-order tolerances alone cannot imply a vanishing gap at the fold.

For unbounded outer coordinates, use the fixed objective $f(s)=\tfrac12\|s-(1,1/2,0)\|^2$. Let $r\to\infty$, $e=(1,0)$ and $I=(1+\phi/2)/(1+\phi^2)$. Again $b=0$, while $m=I\phi-1/2\to-1/2$, $a_r\sim-2/r^2$, and $\kappa_r\sim4/r^3$. Therefore
\begin{equation}
 \varepsilon\sim r^{-2},\qquad\nu\sim2r^{-3},\qquad\Delta(s)\to1/8.
 \label{SI-eq:sstopoutercounter}
\end{equation}
No bound depending only on vanishing $\varepsilon,\nu$ can hold over both unrestricted sheets. Canonicalization changes the derivatives: the inner first derivative in this example does not remain small.

For intensity scaling, set $w=0$, $I=\exp a\downarrow0$ and $f(s)=\tfrac12\|s-(1,0,0)\|^2$. The raw intensity derivative is $I(I-1)$, its second derivative is $I(2I-1)$, and the polarization Hessian is $4I^2\operatorname{Id}_2$ with zero mixed block. Raw first derivatives and the negative part of the full raw Hessian tend to zero, but $\Delta\to1/2$. In contrast, $b=I-1$ does not vanish, so the normalized test rejects the sequence. This explains why a lower intensity/coordinate-derivative bound is needed when converting a raw solver tolerance.

\subsection{Relation to squared slack and KL results}
The scalar slack of the normalized polarization boundary is
\begin{equation}
 I-\|p\|=I\frac{(1-r)^2}{1+r^2}
       =\left(\frac{\sqrt I(1-r)}{\sqrt{1+r^2}}\right)^2.
 \label{SI-eq:sstopslacklink}
\end{equation}
Locally at a nonzero fold this is a squared slack in a regular radial coordinate. This identity connects the fold to squared-variable theory. The proof combines the corresponding chain-rule geometry with the standard dual-residual bound. Ding--Wright \cite[Thm.~3.6, Cor.~3.7]{supp-dingwright} already establish quantitative approximate second-order transfer with active-Jacobian rank conditions and conditioning-dependent constants. Their conclusion is an approximate weak second-order condition, whereas the present strongly convex specialization reports an explicit gap from a single radial lower test.

Ouyang \emph{et al.} \cite[Thm.~3.11, Rem.~3.12]{supp-ouyang} establish local KL exponents for Hadamard difference models under a KL assumption and an error bound, including exponent $3/4$ without strict complementarity. Thus a gradient-to-gap power $4/3$ is known in degenerate-contact problems. Here it is explicit on the stated domain, with uniform constants, a scalar curvature test and supplied derivative errors. Li \emph{et al.} \cite[Thm.~5, Lem.~6.1]{supp-hadamard} give an algorithmic gap guarantee assuming strict saddles, nondegenerate second-order KKT points and globally optimal local minima, and qualitative proximity of approximate stationary sets. Our physical bound assumes strong convexity, but neither strict saddles nor nondegeneracy or a known neighborhood of the optimum.

For a general nonconvex $C^2$ objective, the existing recovery-error bound can instead be combined with a standard projected-gradient residual. With the Euclidean projection $\Pi_{\mathcal C}$ in \eqref{SI-eq:euclideancone}, define $R_\gamma(g)=\gamma^{-1}\|s-\Pi_{\mathcal C}(s-\gamma g)\|$ for $\gamma>0$. Euclidean nonexpansiveness gives $R_\gamma(g)\le R_\gamma(\widehat g)+\|g-\widehat g\|$. This standard inequality gives an approximate stationarity statement for the nonconvex case.

\subsection{Fixed verification and reproduction}
The fixed grid has three SPD metrics, intensities $0.02,1,4$, radii $0,0.25,0.5,0.75,0.9,0.99,0.999,1$, two directions and four gradient-center choices, giving 576 quadratics. Twenty-four additional strongly convex quartics use the ideal metric plus $0.1(v^{\mathsf T}(s-z))^4/4$, $v=(1,0.3,-0.2)$, with the same intensities, four radii and two directions. Their Hessian's actual polarization upper eigenvalue supplies a valid pointwise $L$. Supplied derivative perturbations have budgets zero, $10^{-8}$ or $10^{-4}$. Every result is retained; no configuration is added in response to its outcome. SLSQP supplies independent physical minima, with 600 native successes and maximum feasibility violation $9.64\times10^{-14}$. Computed gap-minus-bound residuals reach $1.24\times10^{-14}$, consistent with that reference accuracy; these are not symbolic proofs. Independent PyTorch derivatives check all 600 objectives, including the prescribed radial directions at the origin, where the coordinate alone does not encode that direction.

Ninety-digit decimal calculations retain 12 first-tolerance and 12 curvature-tolerance examples, eight unbounded-outer examples and 16 vanishing-intensity examples. The last first-tolerance ratio is $0.3968502629915207$, approaching $2^{-4/3}=0.3968502629920499$. The constructions and their limiting calculations above provide the proofs; finite precision checks do not establish the exponents on their own.

The retrospective endpoint check uses all 9\,284 previously stored real-data warm endpoints. Exactly 463 are switched inward, preserving their outputs. Derivative tests are recomputed at the inner coordinates; $\mu=L=1/2$, $\overline I=8$ and zero supplied derivative-error budgets are used. All recorded intensities satisfy that domain. The pointwise bound accepts 8\,861 endpoints and rejects all 295 historical false successes, plus 128 accurate endpoints. The coarse uniform bound alone accepts 980. This check makes no new optimizer calls but acquires 9\,284 full head Hessians, including points where the theorem would not require one. The implementation prioritizes complete coverage of the stored endpoints over monitor cost. Validated rounding-error budgets and evaluation on unseen observations remain outside this retrospective check.

Sources, fixed protocols, all numeric records, a standalone certificate implementation, an independent AD audit and replay scripts are in \nolinkurl{theory/stopping_guarantee/}. Prior experiments and their files remain unchanged.

\section{Outward evaluation of the nominal stopping bound}
\label{SI-sec:svalidated}
Section S20 gives bounds in real arithmetic. Here we evaluate its pointwise expression with outward intervals for the nominal ideal four-analyzer quadratic. This adds numerical enclosure of the intensity, coordinate geometry, and derivative tests. The model remains $f(s)=\tfrac12\|As-y\|^2$, with the ideal $A$ in Section S1 and $\mu=L=1/2$. Each stored binary64 coordinate and observation denotes its exact real value. The quantity certified is the optimization gap for that model and those inputs.

\subsection{Arithmetic and input semantics}
The interval layer follows the standard inclusion principle \cite[Secs.~5.3--5.5]{supp-rump}. For $p$ fractional bits, endpoints are integers divided by $Q_p=2^p$. Additions and subtractions use exact integer arithmetic. Multiplication and division round the lower endpoint downward and the upper endpoint upward using integer floor and ceiling division. Division by an interval containing zero returns an inconclusive result. Square roots use integer square roots of the scaled endpoints, increasing the upper root by one exactly when it is not a perfect square.

The exponential uses a separate enclosure rather than a library approximation. For an endpoint $x\in[-64,64]$, range reduction gives $t=|x|/2^k\le1/8$. With $N=48$, positivity and a geometric bound on the remaining term ratios give
\begin{equation}
 \begin{aligned}
 S_N(t)&=\sum_{j=0}^{N}\frac{t^j}{j!},\\
 S_N(t)&\le e^t\le S_N(t)+
 \frac{t^{N+1}}{(N+1)!}\frac{1}{1-t/(N+2)}.
 \end{aligned}
 \label{SI-eq:intervalexp}
\end{equation}
The sum and remainder are enclosed by integer interval operations. Repeated squaring reverses the range reduction; reciprocation handles negative endpoints. Monotonicity supplies the enclosure for an interval argument, including one that crosses zero. Arguments outside the supported range return inconclusive. All operations determining acceptance use integers; conversion to a decimal or binary64 display occurs afterward.

For an outer input, the implementation uses the exact inner preimage $C(w)=w/\|w\|^2$ algebraically. It evaluates the physical state from the original coordinates, and the derivative tests at this inner preimage. The identity $F(C(w))=F(w)$ preserves the exact physical output. The original squared radius is compared with one using exact rational arithmetic. Thus the inner radius is known to lie in $[0,1]$, and intersecting its computed enclosure with this interval is justified by the map identity. The same reasoning permits the nonnegative support-slack ratio in Section S20. No rounded optimizer restart is introduced.

\subsection{Correctness of an accepted bound}
\begin{proposition}[Validated nominal stopping]
\label{SI-prop:validatedstop}
For finite binary inputs $(a,w,y)$, suppose the supplied interval monitor with target tolerance $10^{-6}$ reports a certified result. Then the mathematical mapped state $s=T_F(a,w)$ satisfies
\begin{equation}
 f(s)-\min_{u\in\mathcal C}f(u)\le10^{-6}.
 \label{SI-eq:validatedstop}
\end{equation}
\end{proposition}
\begin{proof}
The input conversion encloses the exact binary values. Each elementary operation preserves containment by outward integer rounding. Equation~\eqref{SI-eq:intervalexp} encloses $I=e^a$; a successful evaluation also verifies $0<I\le8$. Exact inner-preimage selection gives the domain of the Section S20 pointwise bound. For the ideal quadratic, $e^{\mathsf T}H_{pp}e=\mu=1/2$ exactly.

The implementation encloses $b$, $\|j\|/I$, and $c_r/I$, and hence the sufficient $\varepsilon$ and $\nu$. It evaluates the first-order and one-sided-curvature expressions in \eqref{SI-eq:sstopfirst} and \eqref{SI-eq:sstoplocal} with intervals, including all geometry operations. Minima and maxima preserve containment of the corresponding real values. At the origin, $DF=2\operatorname{Id}_2$ supplies the first-order expression directly. Every accepted bound therefore encloses a valid upper bound on the real gap. If its upper endpoint is $U/Q_p$, the final comparison is the exact integer test $10^6U\le Q_p$, proving \eqref{SI-eq:validatedstop}. An unresolved denominator or domain check returns inconclusive rather than accepting.
\end{proof}

The implementation evaluates a single scalar radial curvature when $r\ge1/2$. It obtains the derivative enclosures analytically from the nominal model and forms no full head Hessian. This model-specific route also has access to physical gradients; the direct physical controls remain applicable.

\subsection{Retrospective verification and precision}
The protocol fixes fractional precisions $32,64,128,256$ and all 9\,284 stored warm endpoints before this arithmetic comparison. States, observations, histories, and optimizer results are unchanged. A separate outward implementation of the direct cone reduction at 256 bits classifies 8\,989 endpoints below the gap threshold and 295 above it, with no threshold-overlap interval. Each accepted stopping result is checked against this independent cone expression.

\begin{table}[!t]
\caption{Outward pointwise stopping with the indicated number of fractional bits on the same 9\,284 endpoints. ``Accurate rejected'' includes inconclusive results. All four precisions accept zero inaccurate endpoints. Times describe one CPU pass; the separate cone-reference calculation takes 7.69 s.}
\label{SI-tab:validatedstop}
\centering\small
\begin{tabular}{@{}rrrrr@{}}\toprule
Bits & Certified & Inconcl. & Accurate rej. & Time (s)\\\midrule
32  & 8\,858 & 3 & 131 & 6.21\\
64  & 8\,861 & 0 & 128 & 6.38\\
128 & 8\,861 & 0 & 128 & 7.32\\
256 & 8\,861 & 0 & 128 & 8.73\\\bottomrule
\end{tabular}
\end{table}

Table~\ref{SI-tab:validatedstop} shows identical decisions at 64--256 fractional bits and agreement with the earlier float64 pointwise decisions. All 295 inaccurate endpoints are rejected; 128 accurate endpoints also remain rejected. At 32 bits, three additional accurate endpoints are inconclusive because their radius enclosures include zero in a required denominator. Their true inner radii are approximately $9.84\times10^{-6}$, $2.05\times10^{-6}$, and $6.67\times10^{-6}$. They are retained in the denominator and the reported rejection count.

At every precision, the monitor makes 52 scalar curvature evaluations and forms zero full head Hessians. The previous coverage implementation acquired 9\,284 full head Hessians. The new accounting concerns this analytic nominal-model implementation; Table~\ref{SI-tab:validatedstop} reports its interval cost separately.

Independent exact-rational checks verify the arithmetic, using a direct Taylor-series exponential oracle at the original argument rather than the implementation's range reduction. Additional Cartesian Jacobian/Hessian checks at exact $I=1$ verify 196 radial-tolerance enclosures on 49 inputs. The frozen stress suite has 43 inputs and 172 precision-specific checks, including exact folds, nearby inner and outer states, low intensity, and unsupported inputs. Unsupported inputs remain inconclusive. The mathematical enclosure argument above supplies the acceptance implication; these checks test its implementation.

\subsection{Reproduction}
The implementation requires only the Python standard library. From the source-package root, run \nolinkurl{theory/validated_stopping/replay.py} with \nolinkurl{--output} naming a new directory. The script reproduces the records and compares their integer endpoints exactly. A minimal folded optimum for \nolinkurl{cli.py} uses raw coordinates $(0,1,0)$ and observations $(1,1/2,0,1/2)$. The README gives complete one-line commands.

The source protocol, integer bound endpoints, independent checks, all four precision runs, and replay comparisons are retained in \nolinkurl{theory/validated_stopping/}. Applying the method to uncertain instrument parameters requires enclosures for that expanded physical model.

\clearpage
\section{Additional comparison and audit tables}
\label{SI-sec:extra_tables}

\begin{table}[H]
\caption{Relation to prior results. Locators for \cite{supp-lifts} refer to arXiv:2207.03512v5.}
\label{SI-tab:priorart}
\centering\small\renewcommand{\arraystretch}{1.12}
\begin{tabularx}{\textwidth}{@{}p{.23\textwidth}p{.34\textwidth}X@{}}\toprule
Prior statement & Established output & Relation to this paper\\\midrule
Wang \emph{et al.}, V-B, (78)--(79) \cite{supp-gcopter} & Rational ball map and propagated first gradient & We evaluate inverse derivative observations for its Stokes specialization.\\
Levin \emph{et al.}, Thm.~2.4 \cite{supp-lifts} & First-critical transfer iff the differential image equals the target tangent cone & Our question reconstructs the physical gradient from signed derivative values, including nonstationary points.\\
Levin \emph{et al.}, Lem.~3.24 and Prop.~3.26 \cite{supp-lifts} & Second-output derivatives modulo the first differential image; kernel-curvature descriptions & Our observation lemma uses the same differential calculus; the Stokes inverse makes its coefficients explicit.\\
Levin \emph{et al.}, Thm.~3.23 \cite{supp-lifts} & Necessary and sufficient conditions for second-critical to stationary transfer & We use signed derivative observations to recover gradient values and propagate input errors.\\\bottomrule
\end{tabularx}
\end{table}

\begin{table}[H]
\caption{Comparison for finite-tolerance stopping. PDF theorem locators for the three external preprints use arXiv:2310.01784v2, 2402.00377v2, and 2112.05273v2, respectively. The columns identify the feasible sets, assumptions, and conclusions relevant to the Stokes specialization.}
\label{SI-tab:stoppingprior}
\centering\small\renewcommand{\arraystretch}{1.12}
\begin{tabularx}{\textwidth}{@{}p{.24\textwidth}p{.37\textwidth}X@{}}\toprule
Prior result & Established guarantee & Present specialization\\\midrule
Ding--Wright, Thm.~3.6, Cor.~3.7 \cite{supp-dingwright} & Approximate squared-slack second-order conditions transfer to approximate weak second-order conditions, with active-Jacobian rank and conditioning dependencies. & Direct gap bound for strongly convex Stokes objectives, with one radial curvature and explicit constants.\\
Ouyang \emph{et al.}, Thm.~3.11, Rem.~3.12 \cite{supp-ouyang} & Local KL exponents for Hadamard difference models; exponent $3/4$ occurs without strict complementarity under error-bound assumptions. & Uniform derivative-tolerance bound on a stated chart domain; $4/3$ is consistent with this known exponent.\\
Li \emph{et al.}, Thm.~5, Lem.~6.1 \cite{supp-hadamard} & Algorithmic gap bound: strict saddles, nondegenerate second-order KKT points, all local minima global. Compactness gives qualitative proximity. & Deterministic stopping bound for strongly convex physical objectives; no strict-saddle or nondegeneracy requirements.\\\bottomrule
\end{tabularx}
\end{table}

\begin{table}[H]
\caption{Common-graph AD: median microseconds at $I=1.2$, $r=0.999$, float64; maximum Euclidean gradient error over the intensity/radius grid for each precision. Four exact-fold first-inverse cases are singular and omitted from error maxima.}
\label{SI-tab:adcost}
\centering\small
\begin{tabular}{@{}lrrr@{}}\toprule
Route & $\mu$s & Error32 & Error64\\\midrule
Direct physical & 62.9 & $1.69\!\times\!10^{-7}$ & $5.00\!\times\!10^{-16}$\\
First inverse & 114.4 & $1.27\!\times\!10^{0}$ & $1.67\!\times\!10^{-9}$\\
Vector HVP & 274.5 & $7.07\!\times\!10^{-7}$ & $9.69\!\times\!10^{-16}$\\
Mixed scalar & 277.6 & $6.75\!\times\!10^{-7}$ & $9.99\!\times\!10^{-16}$\\
\bottomrule
\end{tabular}
\end{table}

\begin{table}[H]
\caption{Zero angle change and tolerance $10^{-6}$. Every row uses the same 60 conditions. A pair is false successes / nonacceptances; the two categories are disjoint. PyTorch acceptance uses the operational gradient test.}
\label{SI-tab:warm}
\centering\small
\begin{tabular}{@{}lrrr@{}}\toprule
Solver & Baseline & Always & Triggered\\\midrule
BFGS, float64 & 12 / 1 & 0 / 0 & 0 / 1\\
L-BFGS-B, float64 & 16 / 0 & 0 / 0 & 0 / 0\\
Trust region, float64 & 0 / 0 & 0 / 0 & 0 / 0\\
PyTorch LBFGS, float32 & 9 / 17 & 0 / 7 & 0 / 7\\
PyTorch LBFGS, float64 & 14 / 0 & 0 / 0 & 0 / 0\\\bottomrule
\end{tabular}
\end{table}

\begin{table}[H]
\caption{Released-data stopping audit. Every count pair is BFGS/L-BFGS-B. A false success is native acceptance with nominal quadratic gap above $10^{-6}$. Counts combine the separately reported sampling strata, not independent scenes. ``Detected'' and ``rejected'' refer to audits of the uncorrected warm-start endpoints. All native failure counts are zero.}
\label{SI-tab:realdata}
\centering\small\renewcommand{\arraystretch}{1.15}
\begin{tabularx}{\textwidth}{@{}Xrrrrr@{}}\toprule
Subset & Entries & Cold false & Warm false & Local detected & Physical rejected\\\midrule
DeepSfP & 576 & 0/0 & 0/0 & 0/0 & 0/0\\
PolarNS means & 384 & 0/7 & 111/64 & 0/0 & 111/64\\
PolarBurstSR & 3\,682 & 0/7 & 16/104 & 0/0 & 16/104\\\bottomrule
\end{tabularx}
\end{table}

\FloatBarrier
\section{Additional empirical controls}
\label{SI-sec:empiricalcontrols}
\subsection{Shared-network fitting and fold incidence}
We fit noisy four-analyzer observations with a shared MLP: four input measurements, two width-32 hidden layers with SiLU activation $x/(1+e^{-x})$, and three head coordinates. The two $32\times32$ fields are the saved weak-polarization renderer reference and a synthetic field with DoLP $0.90$--$0.99$. Each run uses all 1,024 pixels, Gaussian analyzer noise with standard deviation $0.01$, float32 Adam, learning rate $0.003$, and 1,500 updates. Five seeds and two head initializations give 20 fixed runs. These are fixed fitting settings; Supplementary Section S15 specifies the initialization and diagnostics.

We evaluate each unaveraged pixel objective in independent head coordinates at updates $0,100,500,1000,1500$. A candidate requires $|r-1|\leq10^{-3}$ and normalized radial curvature $c_r/I^2<-10^{-7}$. The first threshold specifies a raw-radius neighborhood; the second excludes curvature values within $10^{-7}$ of zero. There are no such candidates at any saved checkpoint. Two high-DoLP pixels are within the radius threshold at update 500, both with nonnegative radial curvature. The closest final distances are $0.957$ for the renderer field and $0.0586$ for the high-DoLP field. The smaller radius thresholds $10^{-4}$ and $10^{-5}$ also give zero candidates.

\begin{figure}[H]
\centering\includegraphics[width=\textwidth]{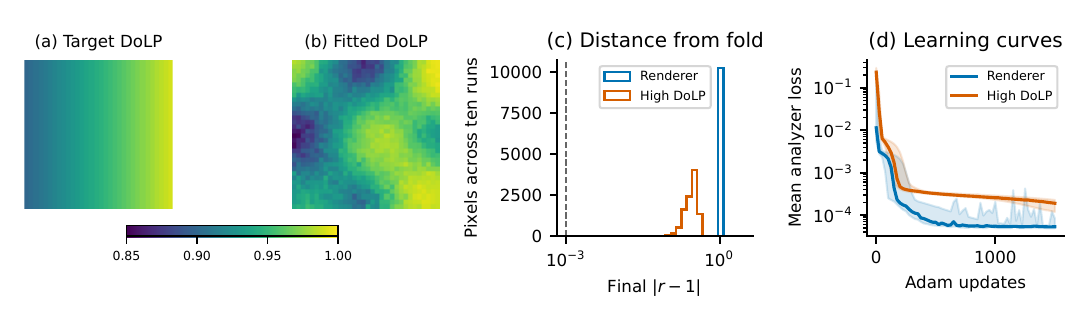}
\caption{Shared-network fitting in float32. (a)--(b) Target and fitted DoLP for the preselected high-DoLP field, standard initialization, seed 0. (c) Final radius-distance histograms over ten runs per field. The dashed line marks the $10^{-3}$ candidate radius threshold. (d) Median mean analyzer losses, with the full range across five seeds and two initializations. No saved checkpoint contains a near-fold negative-curvature candidate under the stated thresholds. These are single-field fits; the pixels are not independent experimental replications.}
\label{SI-fig:training}
\end{figure}

Figure~\ref{SI-fig:training} shows a fixed example, final radius distances, and learning curves. Independent-coordinate perturbations and a constructed stationary-fold control check the diagnostic and quadratic bound separately. The control is excluded from the incidence count. Twenty stored networks reproduce their saved predictions exactly; an independent AD check of 80 pixel Hessian contractions confirms the reported radial curvatures.

\FloatBarrier
\subsection{Complete coordinate-repair comparison}
The following table preserves the strategy and query accounting behind the compact fixed-endpoint versus evolving-history comparison in the main text. These interventions use the existing observations and the fixed study protocols.
\begin{table}[H]
\caption{Frozen coordinate-repair follow-up. Inaccurate means nominal gap above $10^{-6}$, irrespective of native or monitored acceptance. Count pairs are BFGS/L-BFGS-B, each with 4\,642 released-data entries and 984 Fresnel frames. Audited arms reject any remaining inaccurate outputs. Physical controls are single algorithms, without a head-solver pair. FG counts exclude separately recorded physical audits and Hessians.}
\label{SI-tab:chartrepairdetail}
\centering\small\renewcommand{\arraystretch}{1.12}
\begin{tabularx}{\textwidth}{@{}Xrrr@{}}\toprule
Strategy & Released inaccurate & Fresnel inaccurate & Released native FG pairs\\\midrule
Raw warm start & 127/168 & 9/7 & 92\,025/32\,235\\
Cold start & 0/14 & 0/0 & 97\,801/29\,468\\
Switch before each warm solve & 9/16 & 12/10 & 94\,974/32\,527\\
Audited cold restart & 0/2 & 0/0 & 95\,791/33\,232\\
Audited coordinate repair & 7/22 & 0/0 & 95\,640/33\,440\\
Coordinate repair with physical fallback & 0/0 & 0/0 & 95\,635/33\,245\\
Physical projected gradient & 0 & 0 & ---\\
Direct cone solution & 0 & 0 & ---\\\bottomrule
\end{tabularx}
\end{table}
\FloatBarrier

\FloatBarrier

\end{document}